\documentclass[final,5p,times,twocolumn]{elsarticle}
\usepackage{amssymb}
\usepackage{amsmath}
\usepackage{url}
\usepackage{hyperref} 

\usepackage{graphicx}

\usepackage{morefloats}

\usepackage{color} 
\usepackage{soul}

\def\units#1{~\hbox{$\,{\rm #1}$}}

\journal{Astroparticle Physics}

\begin{document}

\begin{frontmatter}


\title{Production of secondary particles and nuclei in cosmic rays collisions with the interstellar gas using the {\tt FLUKA} code}

\author[ad1]{M.~N.~Mazziotta\corref{cor1}}
\ead{mazziotta@ba.infn.it}
\author[ad2]{F.~Cerutti}
\author[ad2]{A.~Ferrari}
\author[ad3,ad4]{D.~Gaggero\fnref{label2}}
\author[ad1,ad5]{F.~Loparco}
\author[ad6]{P.~R.~Sala}

\cortext[cor1]{Corresponding author}

\address[ad1]{Istituto Nazionale di Fisica Nucleare, Sezione di Bari, 70126 Bari, Italy}
\address[ad2]{CERN, Geneva, Switzerland}
\address[ad3]{SISSA, via Bonomea 265, 34136 Trieste, Italy}
\address[ad4]{INFN, Sezione di Trieste, via Valerio 2, 34127 Trieste, Italy}
\address[ad5]{Dipartimento di Fisica ``M. Merlin" dell'Universit\`a e del Politecnico di Bari, I-70126 Bari, Italy}
\address[ad6]{Istituto Nazionale di Fisica Nucleare, Sezione di Milano, 20133 Milano, Italy }

\fntext[label2]{Current address: GRAPPA Institute, University of Amsterdam, Science Park 904, 1090 GL Amsterdam, The Netherlands}

\begin{abstract}

The measured fluxes of secondary particles produced by the interactions of Cosmic Rays (CRs)
with the astronomical environment play a crucial role in understanding the physics of CR transport.
In this work we present a comprehensive calculation of the secondary hadron, lepton, gamma-ray and neutrino 
yields produced by the inelastic interactions between several species of stable or long-lived cosmic rays projectiles 
($p$, $D$, $T$, $^{3}He$, $^{4}He$, $^{6}Li$, $^{7}Li$, $^{9}Be$, $^{10}Be$, $^{10}B$, $^{11}B$, $^{12}C$, $^{13}C$, $
^{14}C$, $^{14}N$, $^{15}N$, $^{16}O$, $^{17}O$, $^{18}O$, $^{20}Ne$, $^{24}Mg$ and $^{28}Si$)
and different target gas nuclei ($p$, $^{4}He$, $^{12}C$, $^{14}N$, $^{16}O$, $^{20}Ne$, $^{24}Mg$, $^{28}Si$ and $^{40}Ar$). 
The yields are calculated using {\tt FLUKA}, a simulation package designed to compute the energy distributions of secondary 
products with large accuracy in a wide energy range.
The present results provide, for the first time, a complete and self-consistent set of all the relevant inclusive cross sections regarding 
the whole spectrum of secondary products in nuclear collisions. We cover, for the projectiles, a kinetic energy range extending 
from $0.1~GeV/n$ up to $100~TeV/n$ in the lab frame. 
In order to show the importance of our results for multi-messenger studies about the physics of CR propagation, we evaluate the propagated spectra 
of Galactic secondary nuclei, leptons, and gamma rays produced by the interactions of CRs with the insterstellar gas, exploiting the numerical 
codes \texttt{DRAGON} and \texttt{GammaSky}. 
We show that, adopting our cross section database, we are able to provide a good fit of a complete sample of CR observables, including: 
leptonic and hadronic spectra measured at Earth, the local interstellar spectra measured by Voyager, and the gamma-ray emissivities from Fermi-LAT collaboration. 
We also show a set of gamma-ray and neutrino full-sky maps and spectra.

\end{abstract}



\end{frontmatter}


\section{Introduction}

According to the currently accepted scenario, Galactic Cosmic Rays (CRs) are accelerated in Supernova Remnants (SNRs) 
and interact with the turbulent interstellar magnetic field: the resulting motion is well described by a diffusion equation. 
During their journey through the Galaxy, the inelastic collisions of hadronic CRs with the interstellar medium (ISM), 
produce lighter particles and secondary radiation: the study of these events 
is very important since it may shed light on the origin of the CRs themselves and on the mechanisms governing their transport.

For example, the production of light nuclei -- such as Boron -- from heavier ones (in particular Carbon and Nitrogen) has been 
extensively studied in recent times, since light nuclei ratios (e.g. B/C, N/O) are often used to constrain the propagation 
models in the Galaxy and in particular the rigidity dependence and normalization of the diffusion coefficient 
(see for instance~\cite{Putze:2010zn,DiBernardo:2009ku,Trotta:2010mx}).

Moreover, the collisions of protons and Helium nuclei with the gas, and subsequent decays of the produced neutral pions ($\pi^0$s), 
are expected to give the most relevant contribution to the gamma-ray diffuse emission in the Galactic plane, since that is the 
region with the largest gas column densities. 
In this context, the high precision gamma-ray maps and spectra measured by the Fermi-LAT instrument in the energy band spanning 
from several tens of~\units{MeV} to several hundreds of~\units{GeV}~\cite{Atwood2009}, as well as the data in the~\units{TeV} range 
from ground Air Cerenkov Telescopes~\cite{hess,veritas,magic,cta}, are very useful and represent a unique opportunity to understand 
the CR transport properties in different regions of the Galaxy~\cite{FermiLAT:2012aa,gradient,variabledelta}.
 
Concerning both the study of gamma-ray point sources and the diffuse emission, the radiation in the $\units{MeV}-\units{TeV}$ energy 
range can also be produced by electron bremsstrahlung and inverse Compton (IC) scattering. The relative contributions of leptonic and hadronic 
processes depend on many parameters (i.e. the environment around the source), often resulting 
into considerable uncertainties in the physical interpretation of the observed gamma-ray emission. For this reason, an accurate 
knowledge of the gamma-ray spectra resulting from hadronic interactions is crucial for the physical interpretation of the emission phenomena. 

Gamma rays are also produced in the interactions of cosmic-ray nuclei with the Earth's atmosphere. 
The Fermi collaboration recently inferred the proton spectrum from a measurement of the spectrum of
gamma rays originated from the Earth limb (i.e. the top of the atmosphere)~\cite{Ackermann:2014ula}. 
This measurement allowed the reconstruction of the CR proton spectrum  
in an energy range from $90\units{GeV}$ up to $6\units{TeV}$. The study of the CR
spectra in this region is a topic of extreme scientific interest, since the 
PAMELA collaboration reported a hardering of the proton and helium spectra around $200\units{GeV}$~\cite{Adriani:2011cu} 
that has been recently confirmed by the AMS02 collaboration \cite{Aguilar:2015ooa,ams02he}.
Clearly, also for such analysis, the precise knowledge of the gamma-ray yields produced in 
the hadronic interactions of CRs with atmospheric nuclei is very crucial.

Charged leptons and antiprotons are another important product of the inelastic hadronic collisions
of CRs with the interstellar gas. The PAMELA collaboration 
measured the positron, electron and antiproton spectra, as well as the spectra of many light
nuclei~\cite{Adriani:2008zr,Adriani:2013uda,Adriani:2013as,Adriani:2013tif,Adriani:2011cu,Adriani:2011xv,Adriani:2010rc}. 
Recently the AMS02 collaboration provided very accurate results on the positron fraction 
(i.e. the ratio $e^+/(e^- + e^+)$)~\cite{Aguilar:2013qda,Accardo:2014lma} and new 
high-precision results on the positron and electron intensities ~\cite{Aguilar:2014mma}. 
They also presented preliminary results on the 
B/C ratio \cite{ams02_icrc2013,ams02_cern2015}. All those datasets were taken during the same period, 
and this circumstance makes the interpretation of the data easier. Simultaneous 
measurements of several particle spectra performed by the same experiment over a wide energy 
range will in fact ensure reduced experimental systematics and will also limit the uncertainties 
arising from the CR propagation in the heliosphere (solar modulation) and in the Galaxy. 

Since the hadronic interactions (in particular the $p-p$ interaction) 
are the main processes responsible of the production of secondary particles 
such as gamma rays, neutrinos, electrons and positrons, and since the decays of pions 
and kaons play a major role in these processes, several parameterizations were developed 
over the years to describe the production of these mesons in 
$p-p$ interactions (see for instance~\cite{stecker1970,dermer1986,moska1998}).
These calculations are mainly based on the inclusive cross section of pion production in $p-p$ collisions,
evaluated from the accelerator data. However, the models may lose accuracy in the high-energy region, where 
experimental data have established a logarithmic increase of the total inelastic cross section with the incident 
proton energy~\cite{pdg}. 

In addition, the contribution of the gamma-ray production from the decays of 
other particles than neutral pions was also found to be not negligible.
Finally, for low-energy primaries ($<10\units{GeV}$), 
the contribution of resonances to the secondary production cannot be neglected.

In addition to the theoretical methods, many parameterizations based on Monte Carlo (MC) 
calculations were proposed~\cite{huang2007,mori1997,kamae2006,kelner2006,mori2009,kache2012}.
These calculations were performed using event generators developed for the high energy particle 
physics (e.g. {\tt PYTHIA}~\cite{pythia}, {\tt DPMJET-III}~\cite{dpmjetfluka}, {\tt SYBILL}~\cite{sybill}, {\tt QGSJET-II}~\cite{qgsjet}). 
The main advantage of the MC approach is that the codes can be used directly to calculate the 
spectra of all secondary species. The MC approach also allows a direct evaluation of the secondary 
particle yields in nucleus-nucleus interactions, instead of using parameterizations and/or scaling 
factors based only on the $p-p$ interactions (see for instance~\cite{mori2009}). 
However, the event generators included in most MC codes are reliable only above 
a few $10\units{GeV/n}$, and parameterizations need to be implemented to describe the 
low energy region~\cite{huang2007,kamae2006,kache2012}. 

A common practice in the literature is to fit different observables (e.g. hadronic and leptonic spectra) using
cross sections obtained from several parameterizations, derived with different methods and under different assumpions: this may
lead to inconsistencies in the determination of CR diffusion models.
In the present work we aim at providing a complete and consistent set of cross sections for the secondary production
of hadrons and leptons. We perform our study with the {\tt FLUKA} MC simulation code~\cite{fluka2,flukaweb,bohlen,batt2015}.

The paper is organized as follows:
first we describe the main features of {\tt FLUKA}, then
we present and discuss the main results regarding the secondary production cross sections for the interactions 
between several species of CR projectiles 
and different target nuclei.
Finally, we discuss the impact of our results on CR propagation models, using a custom version of the propagation code {\tt DRAGON}. 
In particular, we present a set of three models -- adopting the 
new cross sections -- tuned on a complete set of CR observables in a wide energy range.

\section{Monte Carlo simulations with {\tt FLUKA}}

{\tt FLUKA} is a general purpose MC code for the simulation of hadronic and electromagnetic interactions. 
It is used in many applications, and is continuously checked using the available data from low energy nuclear physics, high-energy 
accelerator experiments and measurements of particle fluxes in the atmosphere. 
Hadronic interactions are treated in {\tt FLUKA} following a theory-driven approach. The general phenomenology 
is obtained from a microscopical description of the interactions between the fundamental constituents, 
appropriate for the different energy regions. 
Below a few $\units{GeV}$, hadron-nucleon interaction model is based on resonance 
production and decay of particles, while for higher energies the Dual Parton Model (DPM) is used,
implying a treatment in terms of quark chain formation and hadronization. 
The extension from hadron-nucleon to hadron-nucleus interactions is done in the framework of
the PreEquilibrium Approach to NUclear Thermalization model ({\tt PEANUT})~\cite{peanut1,peanut2},
including the Gribov-Glauber multi-collision mechanism followed by the pre-equilibrium stage and eventually
equilibrium processes (evaporation, fission, Fermi break-up and gamma deexcitation). The models 
used in {\tt FLUKA} are benchmarked against the available data from experiments. More details 
about the {\tt FLUKA} package can be found in the manual~\cite{fluka2,flukaweb} and a description of hadronic interactions
model used in {\tt FLUKA} can be found in Ref.~\cite{ferrari1996}.

{\begin{figure}[ht!]
\includegraphics[width=1\columnwidth,height=0.25\textheight,clip]{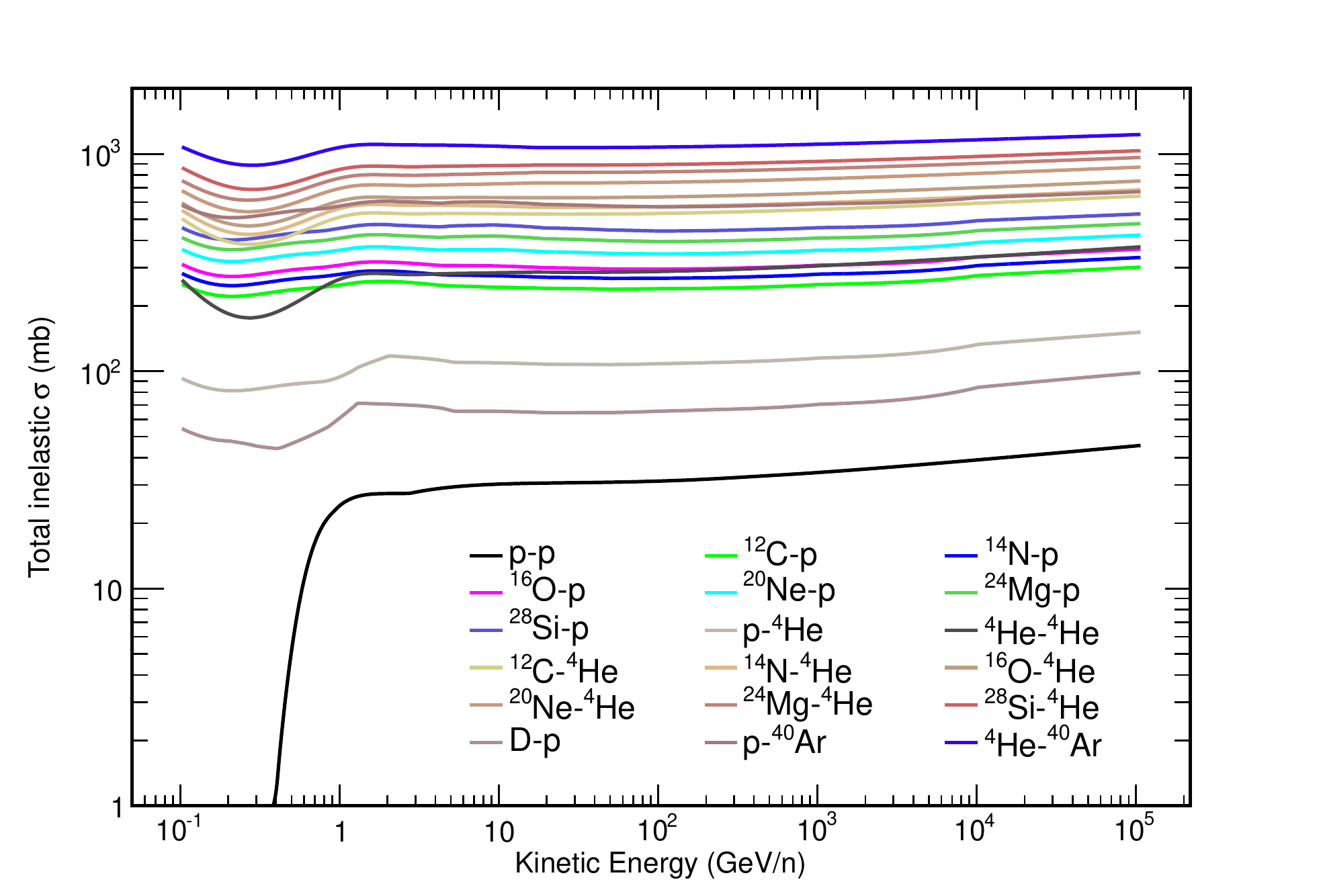}
\caption{Total inelastic cross sections as a function of the energy per nucleon of the incoming projectile.
The plot shows the cross sections for all the projectile-target pairs studied in the present work.}
\label{FigXsecInel}
\end{figure}

{\tt FLUKA} can simulate with high accuracy the interaction and propagation in matter of about 60 different 
species of particles, including photons and electrons from $1\units{keV}$ to thousands of $\units{TeV}$, 
neutrinos, muons of any energy, hadrons of energies up to $20\units{TeV}$ (up to $10\units{PeV}$ when 
it is interfaced with the {\tt DPMJET} code~\cite{dpmjetfluka}) and all the corresponding antiparticles, neutrons down to 
thermal energies and heavy ions. 

In addition, {\tt FLUKA} can handle even very complex geometries, using an 
improved version of the well known Combinatorial Geometry (CG) package. The {\tt FLUKA} CG has been designed 
to track correctly also charged particles, even in the presence of magnetic fields.
The predictions of {\tt FLUKA} have been checked with a large set of experimental data collected in accelerator 
experiments. Few examples of applications in CR physics can be found 
in~\cite{batt2002,batt2003,batt2005}.

\begin{figure}[hb!]
\includegraphics[width=1\columnwidth,height=0.25\textheight,clip]{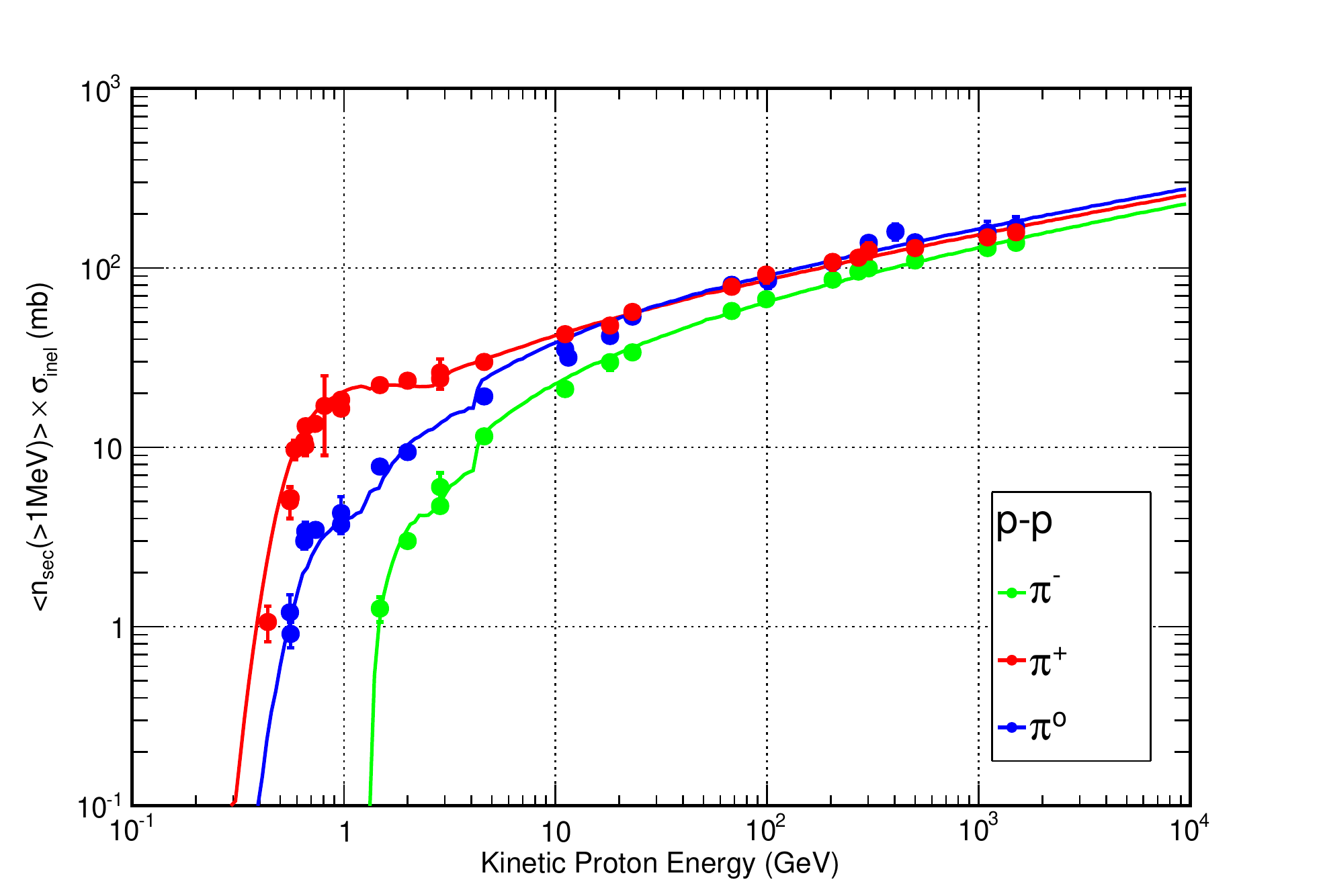}
\caption{Inclusive cross sections for the production of $\pi^0$ (blue), $\pi^{+}$ (red) and $\pi^{-}$ (green)
in $p-p$ collision as function of the incoming proton kinetic energy. 
Lines: {\tt FLUKA} simulation; points: data from Ref.~\cite{dermer1986}.}
\label{FigXsecPion}
\end{figure}

\begin{figure}[hb!]
\includegraphics[width=1\columnwidth,height=0.25\textheight,clip]{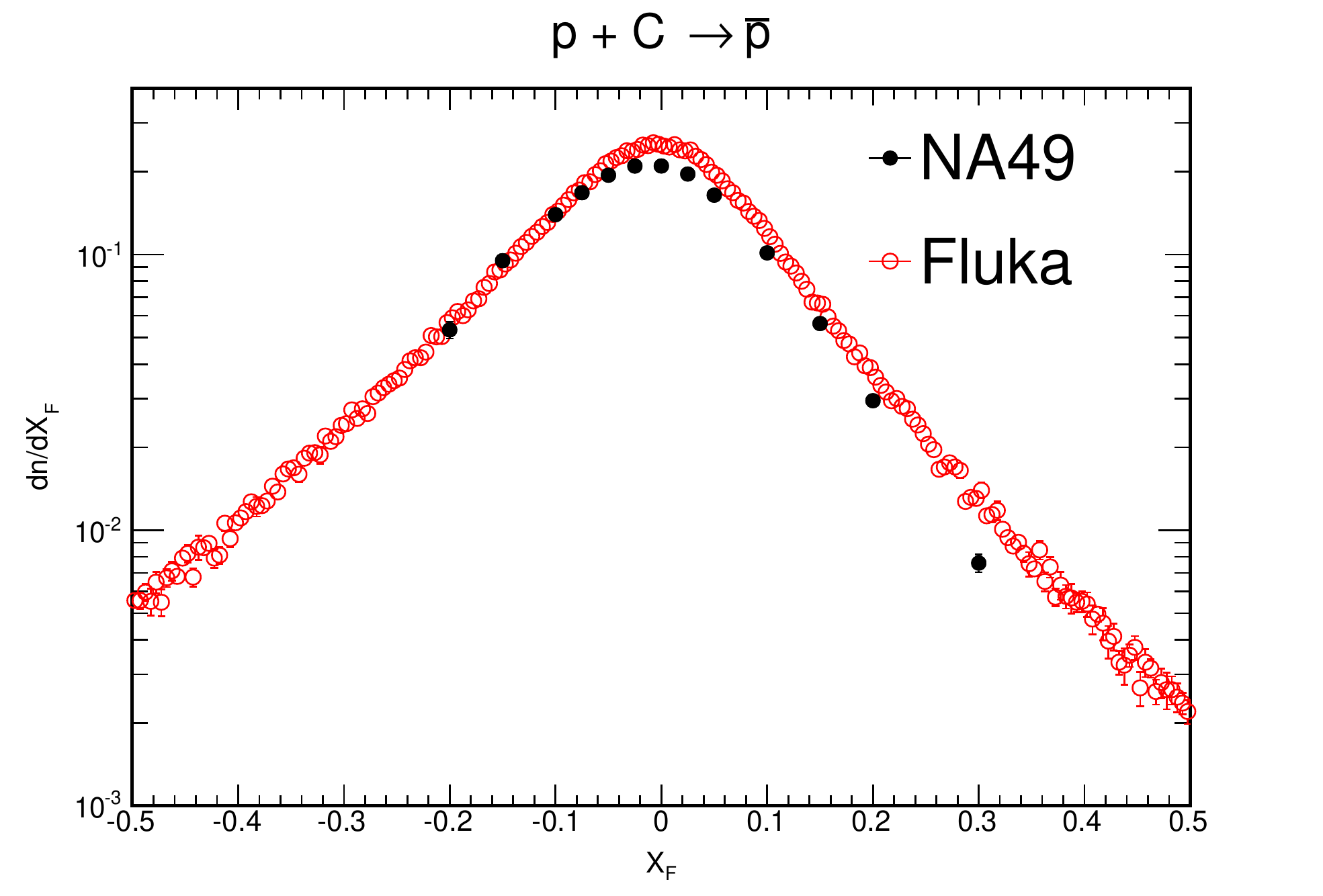}
\caption{Feynman X ($X_{F}$) distribution for $\bar{p}$ production in proton interactions on carbon at $158 \units{GeV/c}$
beam momentum. The NA49 data~\cite{na49} (black points) are compared with the predictions by {\tt FLUKA} (red points).}
\label{Figna49}
\end{figure}


\begin{figure*}[!ht]
\begin{tabular}{cc}
\includegraphics[width=1\columnwidth,height=0.23\textheight,clip]{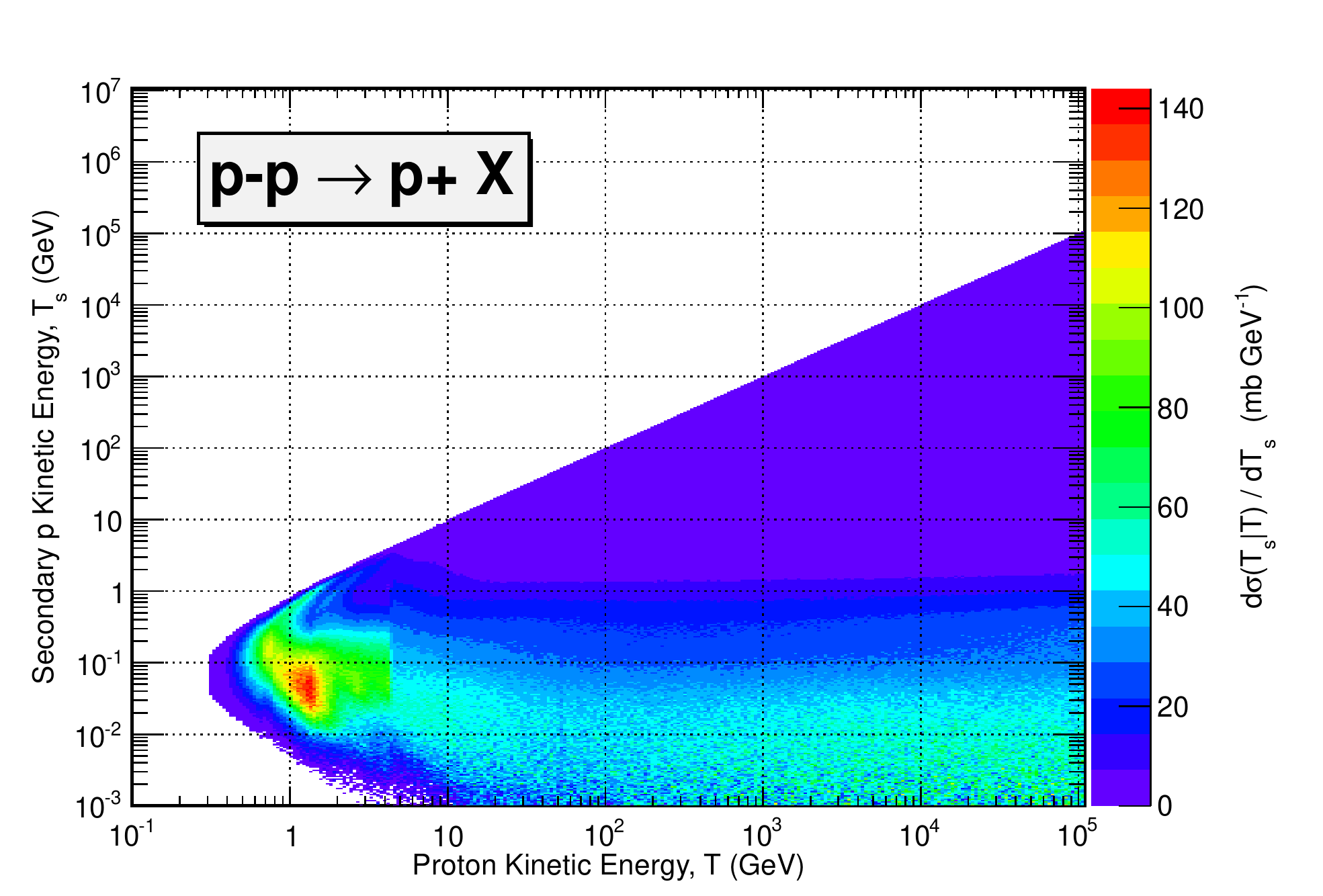} &
\includegraphics[width=1\columnwidth,height=0.23\textheight,clip]{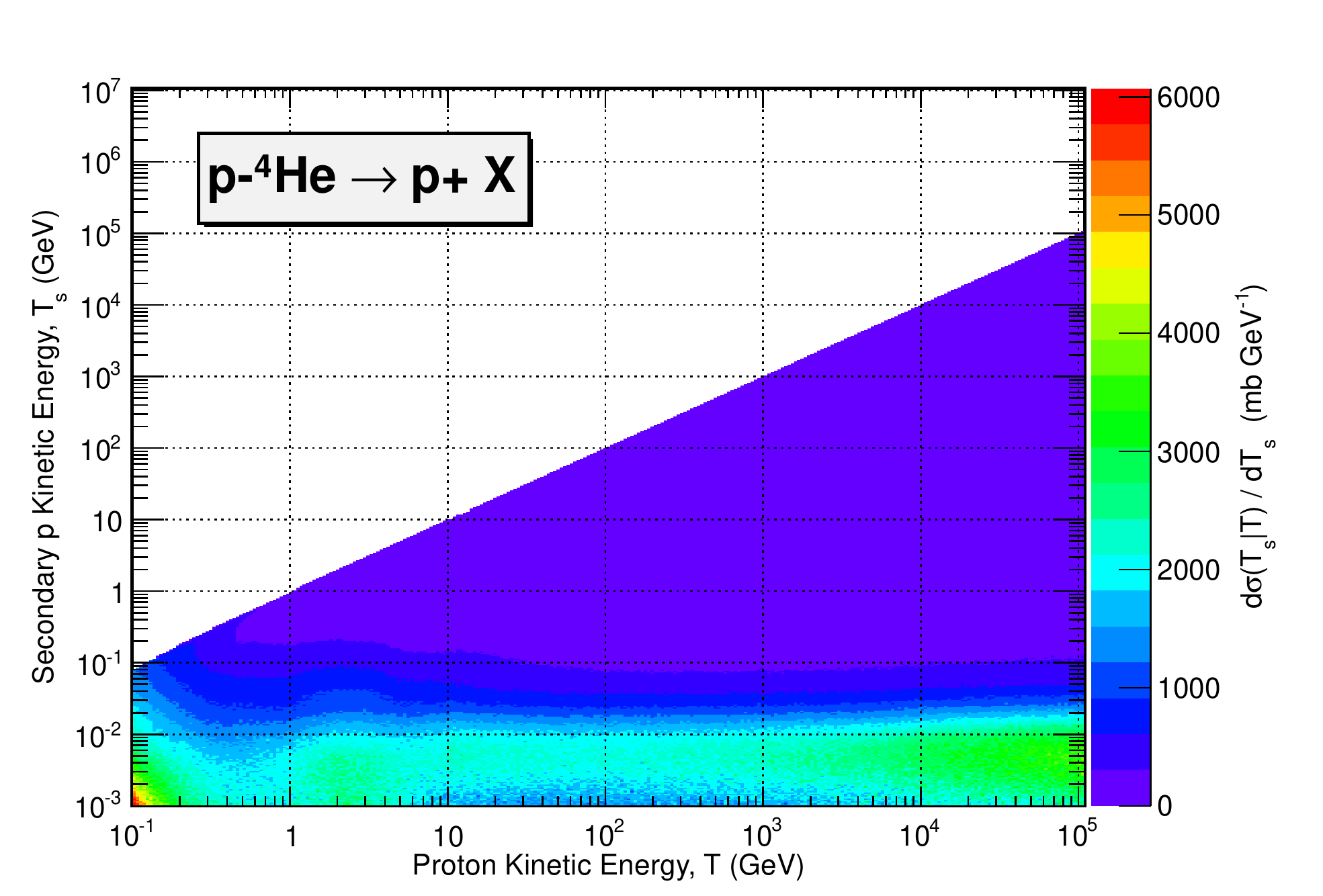} \\
\includegraphics[width=1\columnwidth,height=0.23\textheight,clip]{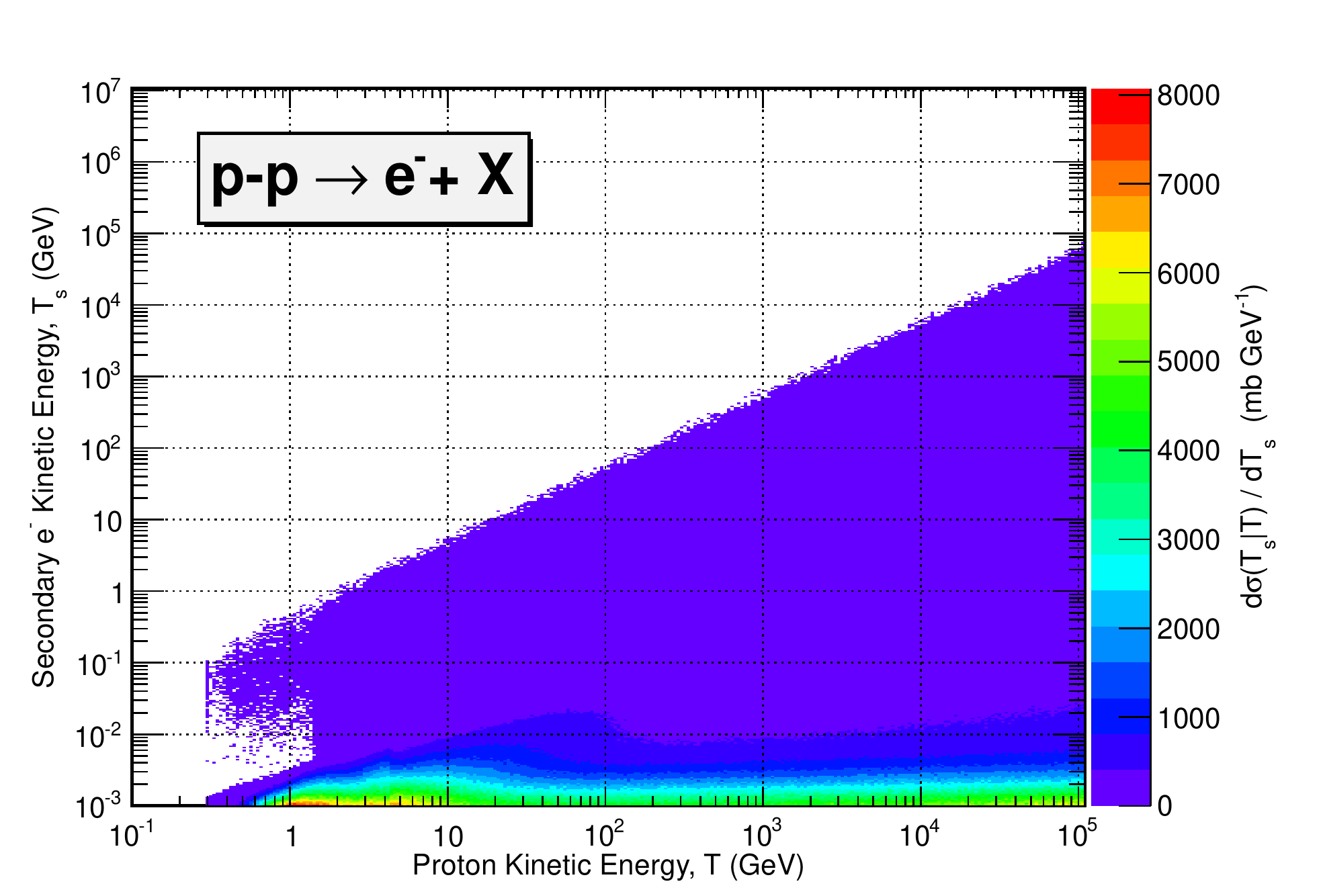} &
\includegraphics[width=1\columnwidth,height=0.23\textheight,clip]{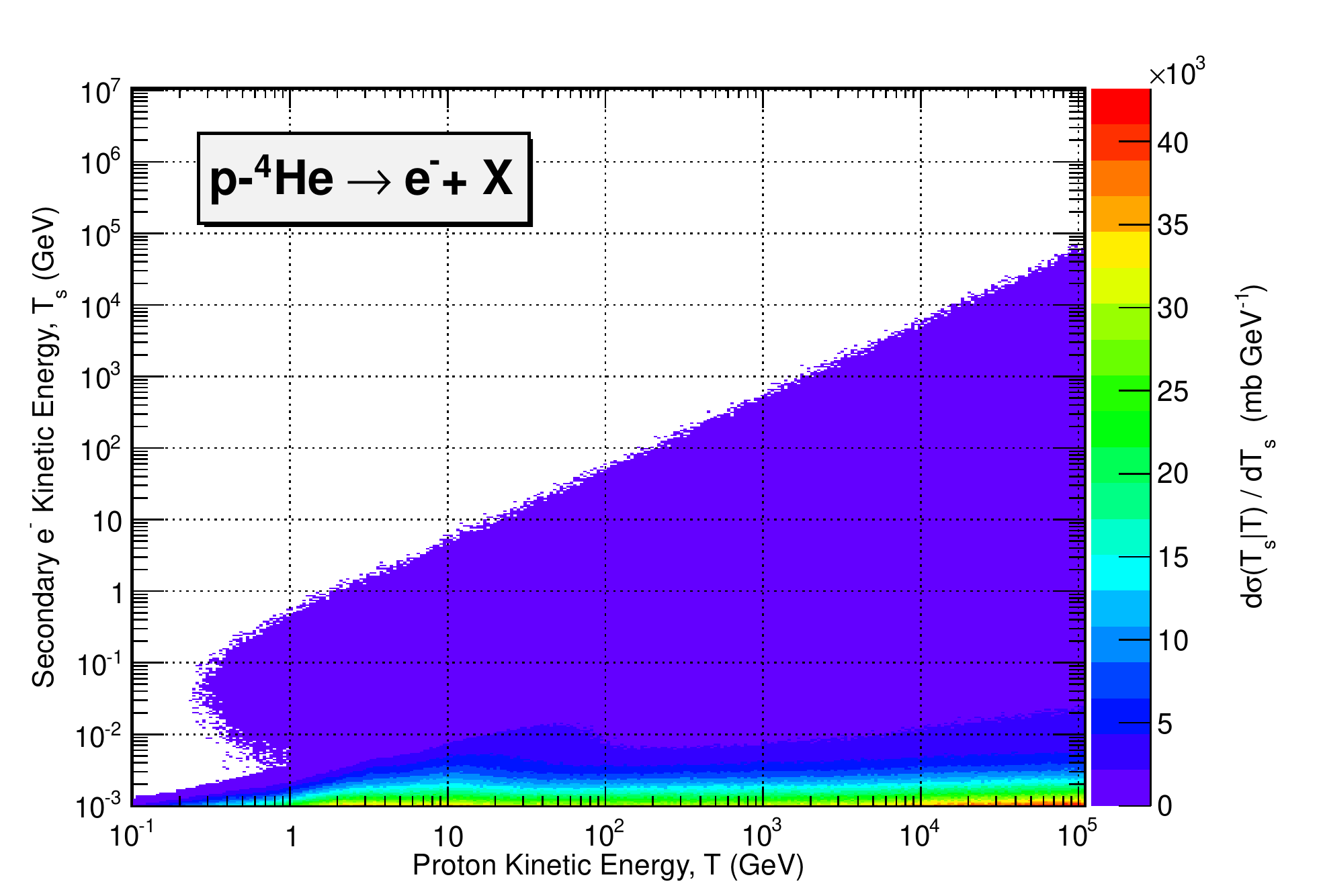} \\
\includegraphics[width=1\columnwidth,height=0.23\textheight,clip]{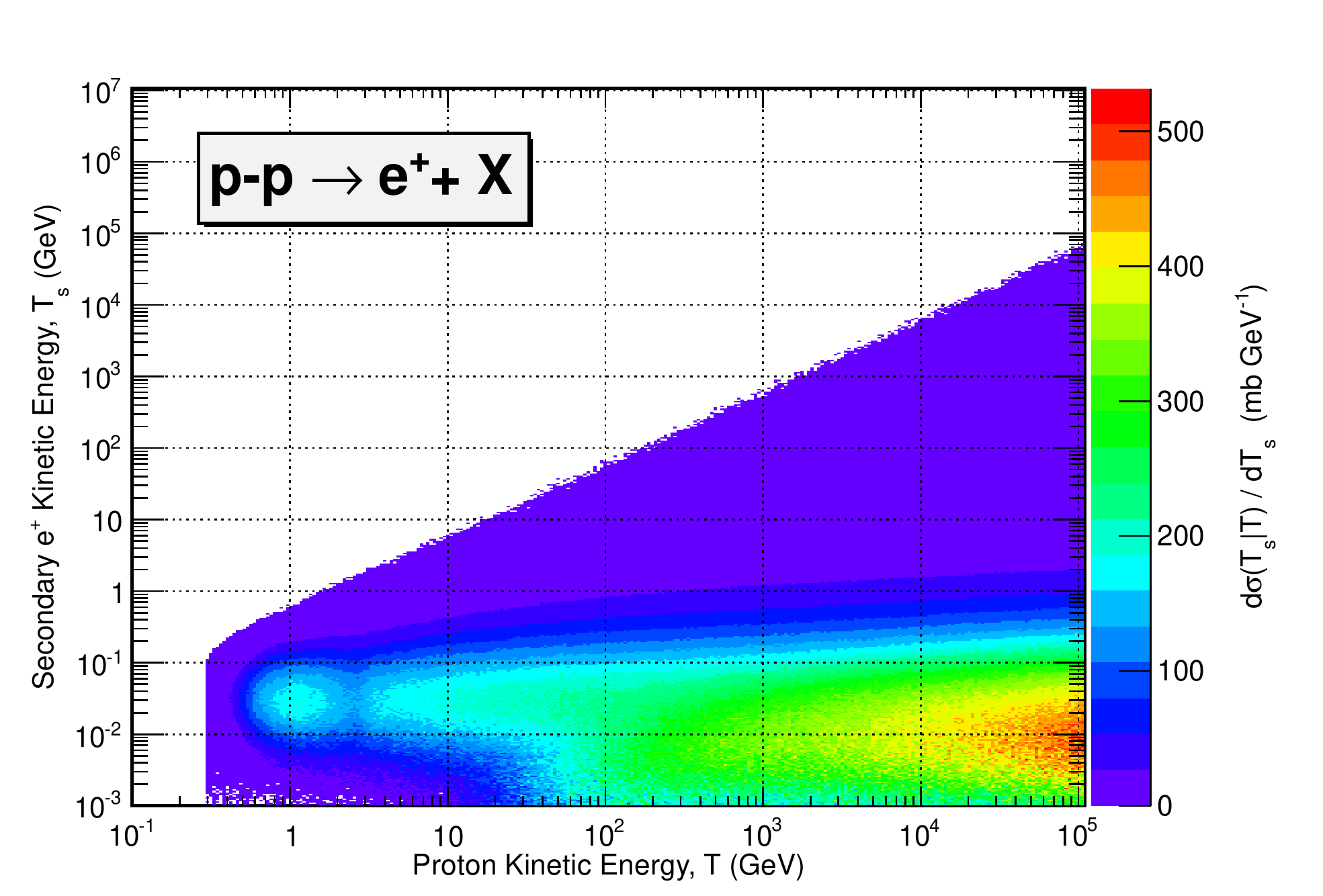} &
\includegraphics[width=1\columnwidth,height=0.23\textheight,clip]{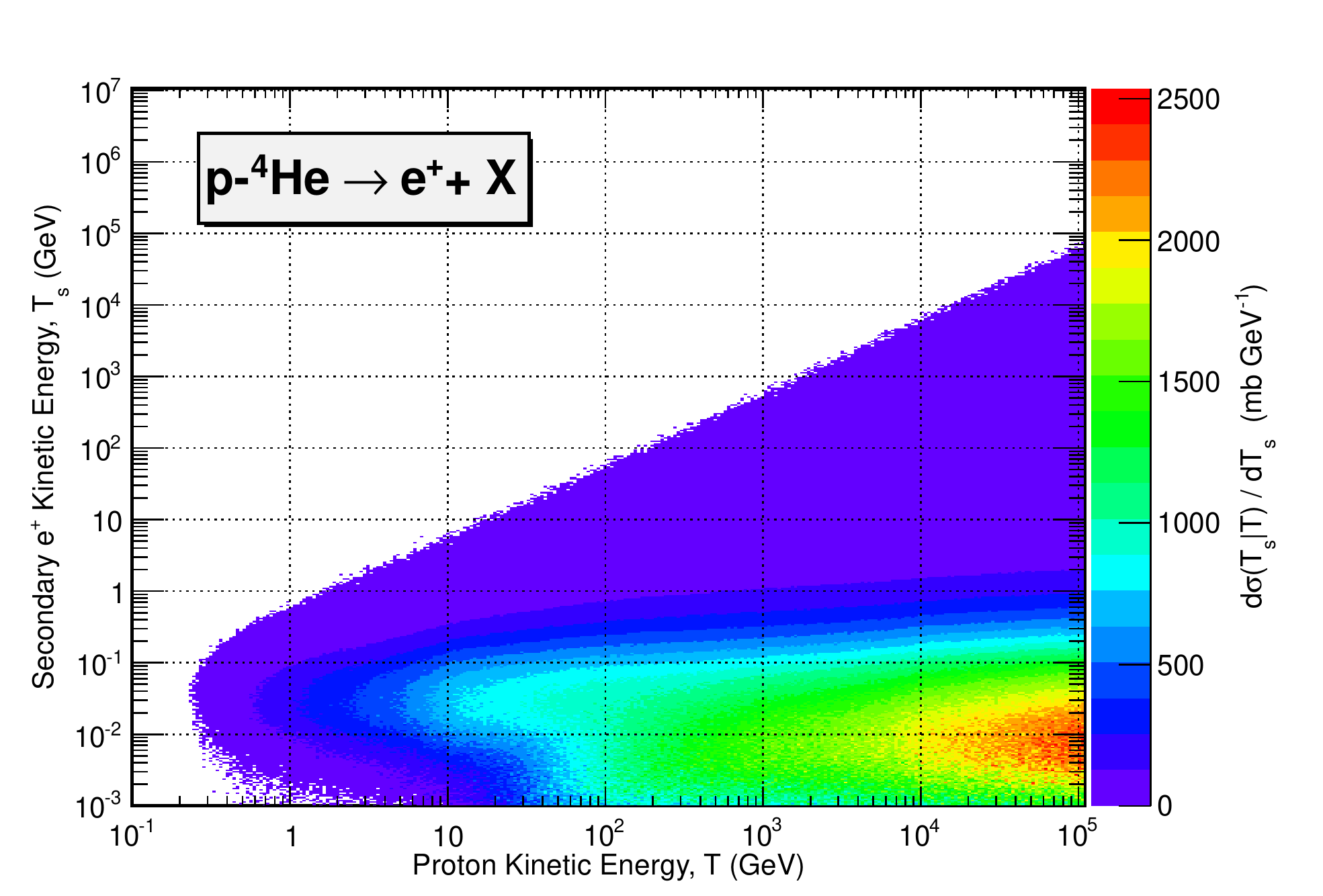} \\
\includegraphics[width=1\columnwidth,height=0.23\textheight,clip]{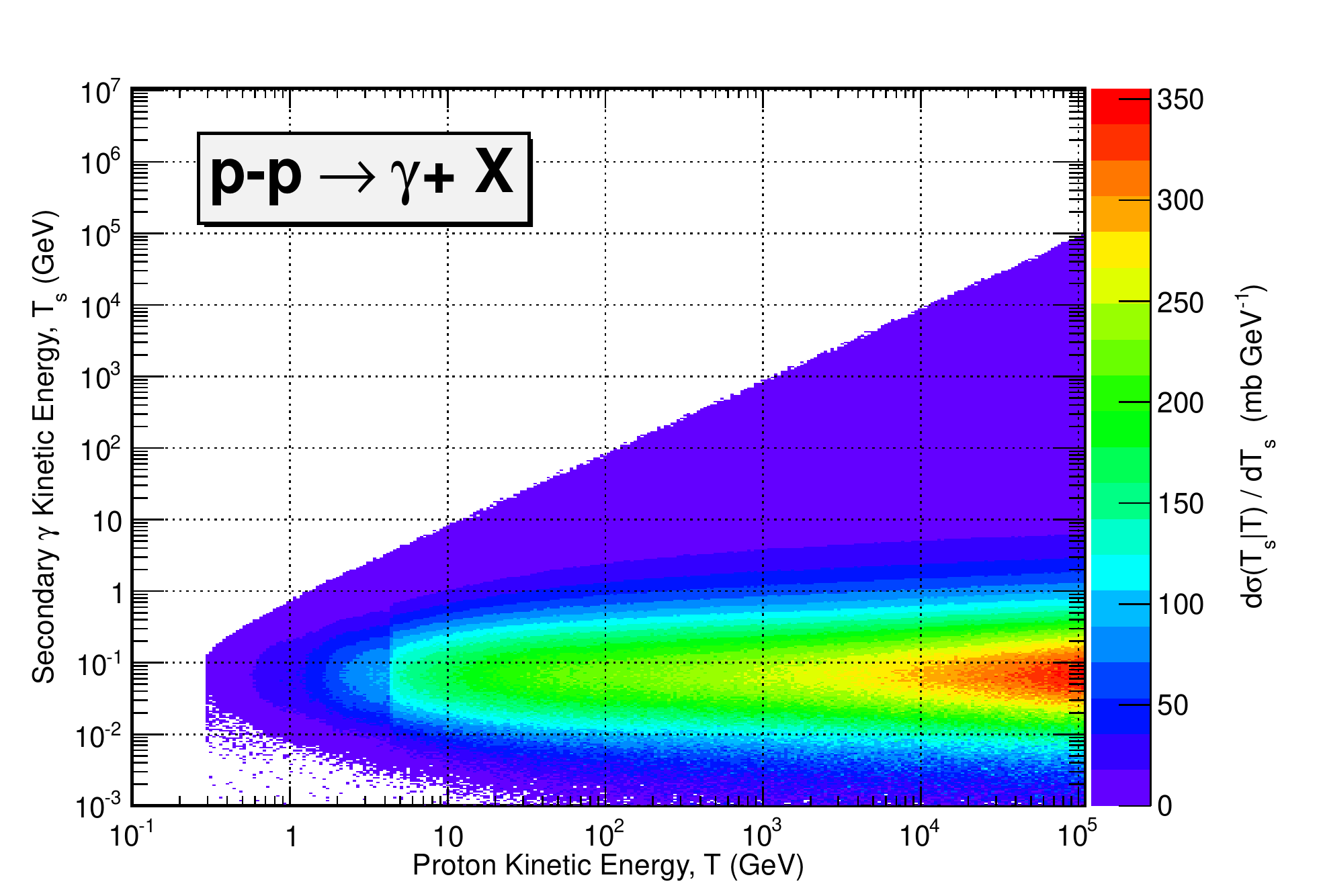} &
\includegraphics[width=1\columnwidth,height=0.23\textheight,clip]{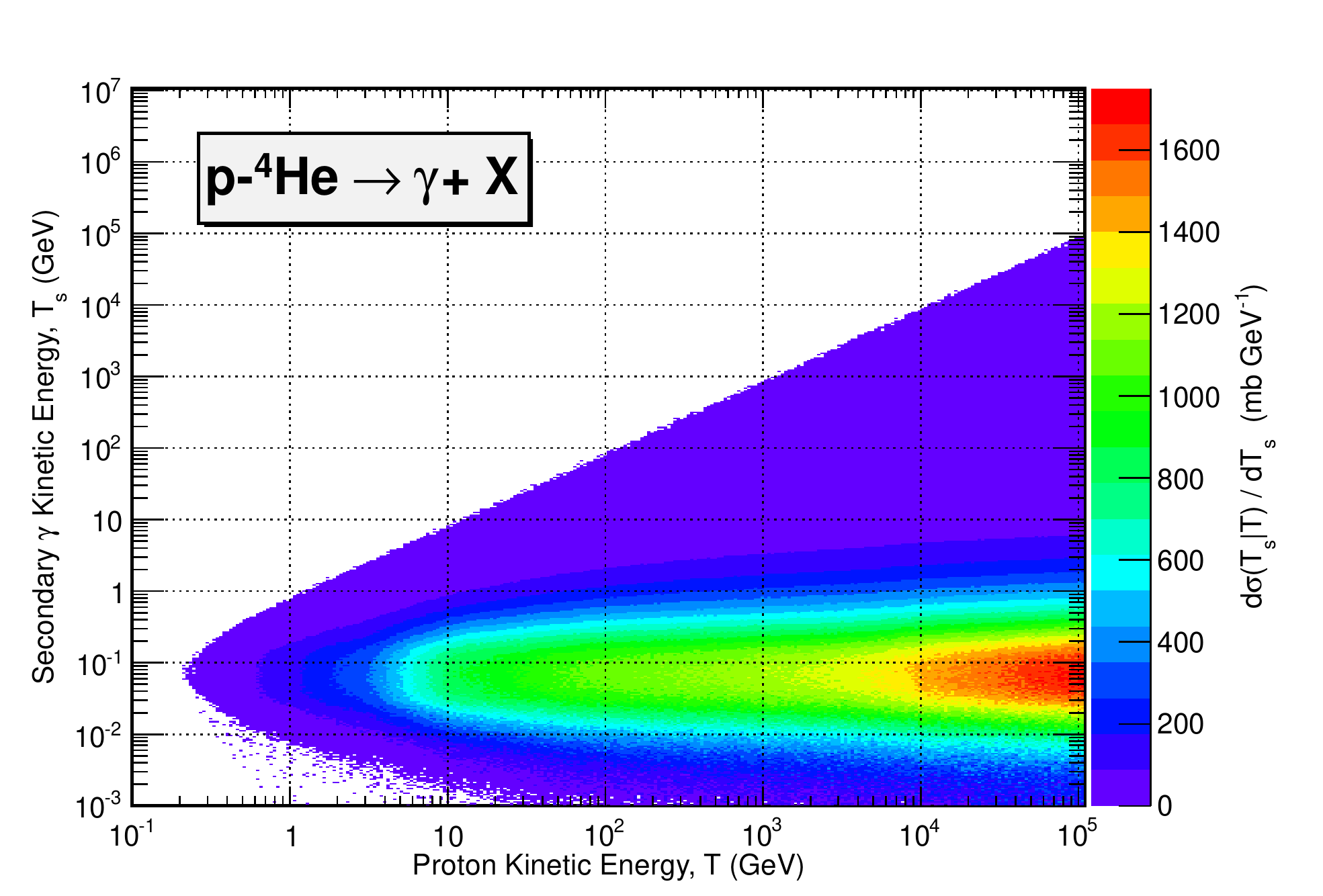} \\
\end{tabular}
\caption{Differential inclusive secondary cross sections for the production of $p$, $e^-$, $e^+$ and $\gamma$  
in $p-p$ (left) and $p-^{4}He$ (right) interactions. The values of the cross sections (color scales) are in $\units{mb~GeV^{-1}}$.}
\label{FigpHe2DXsec}
\end{figure*}

\begin{figure*}[!ht]
\begin{tabular}{cc}
\includegraphics[width=1\columnwidth,height=0.23\textheight,clip]{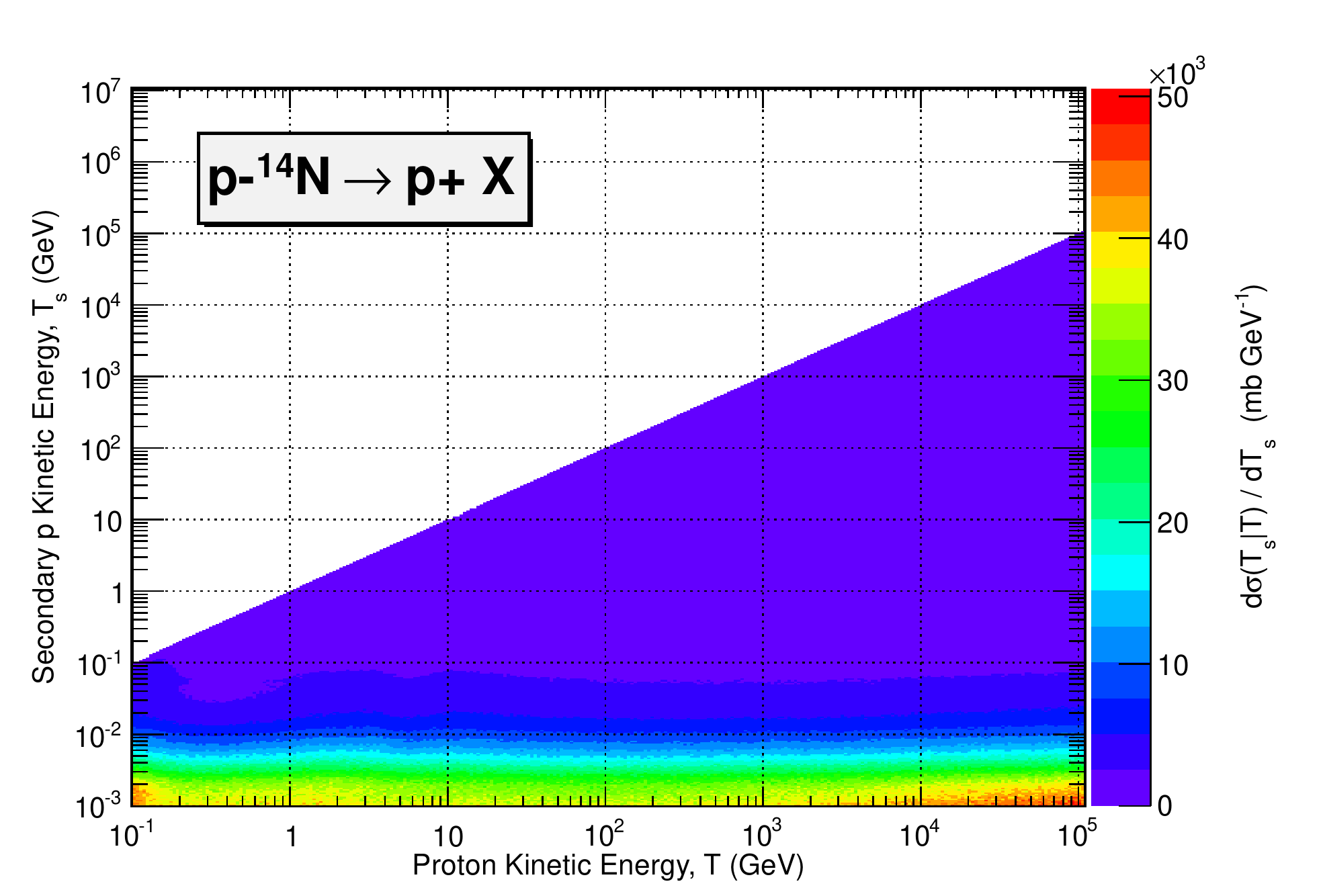} &
\includegraphics[width=1\columnwidth,height=0.23\textheight,clip]{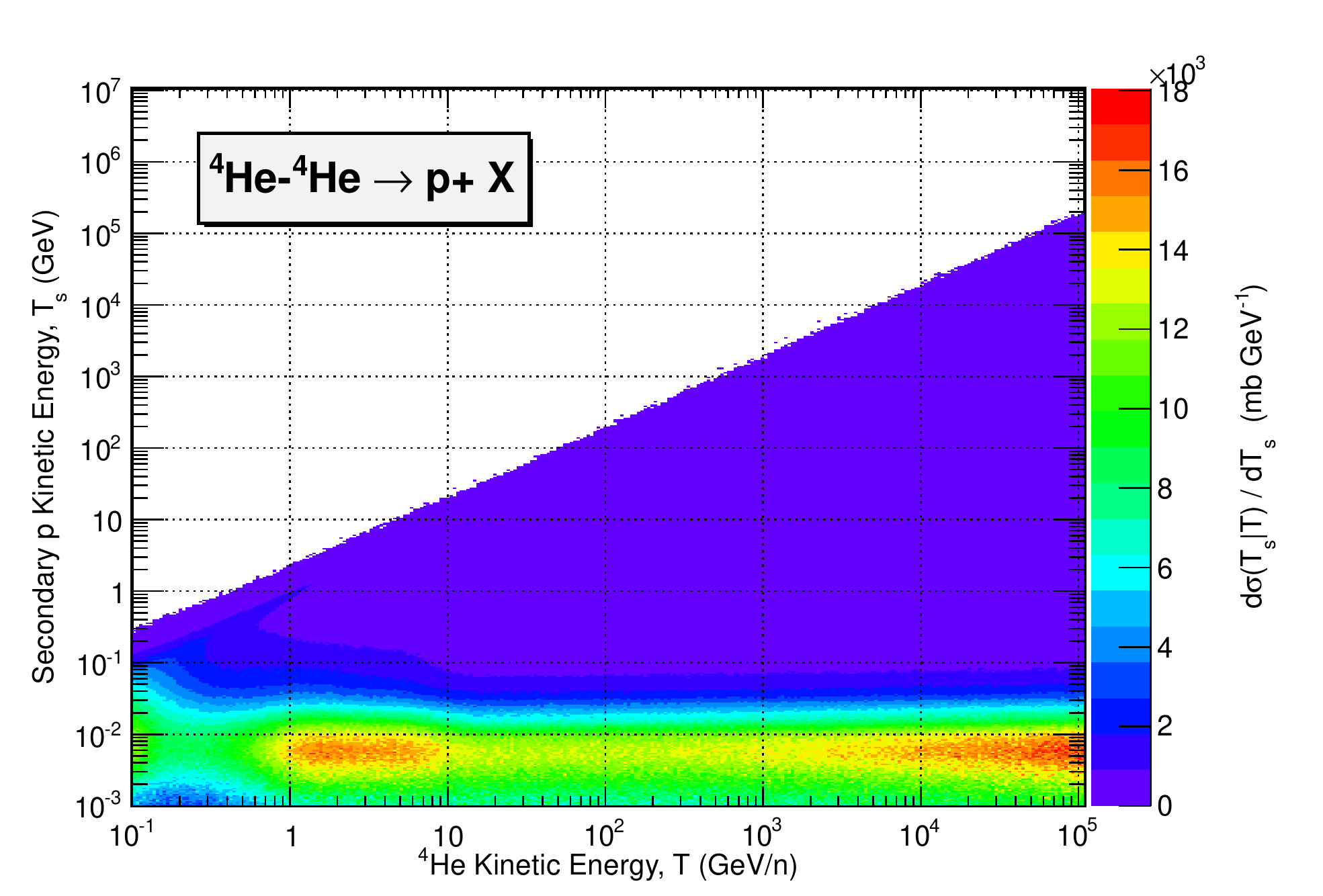} \\
\includegraphics[width=1\columnwidth,height=0.23\textheight,clip]{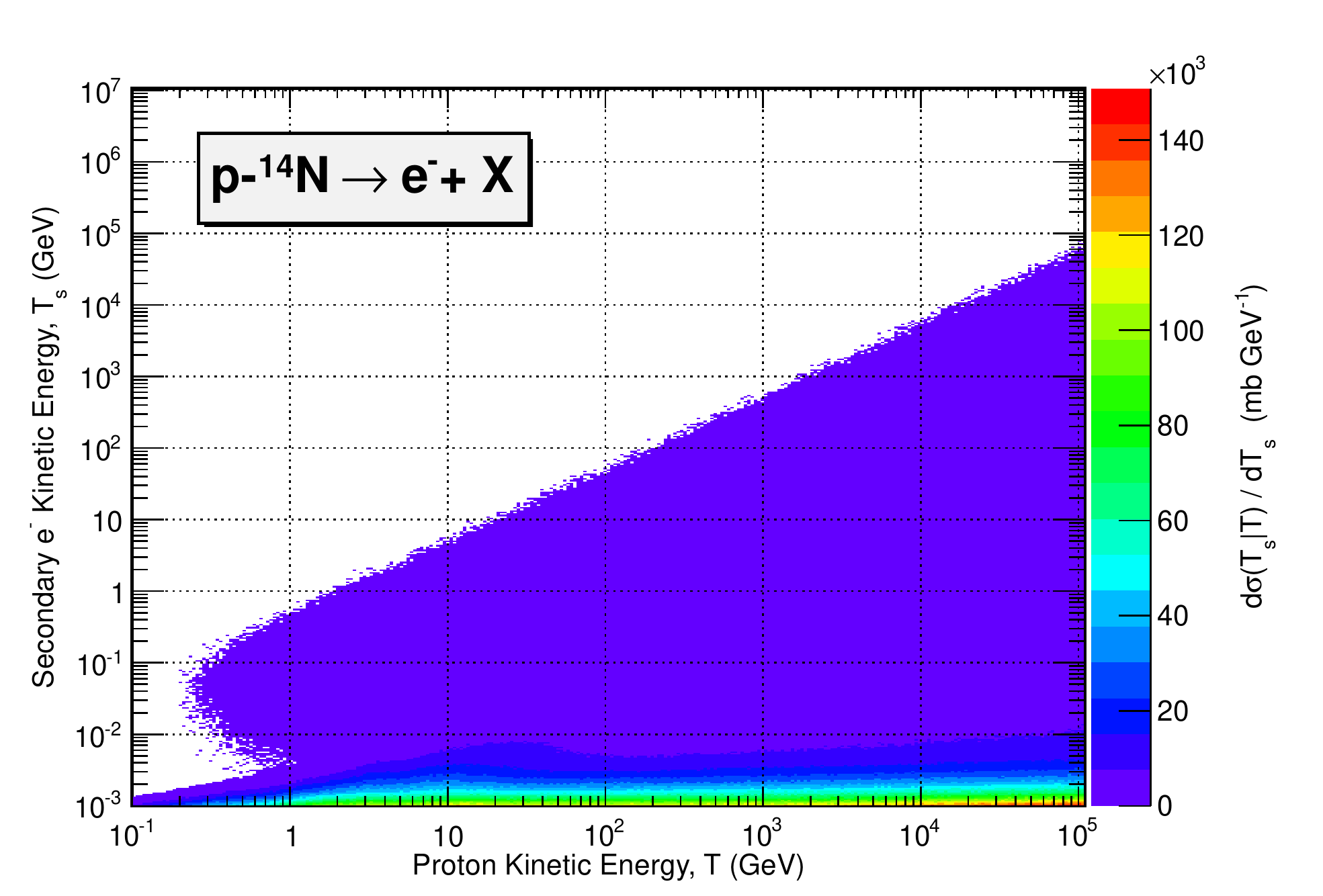} &
\includegraphics[width=1\columnwidth,height=0.23\textheight,clip]{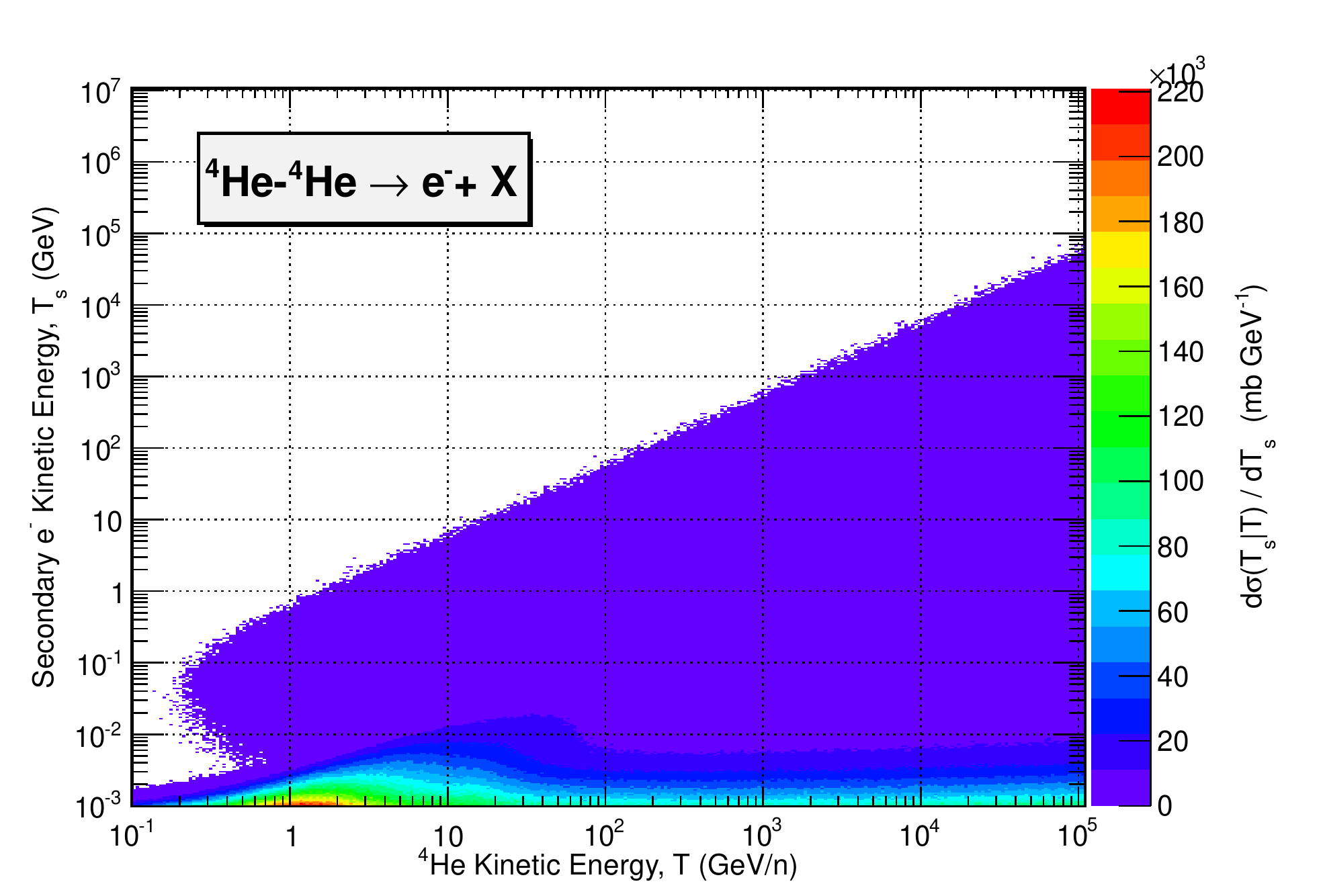} \\
\includegraphics[width=1\columnwidth,height=0.23\textheight,clip]{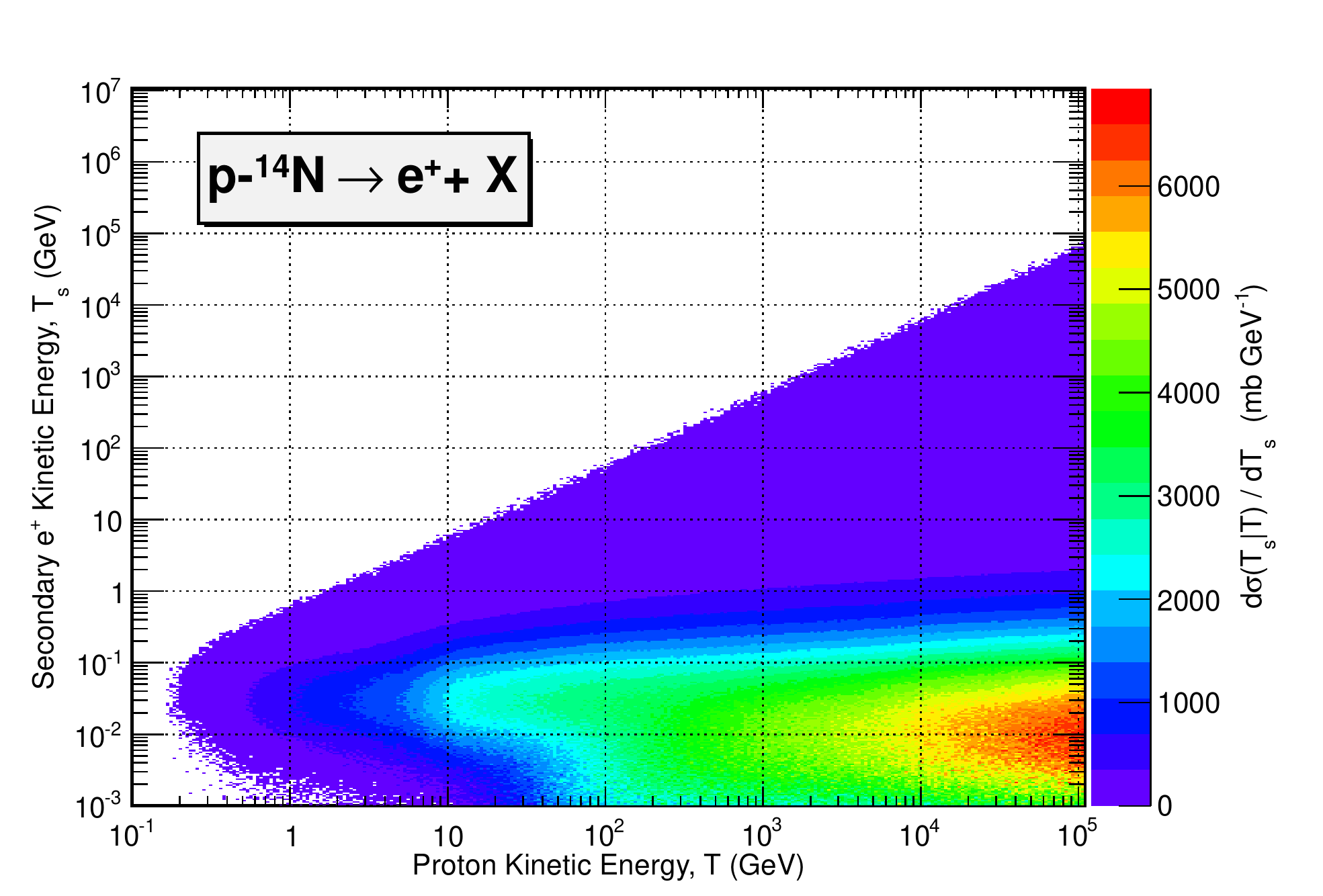} &
\includegraphics[width=1\columnwidth,height=0.23\textheight,clip]{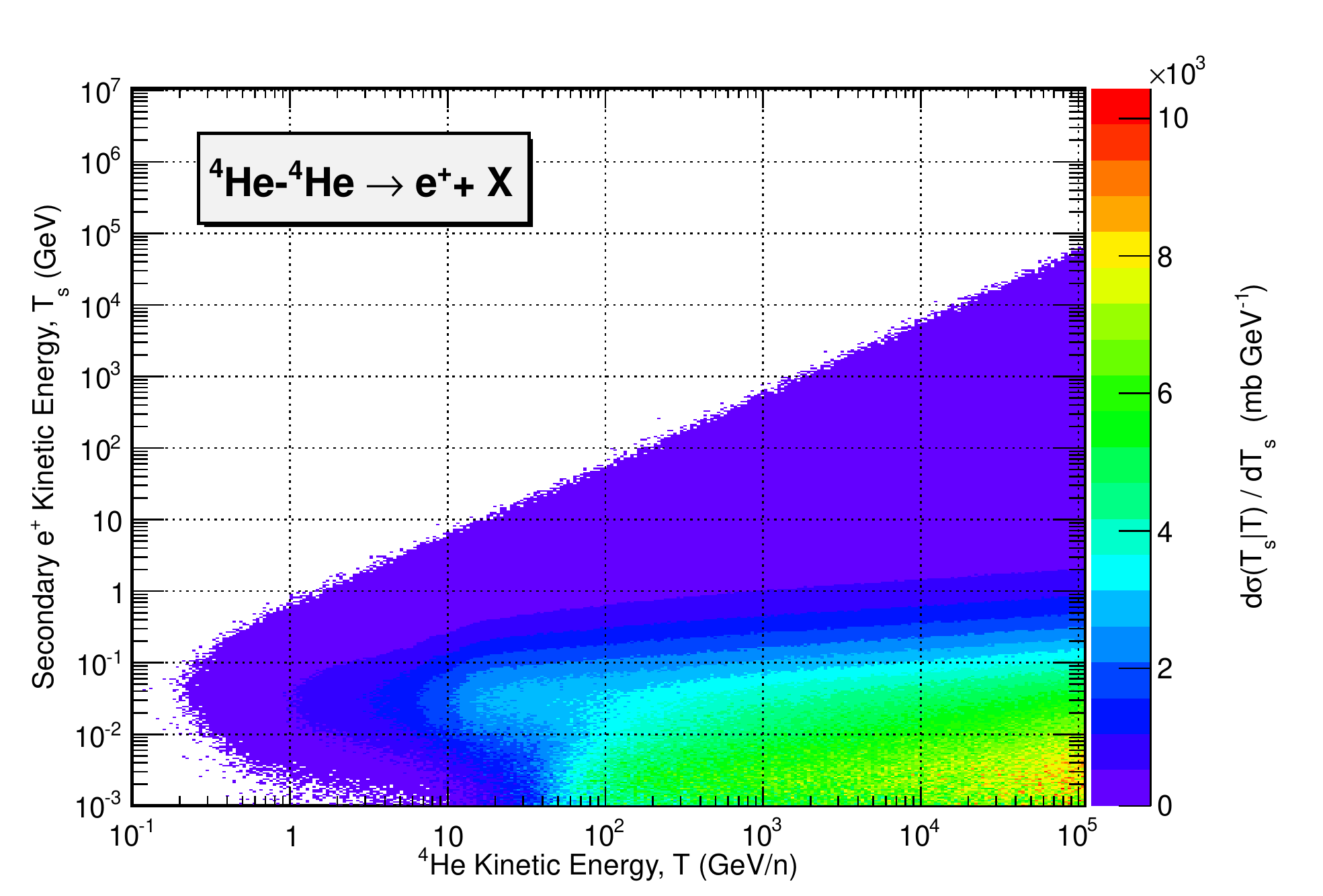} \\
\includegraphics[width=1\columnwidth,height=0.23\textheight,clip]{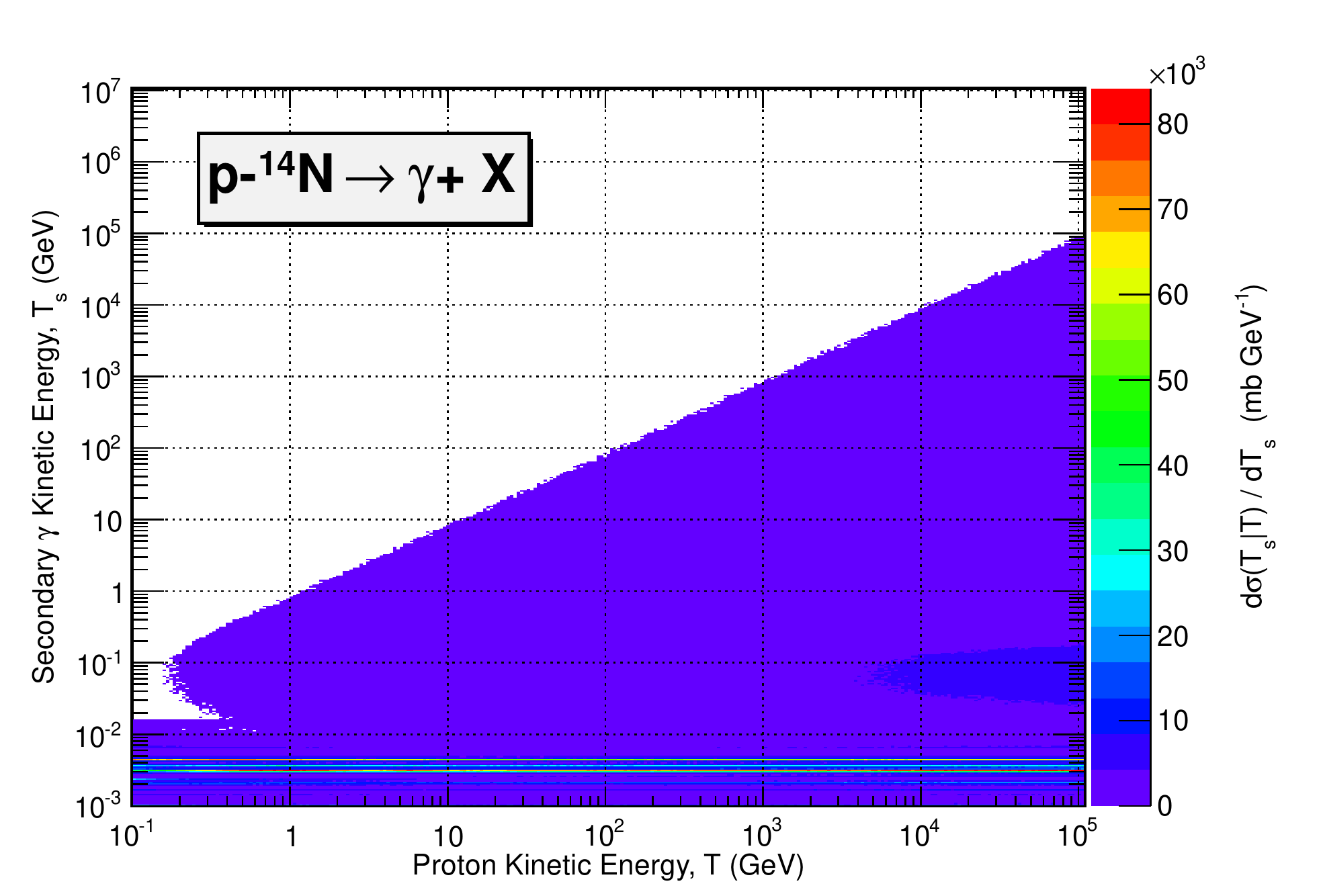} &
\includegraphics[width=1\columnwidth,height=0.23\textheight,clip]{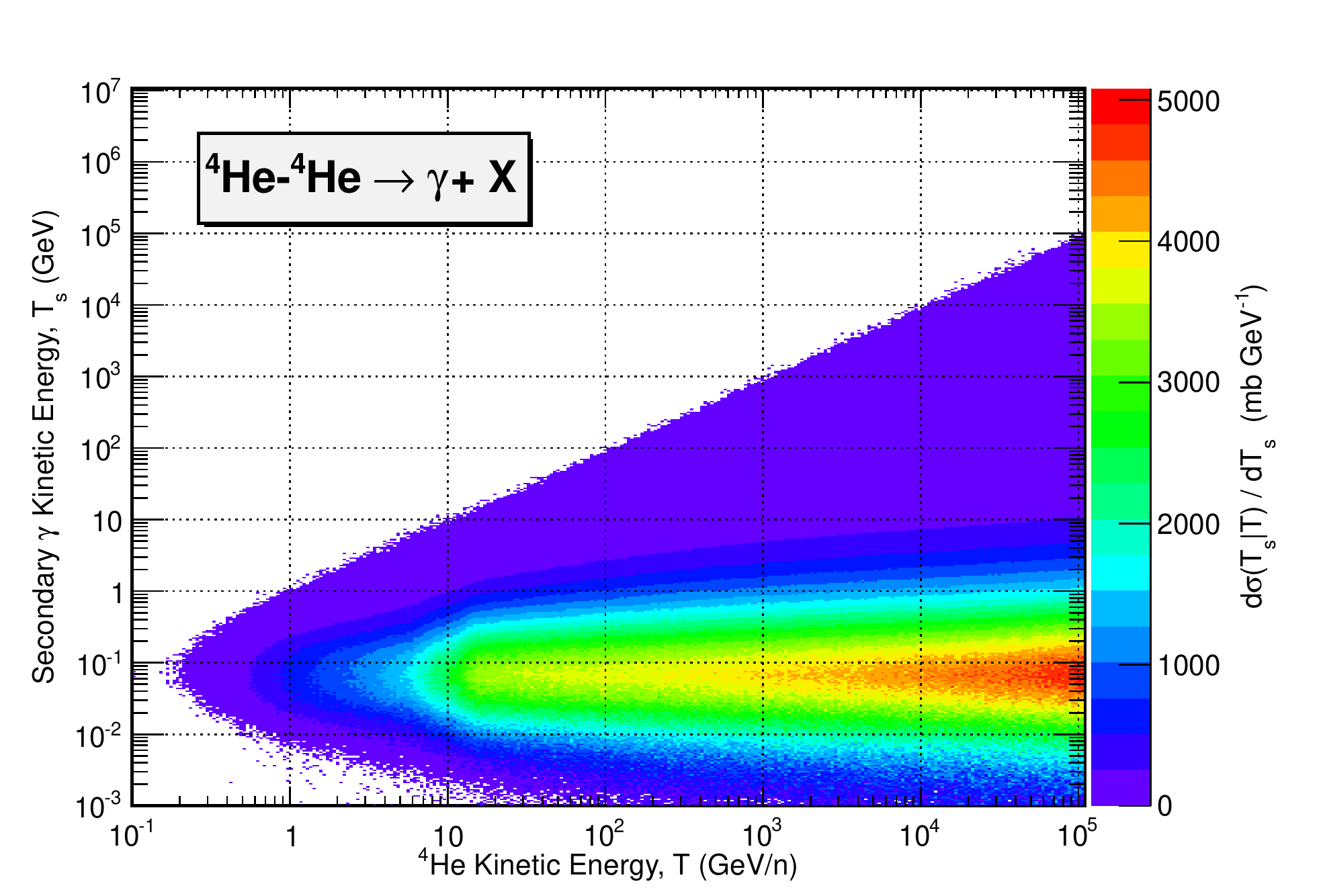} \\
\end{tabular}
\caption{Differential inclusive secondary cross sections for the production of $p$, $e^-$, $e^+$ and $\gamma$ 
in $p-^{14}N$ (left) and  $^{4}He-^{4}He$ (right) interactions. The values of the cross sections (color scales) are in $\units{mb~GeV^{-1}}$.}
\label{FigpNHeHe2DXsecDPM}
\end{figure*}

\section{Secondary particle production in CR interactions}

In this work we use a developement version of {\tt FLUKA}~\cite{bohlen,ballarini,batt2007,batt2015} 
to evaluate the yields of secondary particles and spallation nuclei in CR interactions.  
In case of nucleus-nucleus interactions {\tt DPMJET-III} and a modified version~\cite{ander2004} of 
{\tt RQMD}~\cite{Sor89,Sor89a,Sor95} are used as external event generators. Both generators have been
satisfactorily benchmarked in their respective range of application~\cite{sihver,braun}.

Our focus is on the interactions of CRs with the ISM and with the Earth's atmosphere.
As it is well known, the ISM consists of $90\%$ Hydrogen, $9.9\%$ Helium and $0.1\%$ atoms of heavier elements synthesized in stellar interiors~\cite{Ferriere:2001rg,ramaty1979}, while the atmosphere is composed of 78\% of $N_2$, 21\% of $O_2$ and small fractions of other gases (mainly Argon). 
Primary CRs, on the other hand, are accelerated in astrophysical sources and their hadronic component consists of protons, Helium, and other heavier nuclei ~\cite{pdg}. Therefore, we simulate the interactions of the following stable and long lived CR projectiles: $p$, $D$, $T$, $^{3}He$, $^{4}He$, $^{6}Li$, $^{7}Li$, $^{9}Be$, $^{10}Be$, $^{10}B$, $^{11}B$, $^{12}C$, $^{13}C$, $^{14}C$, $^{14}N$, $^{15}N$, $^{16}O$, $^{17}O$, $^{18}O$, $^{20}Ne$, $^{24}Mg$, and $^{28}Si$
impinging on $p$, $^{4}He$, $^{12}C$, $^{14}N$, $^{16}O$ $^{20}Ne$, $^{24}Mg$, $^{28}Si$, and $^{40}Ar$.
%
For each interaction we simulate a sample of events, each corresponding to a value 
of the projectile kinetic energy $T$ (the number of simulated events is $10^{5}$ for each input energy value). 
The values of $T$ are taken from a grid of 285 values equally spaced 
on a logarithmic scale, ranging from $0.1 \units{GeV/n}$ up to $10^{5} \units{GeV/n}$.

The goal of our study is to evaluate, for each interaction process, the set of differential inclusive cross sections $d \sigma(T_{s} | T)/dT_{s}$, 
where $T_{s}$ is the kinetic energy of the given secondary and $T$ is the kinetic energy 
of the projectile. The differential cross section can be expressed as:

\begin{equation}
\frac{d \sigma(T_{s} | T)}{dT_{s}} = \sigma_{inel}(T) \times \frac{dn(T_{s} | T)}{dT_{s}}
\end{equation}
where $\sigma_{inel}(T)$ is the total inelastic cross section and 
$dn(T_{s} | T)/dT_{s}$ is the differential multiplicity spectrum of the secondary particle 
to be studied. 

\begin{figure*}[!ht]
\begin{tabular}{cc}
\includegraphics[width=1\columnwidth,height=0.25\textheight,clip]{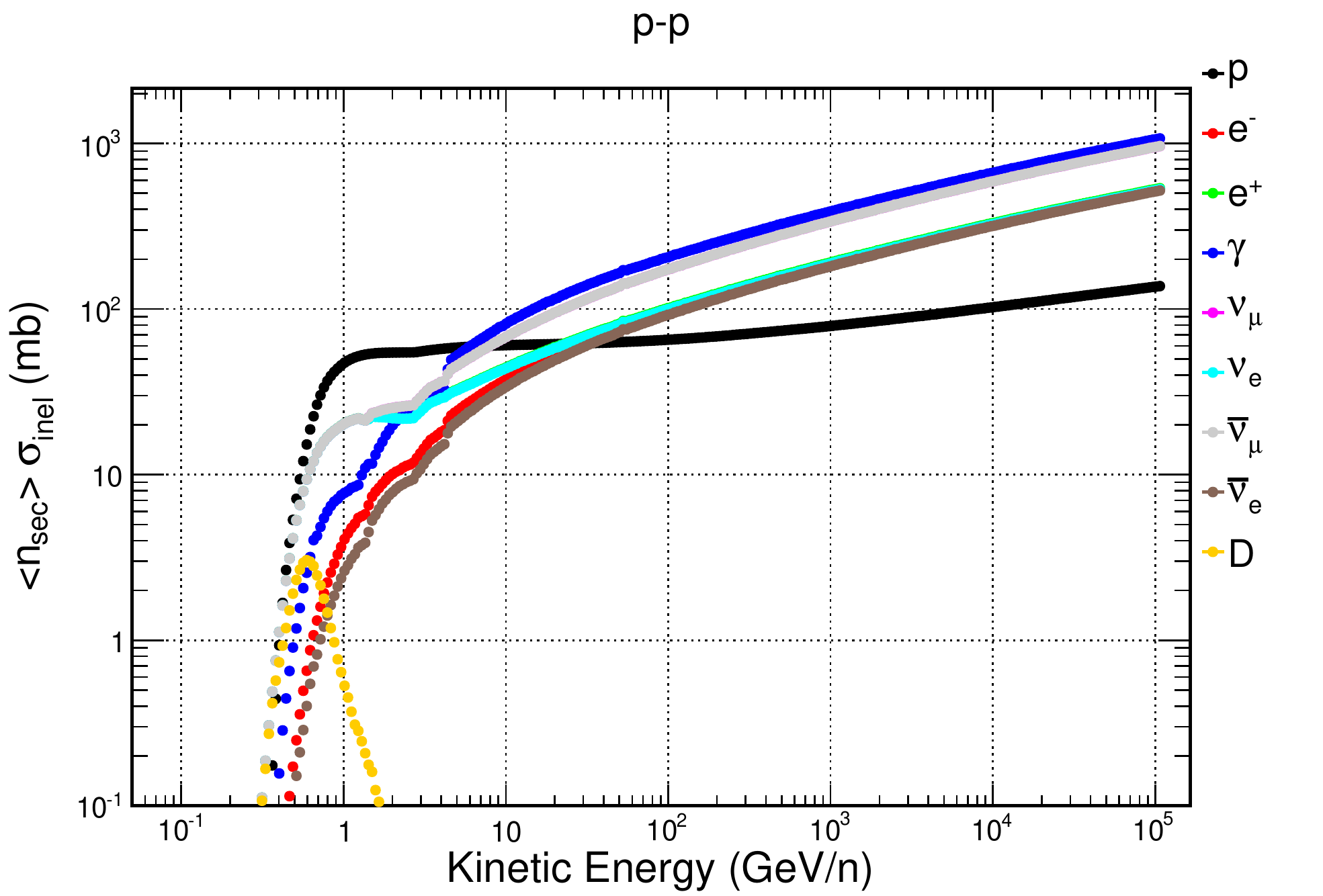} &
\includegraphics[width=1\columnwidth,height=0.25\textheight,clip]{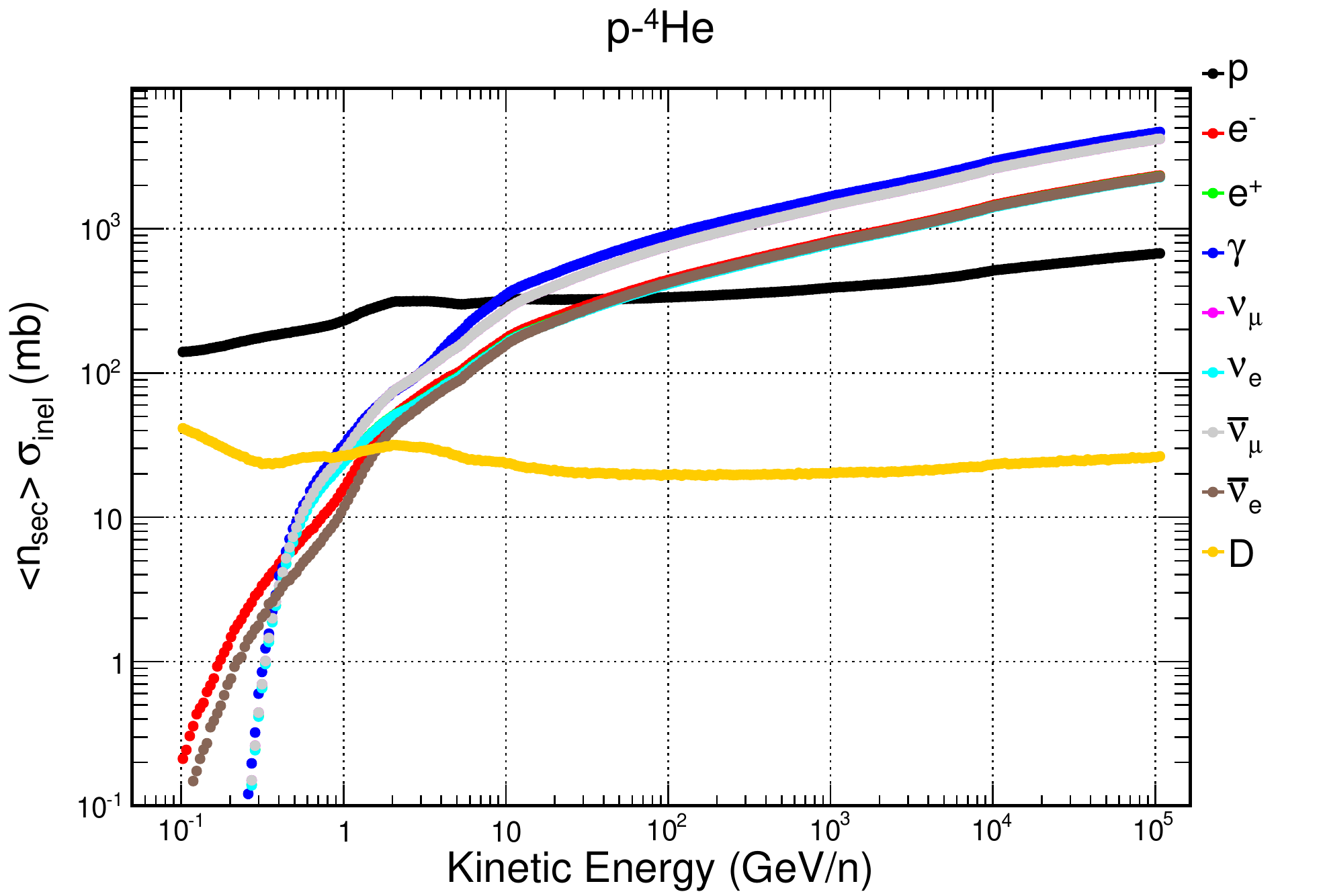} \\
\includegraphics[width=1\columnwidth,height=0.25\textheight,clip]{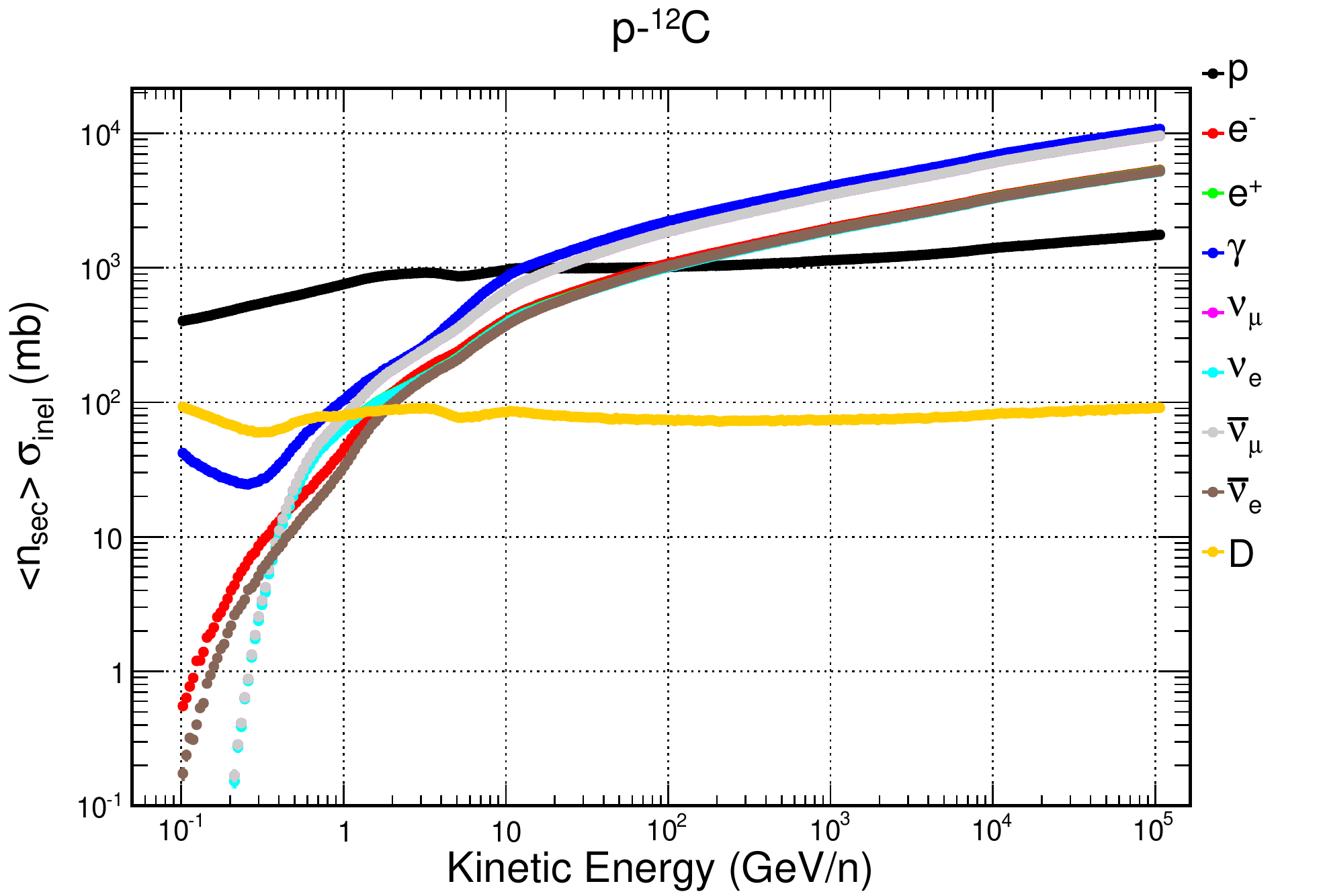} & 
\includegraphics[width=1\columnwidth,height=0.25\textheight,clip]{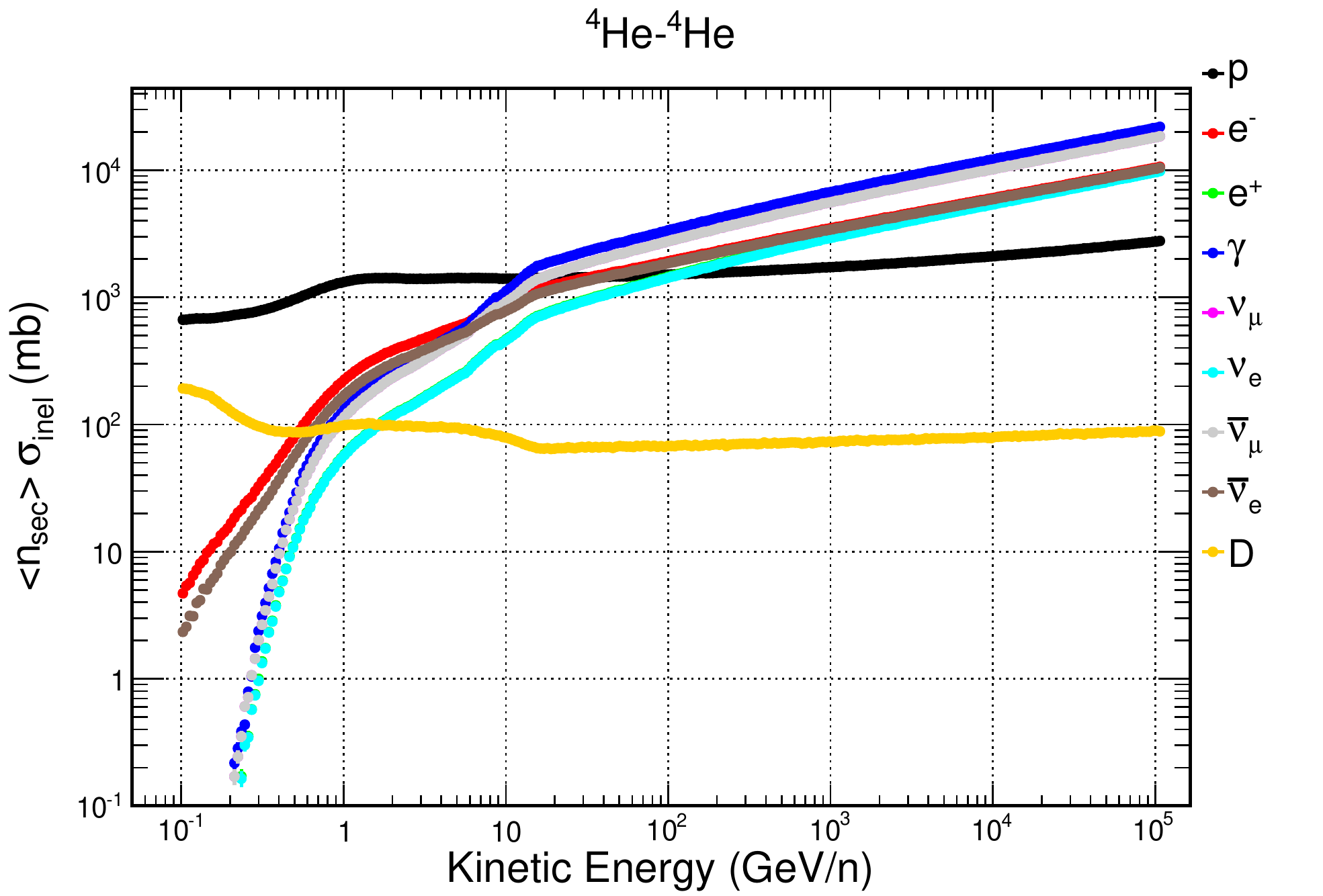} \\
\end{tabular}
\caption{Inclusive cross sections for the production of protons (black), electrons (red), 
positrons (green), gamma rays (blue), electron neutrinos (cyan), 
electron antineutrinos (grey), muon neutrinos (magenta), muon antineutrinos (brown)
and Deuterons (orange) in the collisions of $p-p$, $p-^{4}He$, $p-^{12}C$ and $^{4}He-^{4}He$.}
\label{FigHeOthXsec}
\end{figure*}
\begin{figure*}[!ht]
\begin{tabular}{cc}
\includegraphics[width=1\columnwidth,height=0.25\textheight,clip]{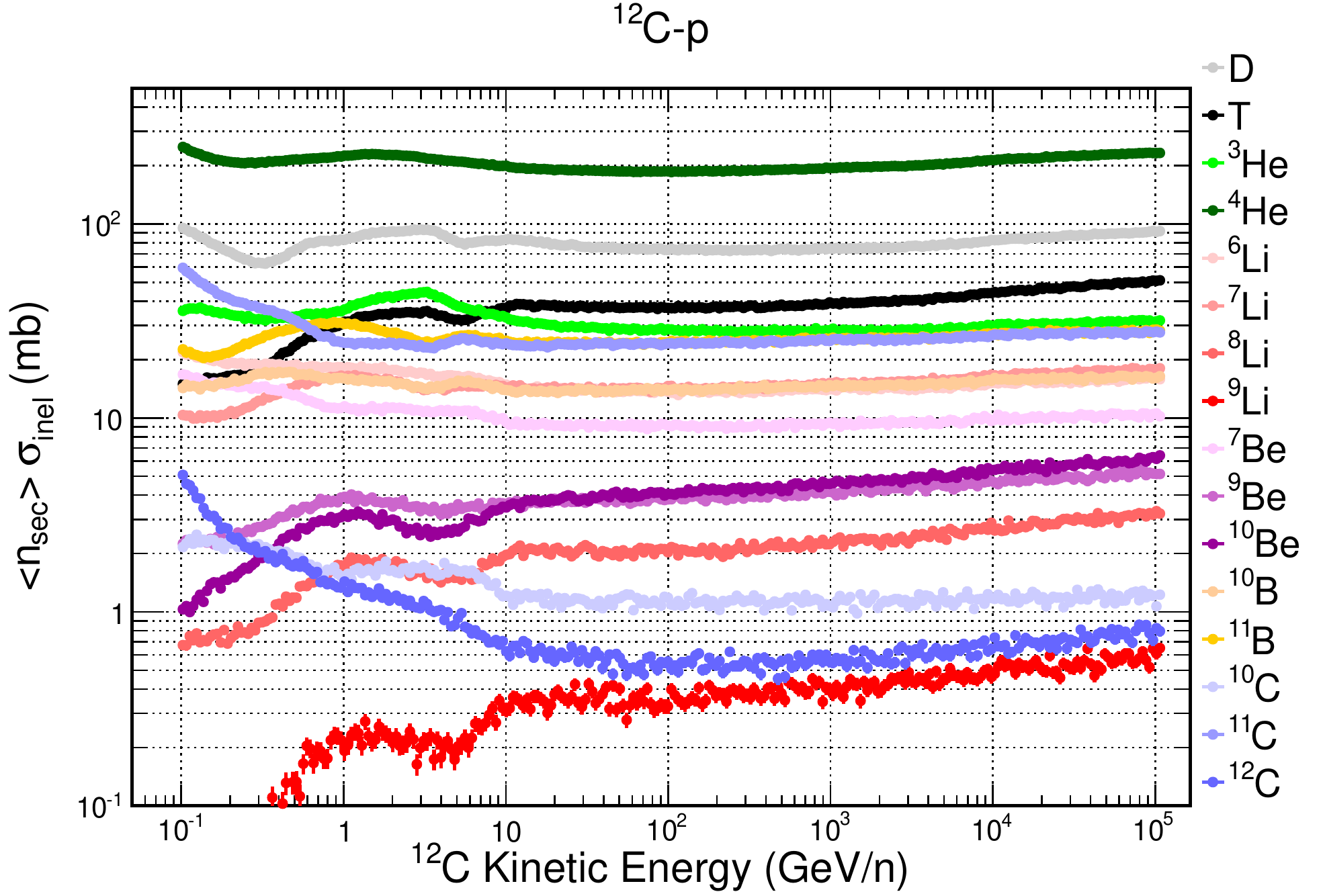} &
\includegraphics[width=1\columnwidth,height=0.25\textheight,clip]{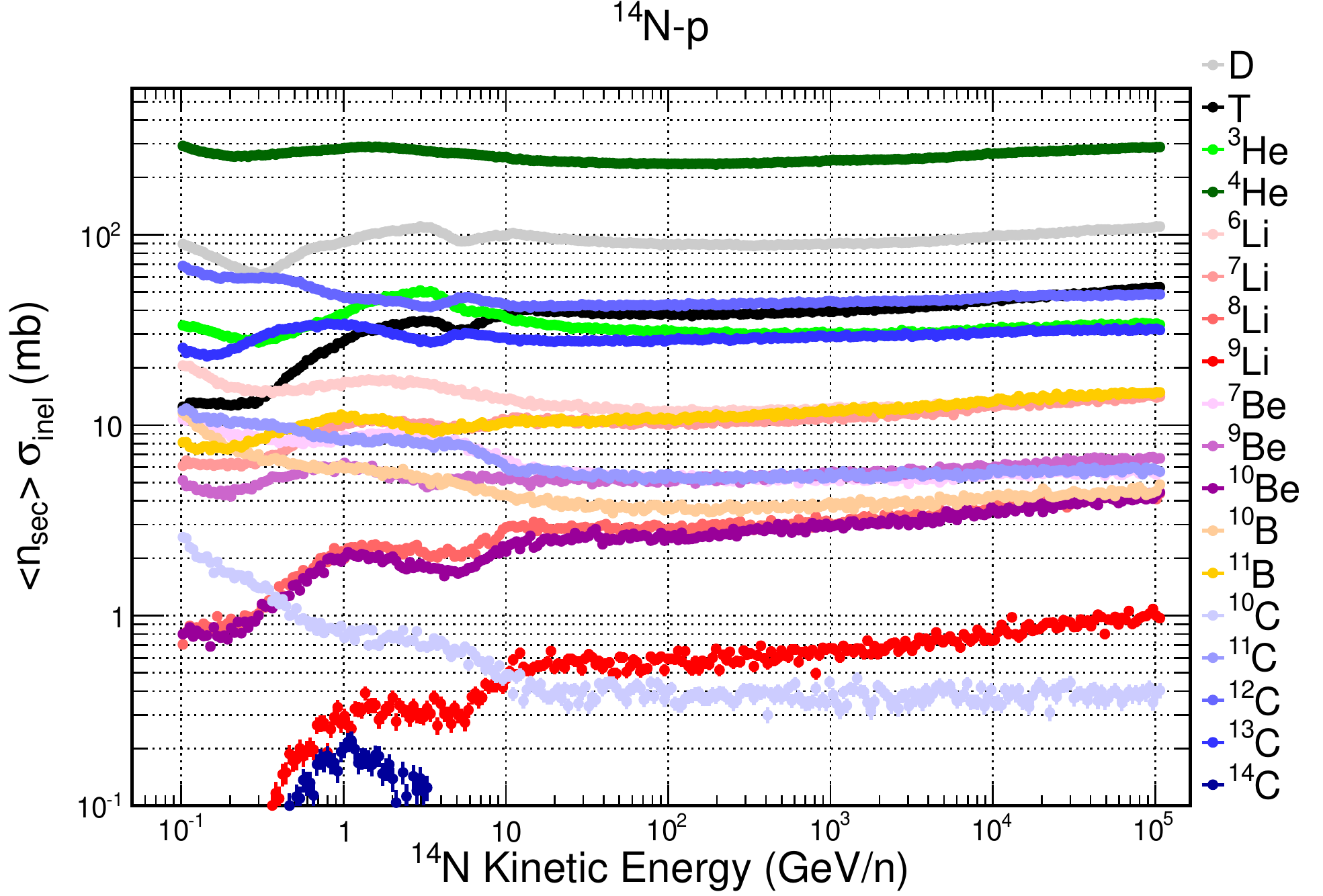} \\
\includegraphics[width=1\columnwidth,height=0.25\textheight,clip]{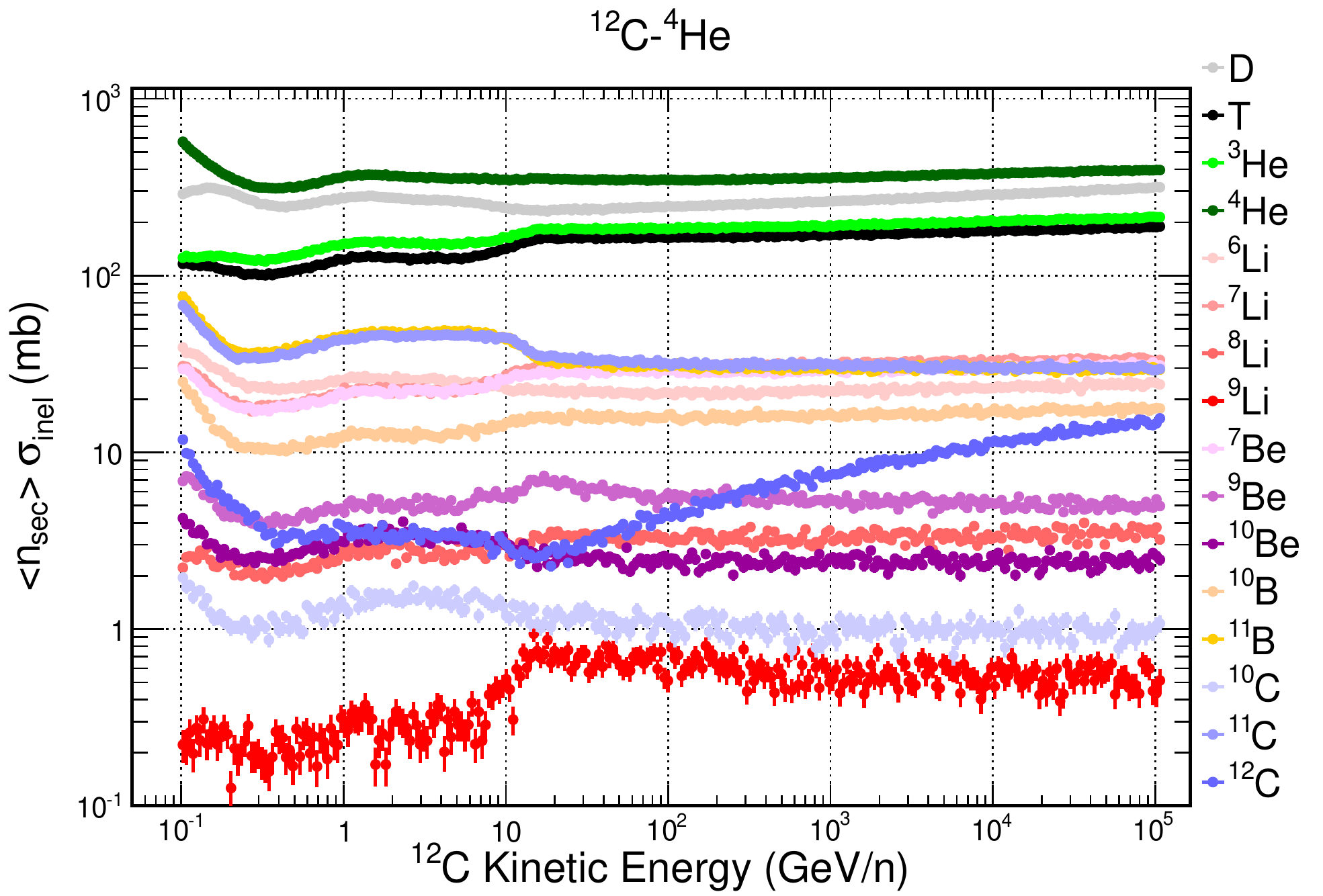} &
\includegraphics[width=1\columnwidth,height=0.25\textheight,clip]{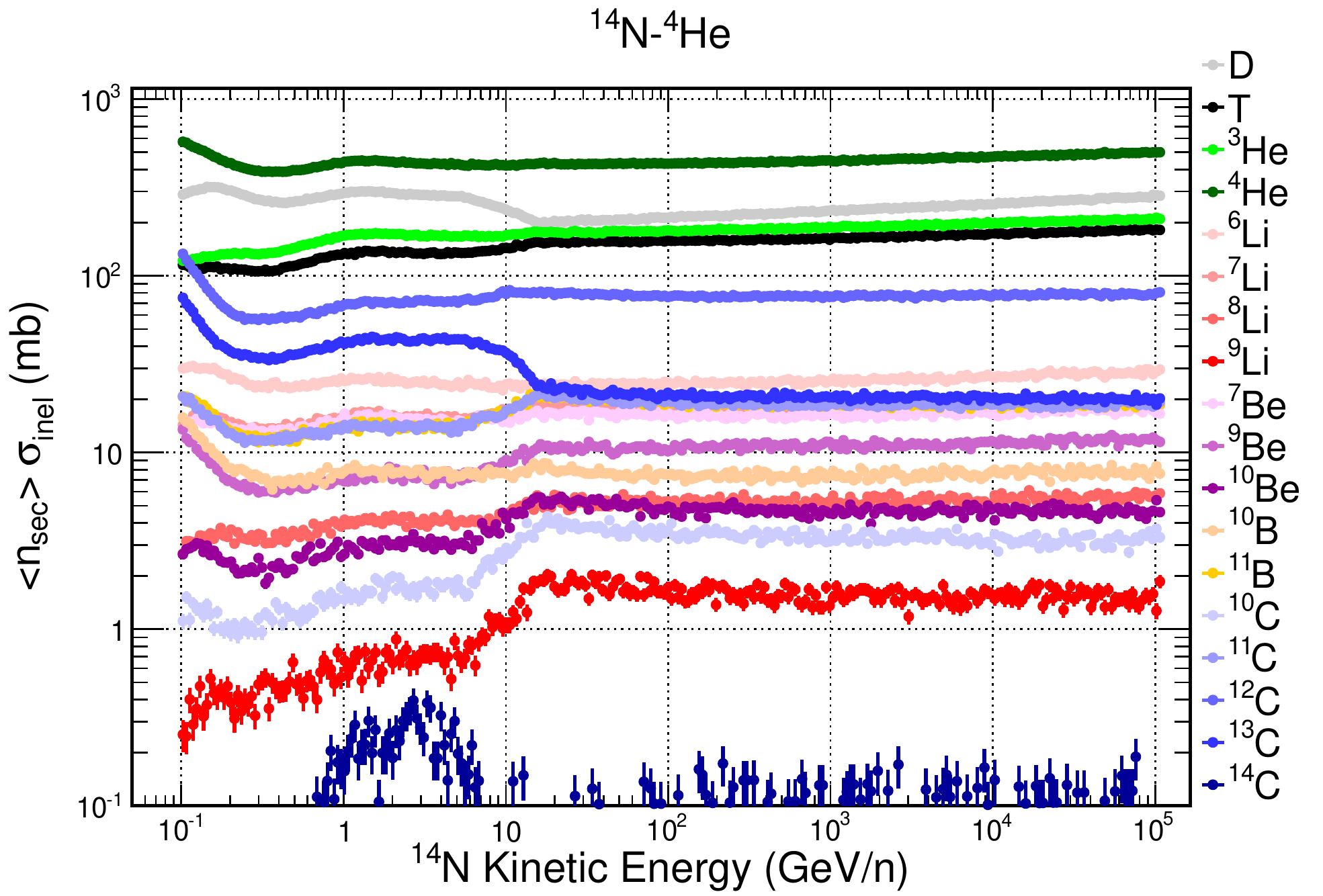} \\
\end{tabular}
\caption{Inclusive cross sections for the production of spallation nuclei in collisions of
$^{12}C$ and $^{14}N$ with p and $^{4}He$ nuclei.
The plots show the cross sections for the production of Deuteron (gray markers), Triton (black markers) 
and for the isotopes of $He$ ($^{3}He$ and $^{4}He$, green markers), 
$Li$ ($^{6}Li$, $^{7}Li$, $^{8}Li$ and $^{9}Li$, red markers), $Be$ ($^{7}Be$ and $^{9}Be$, magenta markers), 
$B$ ($^{10}B$ and $^{11}B$, orange markers) and 
$C$ ($^{10}C$, $^{11}C$, $^{12}C$, $^{13}C$ and $^{14}C$, blue markers). 
Lighter (darker) color shades correspond to lighter (heavier) isotopes.}
\label{FigSiHeXsecDPM}
\end{figure*}
Since we are interested in the production of stable particles, after each collision
all the secondary products are followed inside a spherical volume filled with vacuum with a
very large radius to allow the decay of all unstable particles apart the radionuclides within 
the volume of the sphere.

The inclusive cross sections calculated in this way will rule the CR propagation within  
the astronomical environment. All stable secondaries produced in each interaction will
be propagated together with the primaries and will possibly undergo further interactions  
according to the corresponding cross sections. In a similar way, unstable nuclei produced 
in spallation processes will be able either to interact or decay according to their 
cross sections and lifetimes.


The simulation with {\tt FLUKA} is performed with the following settings:
\begin{itemize} 
\item the {\tt PEANUT} package is activated in the whole energy range for any reaction involving a proton as projectile or
target;
 \item the minimum kinetic energy for {\tt DPMJET-III} is set to $10 \units{GeV/n}$ (applying only to reactions between two
nuclei heavier than a proton);
 \item the minimum kinetic energy for {\tt RQMD} is set to $0.100 \units{GeV/n}$ (applying only to reactions between two
nuclei heavier than a proton).
\end{itemize} 

The first quantities we want to investigate are the total inelastic cross sections for the collisions
we are focusing on, and -- as an important cross check -- the pion production cross section (for the proton-proton case)
compared to current data.

Fig.~\ref{FigXsecInel} shows the total inelastic cross section calculated in the {\tt FLUKA} code for the
collision processes investigated in the present work. Our cross section model was validated with the most recent LHC data, 
as shown in Fig. 4 of Ref.~\cite{fed2015}.

Fig.~\ref{FigXsecPion} shows the pion inclusive cross sections in $p-p$ collisions calculated from the {\tt FLUKA} 
simulation at the interaction level compared to current data.
It is worth here to point out that some short-lived particles produced in the interaction (i.e. $\eta$) 
promptly decay in this phase, and the decayed products cannot be separated
(i.e. $\eta \rightarrow \pi$). Therefore the calculated inclusive cross sections shown 
in Fig.~\ref{FigXsecPion} include these products.

As mentioned above, the {\tt FLUKA} results are compared against the available data from experiments: 
the results of these comparisons are already available in the papers cited in the current work. In particular, 
in Ref.~\cite{bohlen} the double differential $\pi^{+}$ production cross sections for $158 \units{GeV/c}$ 
protons on carbon as a function of transverse momentum $p_{T}$ for different values of the Feynman $X$
are compared with experimental data~\cite{Alt:2006fr}.
In the same reference, a comparison of the $p_{T}$ integrated yields of pions (for $158 \units{GeV/c}$ protons 
on carbon~\cite{Alt:2006fr}) and kaons (for $158 \units{GeV/c}$ protons on proton~\cite{Anticic:2010yg}) with the data is also shown. An agreement 
at the level of a few percent is obtained in the relevant range of Feynman $X$, both for pion and kaon production. 
As a further check, in Fig.~\ref{Figna49} we show the anti-proton production spectrum as a function of the 
Feynman $X$ for $158 \units{GeV/c}$ proton on carbon compared with the experimental data~\cite{na49}. 
Also in this case a good agreement is found with measured data.

With these results in hand, we can present in full detail all the results of our analysis
for all stable particles (including radionuclides whose decay is not taken into account at this stage).
First of all we turn our attention on the light products of $p-p$, $p-He$ and $He-He$ collisions. 
The species produced in these interactions include a wide range of charged particles and antiparticles, plus gamma rays and neutrinos. 
Studying these secondary products is extremely important for several applications of CR physics, and also to set contraints 
on possible exotic contributions (Dark Matter annihilation or decay).

In Figs.~\ref{FigpHe2DXsec},~\ref{FigpNHeHe2DXsecDPM},~\ref{FigpHe2DXsecApx},~\ref{FigpNHeHe2DXsecDPMApx} we show the differential inclusive cross sections for the 
production of several light secondaries (protons, electrons and positrons, gamma rays, electron neutrinos and antineutrinos,
muon neutrinos and antineutrinos, deuterons) in $p-p$, $p-^{4}He$, $^{4}He - ^{4}He$ and $p - ^{14}N$ interactions
as a function of the kinetic energy per nucleon of the projectile and of the kinetic energy of the secondary particle.

As shown in Fig.~\ref{FigpNHeHe2DXsecDPM}, the inclusive cross section for the production of
gamma rays in $p - ^{14}N$ collisions exhibits some horizontal lines in the \units{MeV} region, due to nuclear de-excitation processes.
This feature is common to all collisions involving heavy nuclei. 


The differential inclusive cross sections for the production of electrons, positrons, electron neutrinos 
and electron antineutrinos shown in Figs.~\ref{FigpHe2DXsec},~\ref{FigpNHeHe2DXsecDPM},~\ref{FigpHe2DXsecApx} 
and ~\ref{FigpNHeHe2DXsecDPMApx} are evaluated taking into account the decay of neutrons and antineutrons.
These decays produce low-energy secondaries and provide the main contribution to the inclusive cross sections
for secondary energies below a few tens of \units{MeV}. Therefore in this region the electron (positron) 
and the antineutrino (neutrino) cross section will have similar behaviors.

In Figs.~\ref{FigHeOthXsec},~\ref{FigHeOthXsecApx1} and ~\ref{FigHeOthXsecApx2} we show the total inclusive cross sections for the production of 
protons, electrons and positrons, gamma rays, electron neutrinos and antineutrinos,
muon neutrinos and antineutrinos and Deuteron, as a function of the kinetic energy per nucleon of the projectile in several interactions
involving different projectiles and targets. 
As discussed above, the inclusive cross section for the production 
of electron antineutrinos is similar to that for the production of electrons, and the cross section for the
production of positrons is similar to that for the production of electron neutrinos. 
 
In the case of heavy projectile and/or target, the yields of secondary nuclei are not negligible.
Some secondary nuclei are unstable: this is for instance the case of Triton, that decays 
into $^3He + e^- + \bar{\nu}_e$ with a half-life of $12.32 \units{years}$.
In our simulation the decay of unstable nuclei produced in spallation processes is not included
and must be taken into account when describing CR propagation.

In Fig.s~\ref{FigSiHeXsecDPM} and ~\ref{FigSiHeXsecDPMApx} we show the most relevant results among all the calculations we
performed for the heavy nuclei. 
We plot the inclusive cross sections related to the the interactions of $^{12}C$, $^{14}N$, $^{16}O$, $^{20}Ne$, $^{24}Mg$ 
and $^{28}Si$ projectiles with protons and $^{4}He$ nuclei, yielding deuterons, tritons and all the isotopes
of $He$, $Li$, $Be$, $B$ and $C$.

Concerning ion fragmentation in proton-nucleus collisions, several different verifications of the {\tt FLUKA} results 
have been already published elsewhere. For instance, Ref.~\cite{bohlen} contains the excitation function 
of $^{11}C$ for $p + C$ reactions up to $1\units{GeV}$, while Ref.~\cite{batt2015} shows 
the mass spectrum of fragmentation products from $Pb + p$ reactions at 1 $\units{GeV/n}$. 
Furthermore, Ref.~\cite{Ballarini:2004ai} covers the high energy side, with mass spectra 
for $p+Ag$ and $p+Au$ reactions at $300 \units{GeV}$ and $800 \units{GeV}$ respectively. 
Finally, Ref.~\cite{ferrari1997} reports the excitation functions of several fragmentation products 
from $p + ^{59}Co$ reactions up to $700 \units{MeV}$.
Concerning fragmentation in nucleus-nucleus reactions, Ref.~\cite{sihver} contains many charge-changing 
cross sections for different reactions up to $1 \units{GeV/n}$, while Ref.~\cite{braun} does the same 
for $^{208}Pb$ induced reactions at $158 \units{GeV/n}$.

We remark that this is the first time in which such a complete sample of nuclear cross sections are computed in a consistent way
with a single numerical code. Due to the completeness and accuracy of these computations, we point out that the dataset we 
produced can be used by the community working on CR physics to constrain the properties of CR transport in a more solid way, 
as we will show in the Section~\ref{dragon}.

\begin{figure}[!ht]
\includegraphics[width=1\columnwidth,height=0.25\textheight,clip]{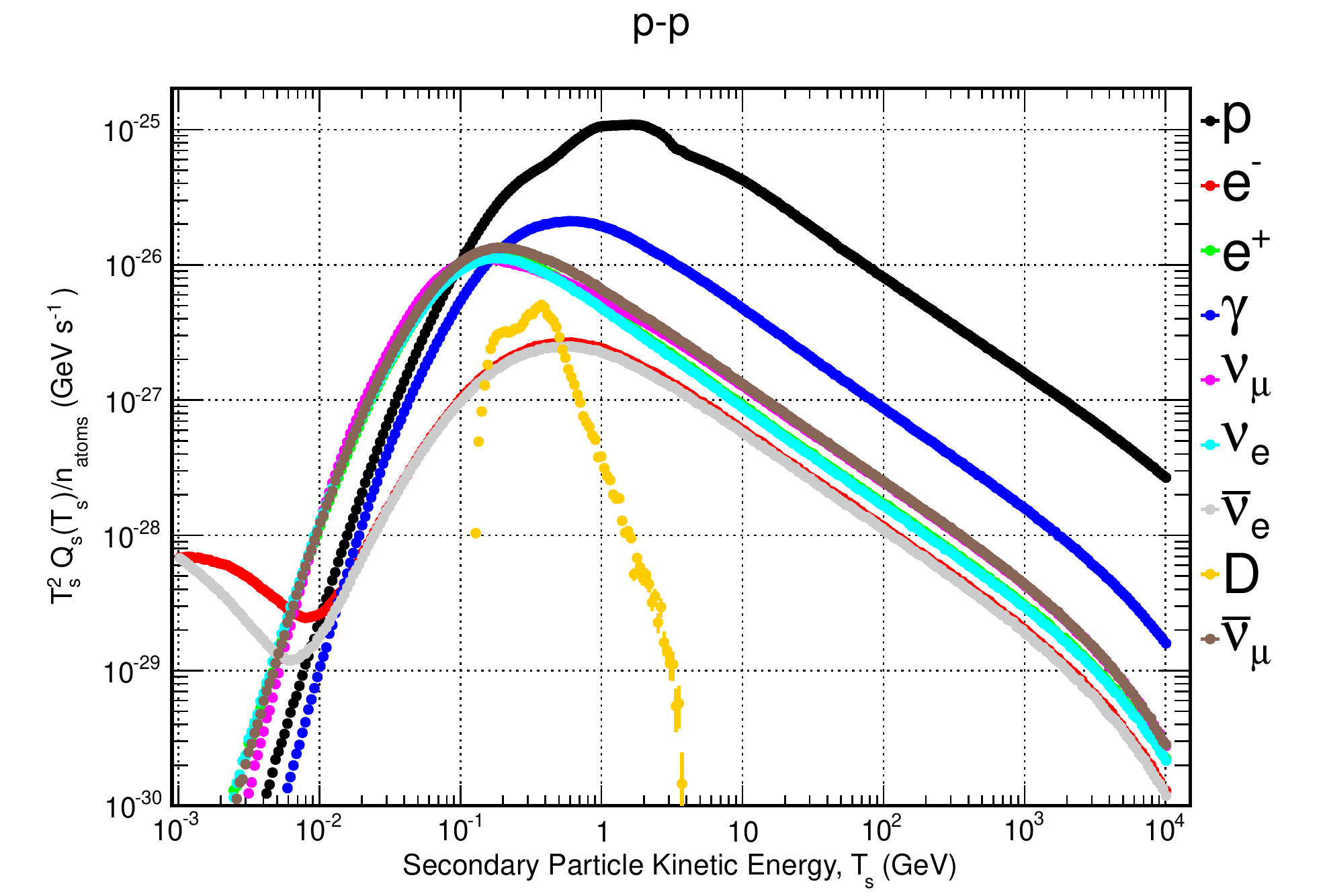} 
\includegraphics[width=1\columnwidth,height=0.25\textheight,clip]{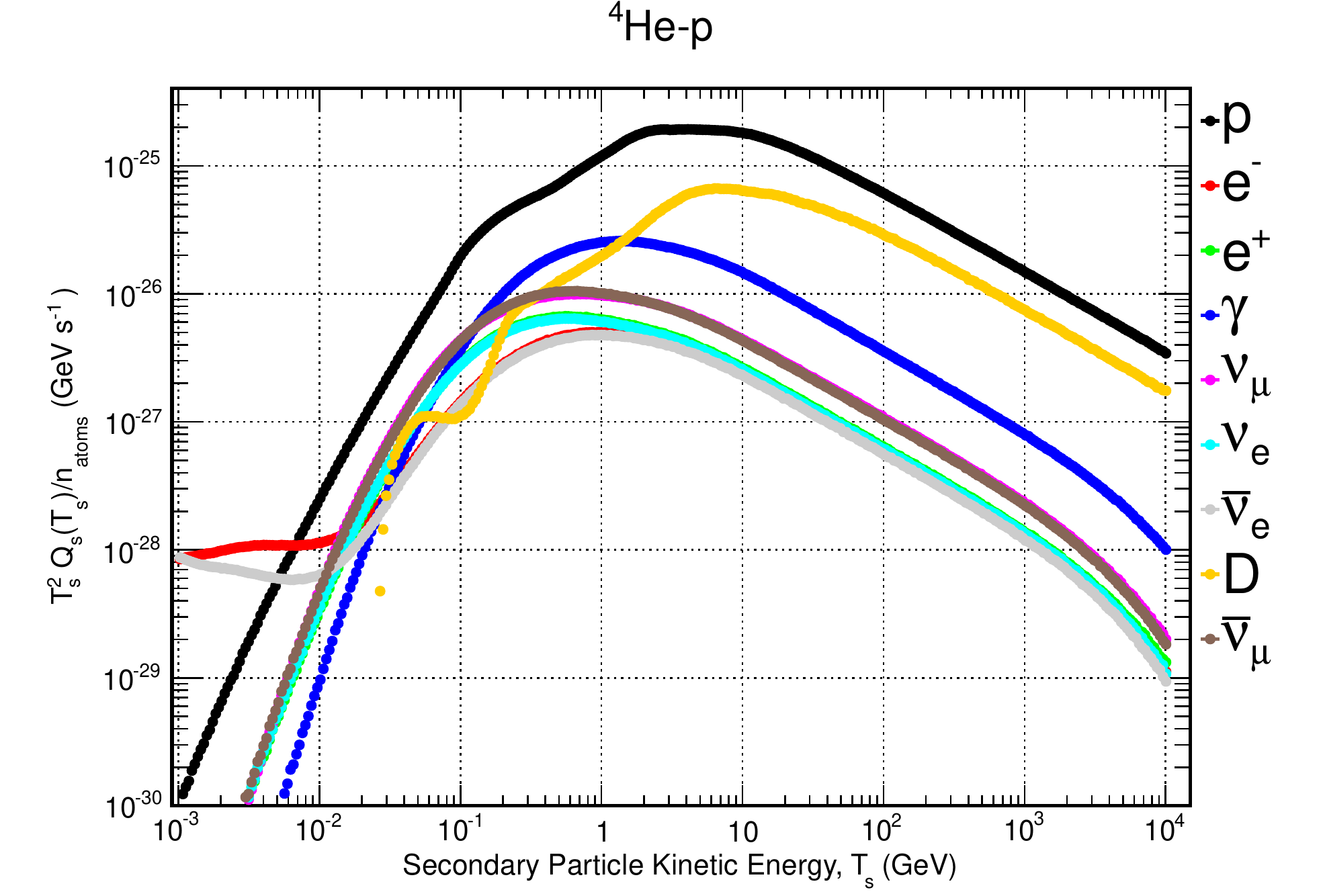}
\caption{Secondary spectra for proton-proton interaction (top panel) and $^4He$-proton interaction
(bottom panel). Protons (black), electrons (red), 
positrons (green), gamma rays (blue), electron neutrinos (cyan), 
electron antineutrinos (grey), muon neutrinos (magenta), muon antineutrinos (brown)
and Deuterons (orange) The CR proton and Helium intensity is given by the Eq.~\ref{Eqpintensity}.}
\label{FigppHepSecEmiss}
\end{figure}

\section{Evaluation of the energy spectra of secondary cosmic rays}

As a first application, we evaluate the energy spectra of several secondary species
produced in the interactions of primary CRs with the ISM. We will consider a simplified 
CR model in which the differential particle momentum density of any 
primary species is given by a power law:

\begin{equation}
n(p) = k_{0} \left( \frac{p}{p_{0}} \right)^{-\alpha}
\end{equation}
where $\alpha$ is the spectral index and $k_{0}$ is a normalization constant.
In writing the previous equation we introduced a momentum scale 
$p_{0}=1~\units{GeV}$. Therefore $k_{0}$ will be expressed in the same units 
as $n(p)$, i.e. in $\units{GeV^{-1} m^{-3}}$.

The spectral differential intensity in momentum is obtained 
by multiplying $n(p)$ by the factor $\beta c / 4\pi$, where $\beta c$ 
is the particle velocity: 

\begin{equation}
J(p) = \frac{\beta c}{4\pi} n(p) = k \beta \left( \frac{p}{p_{0}} \right)^{-\alpha}.
\label{eqlispowerlaw}
\end{equation}
where $k=k_{0}c/4\pi$ ($k$ will be expressed in the
same units as $J(p)$, i.e. in $\units{GeV^{-1} m^{-2} s^{-1} sr^{-1}}$). 

The differential intensity in momentum can be converted into a differential intensity 
in kinetic energy taking into account that:

\begin{equation}
J(T) = \frac{dp}{dT}~J(p)
\end{equation}
\begin{equation}
\frac{dp}{dT} = \frac{T+m}{\sqrt{T(T+2m)}} = \frac{1}{\beta}
\end{equation}
where $m$ is the rest mass of the particle. 
The CR spectrum as a function of the kinetic energy is therefore given by:

\begin{equation}
J(T) = k \left[ \frac{T(T+2 m)}{p_{0}^{2}} \right]^{-\frac{\alpha}{2}}
\label{eqlis}
\end{equation}
 
In performing the calculations we assume that the energy spectrum of each primary species is given by:

\begin{equation}
 J(T) = k \left[ \frac{T(T+2 m)}{p_{0}^{2}} \right]^{-\frac{\alpha}{2}}  e^{-\frac{T}{T_{max}}}~\varTheta(T_{max}-T).
 \label{Eqpintensity}
\end{equation}
The term $\varTheta(T_{max}-T)$ has been introduced in the formula to take into account that in our simulation we generated 
primary CRs with energies up to a maximum value $T_{max}$.
We simulated both proton and Helium primaries. 
For the protons we set $k=2.2\units{GeV^{-1}~cm^{-2}~s^{-1}~sr^{-1}}$~\cite{Dermer2012} and
$\alpha=2.75$, while for the Helium nuclei we set 
$k=1.0\units{GeV^{-1}~cm^{-2}~s^{-1}~sr^{-1}}$ and
$\alpha=2.66$. The cutoff energy value was set at $T_{max}=10^{5}\units{GeV}$.

The energy spectrum of the secondary particle $s$ is usually expressed in terms of
the production function $Q_{s}$, that is defined as:

\begin{equation}
\frac{Q_s(T_s)}{n_{atoms}} = 4 \pi \int J(T) \frac{d \sigma(T_s | T)}{dT_s} dT
 \label{Eq:emiss}
\end{equation}
where $n_{atoms}$ is the number of target atoms per unit volume. The ratio
$Q_{s}(T_{s})/n_{atoms}$ is expressed in units of $\units{GeV^{-1}~s^{-1}}$.
Fig.~\ref{FigppHepSecEmiss} shows the energy spectra of secondaries
produced by proton and Helium primaries colliding with proton targets.

\begin{figure*}[!ht]
\begin{center}
\begin{tabular}{cc}
\includegraphics[width=1.\columnwidth,height=0.25\textheight,clip]{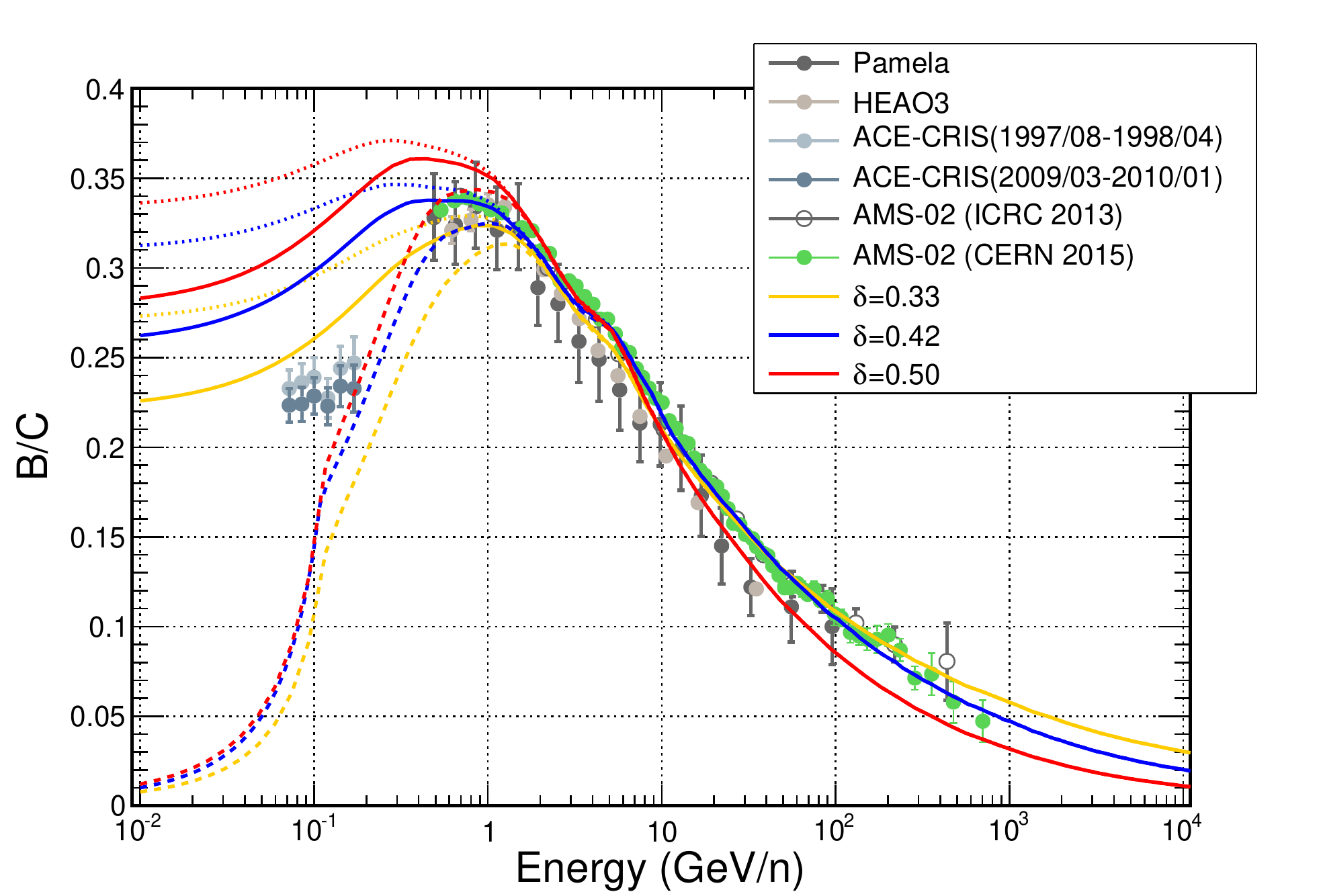} &
\includegraphics[width=1.\columnwidth,height=0.25\textheight,clip]{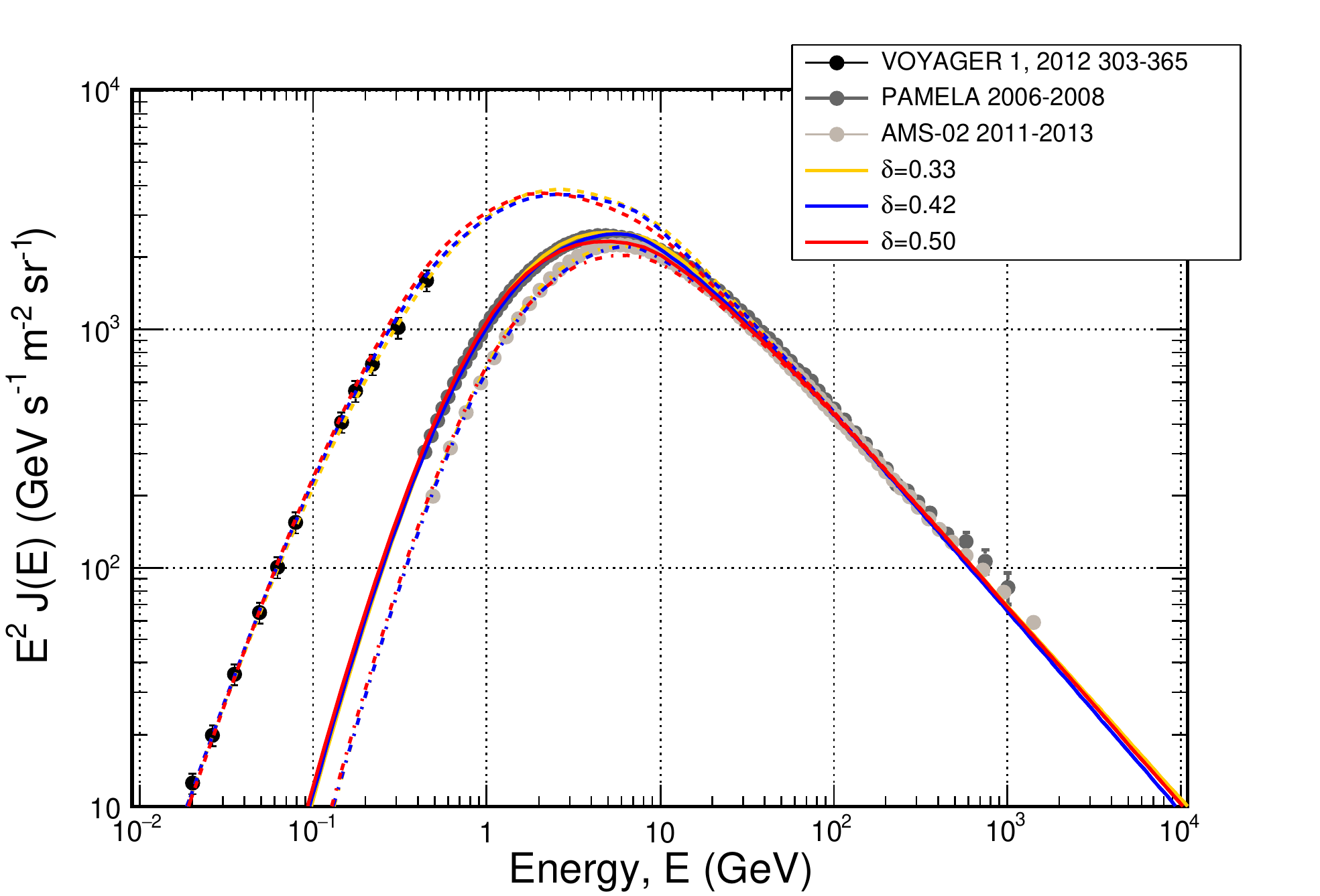} \\
\includegraphics[width=1.\columnwidth,height=0.25\textheight,clip]{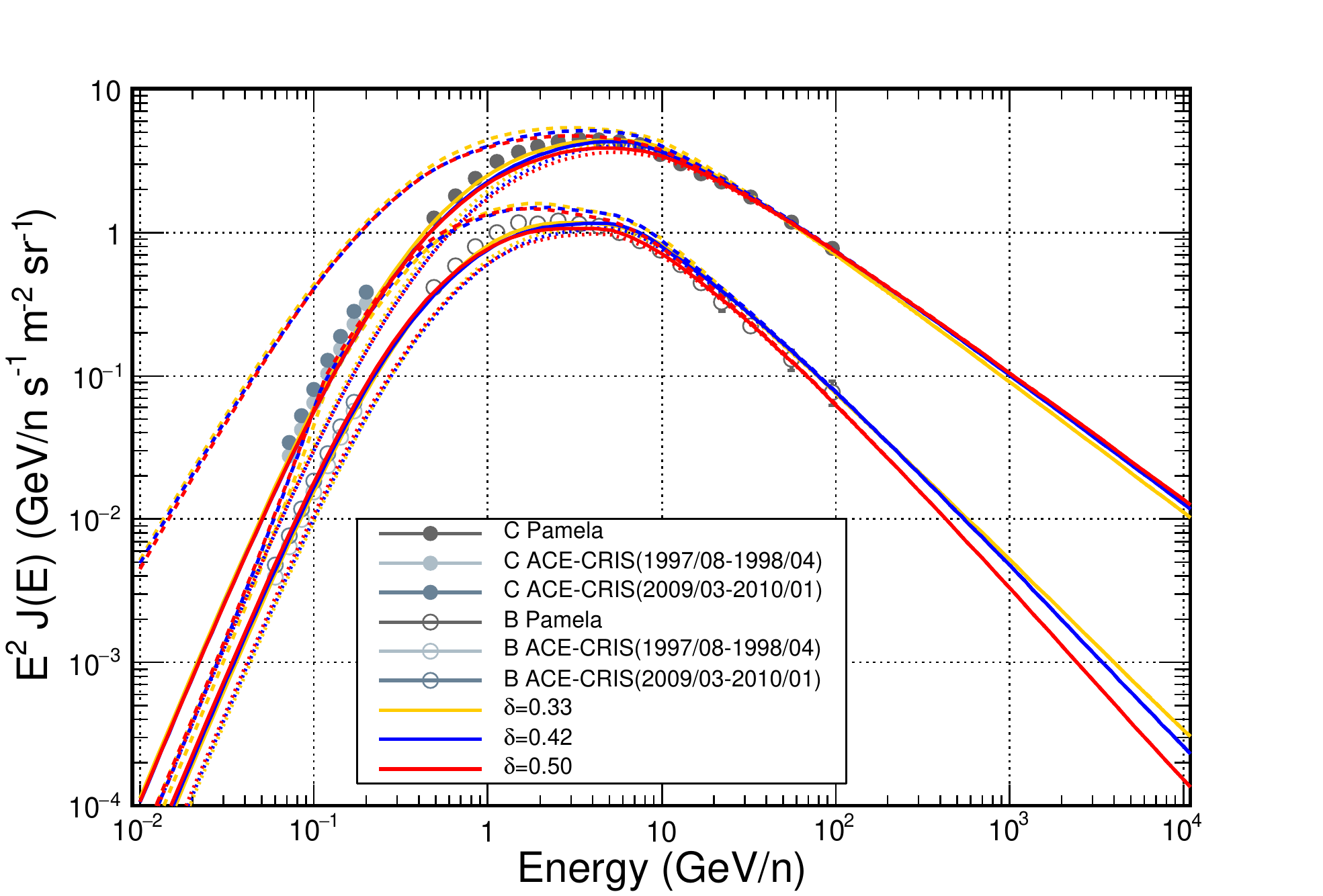} &
\includegraphics[width=1.\columnwidth,height=0.25\textheight,clip]{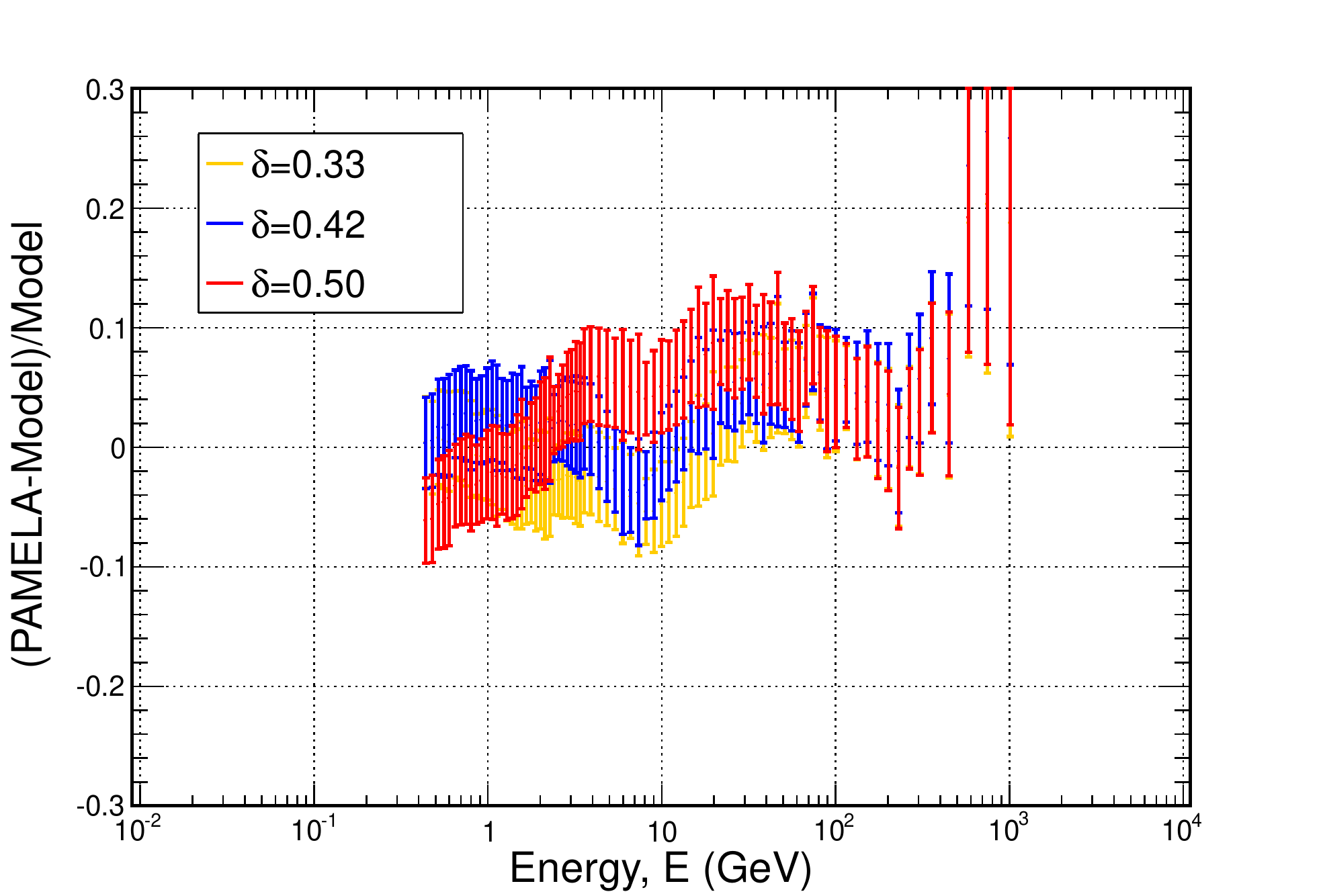} \\
\includegraphics[width=1.\columnwidth,height=0.25\textheight,clip]{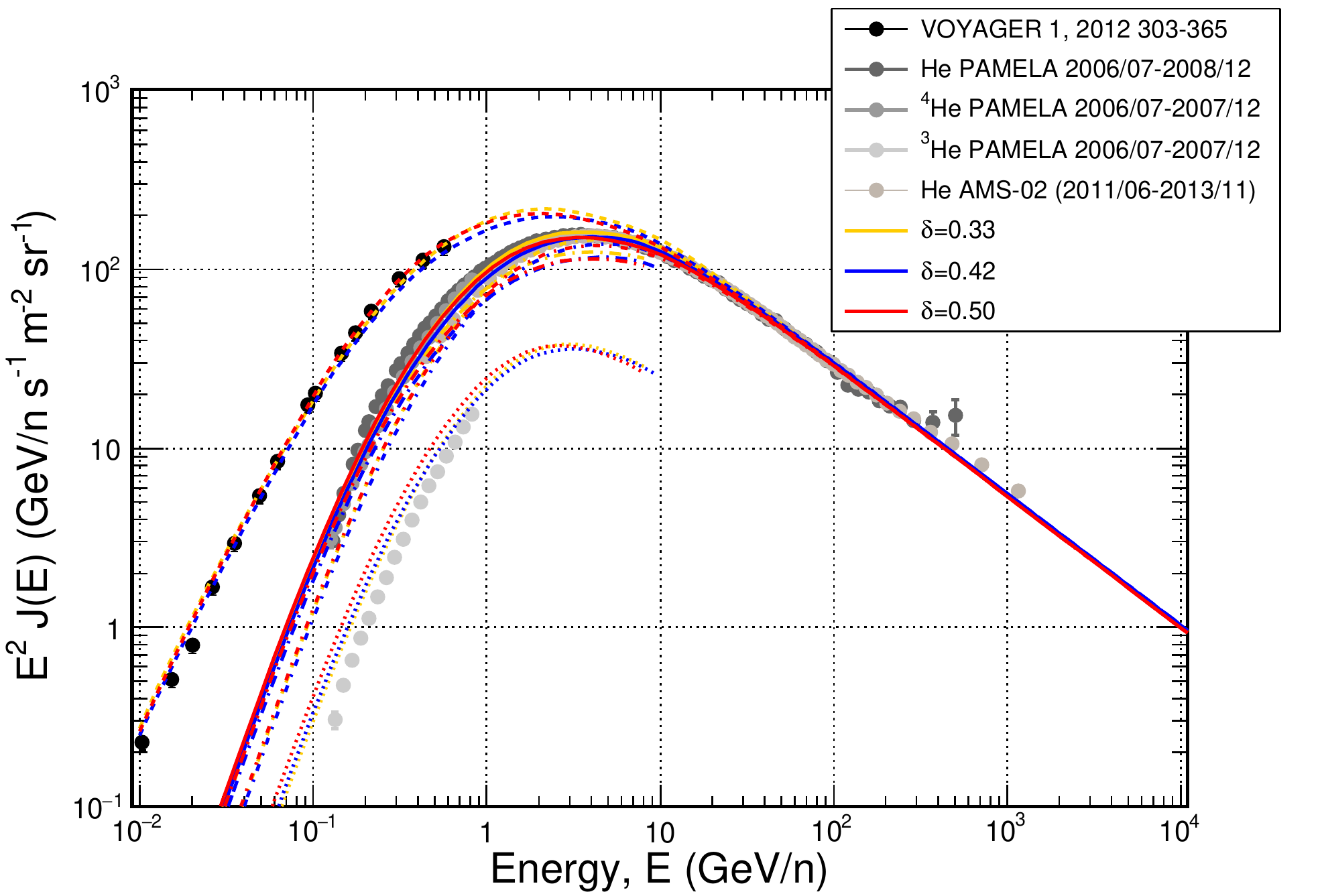} &
\includegraphics[width=1.\columnwidth,height=0.25\textheight,clip]{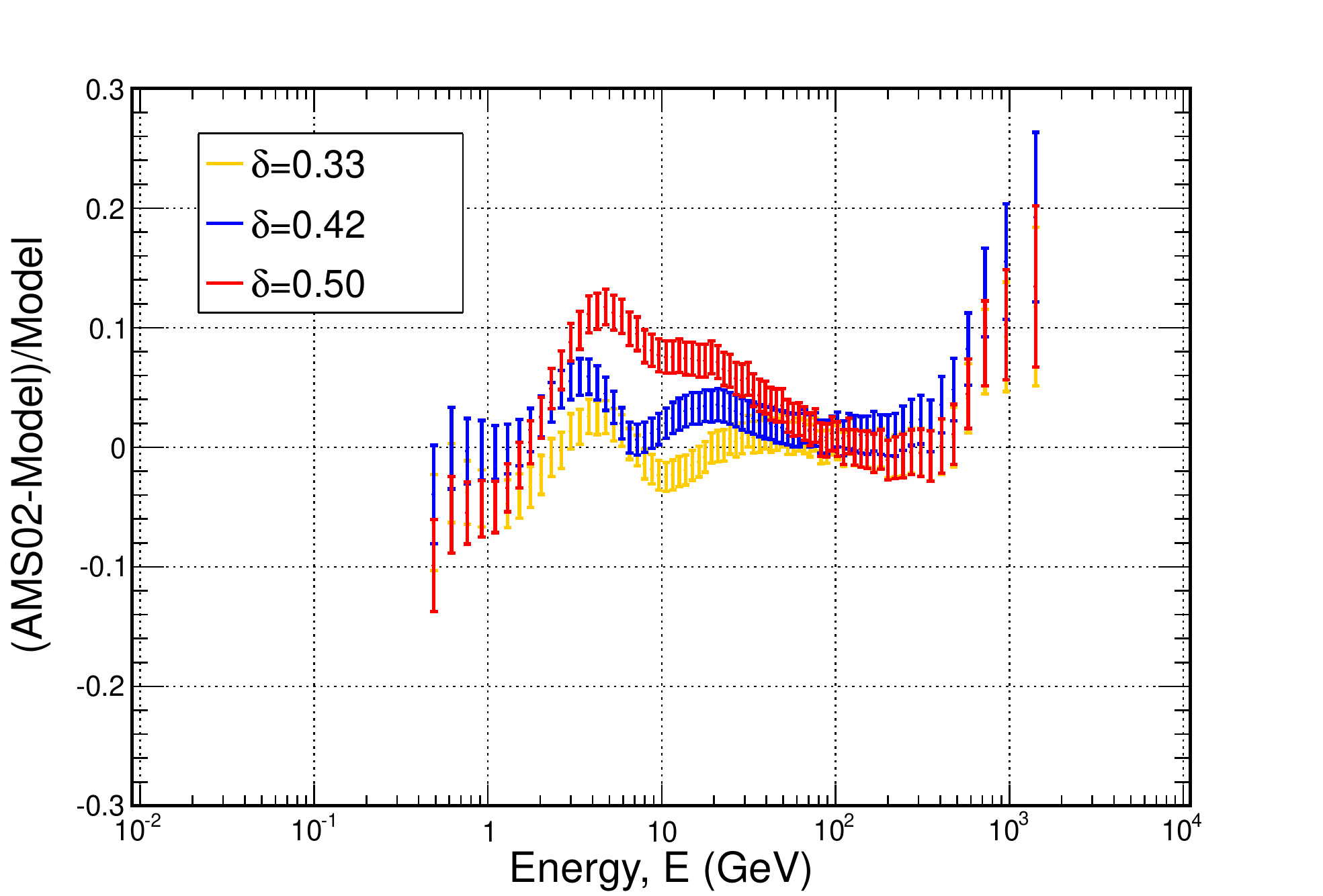} 
\end{tabular}
\end{center}
\caption{DRAGON results. The models are: $\delta=0.33$ (KOL), $\delta=0.42$ and $\delta=0.50$ (KRA). 
Left, from top to bottom: Boron to Carbon ratio, Carbon and Boron and Helium. Right, from top to bottom, proton comparison with PAMELA and AMS02 data. Dashed line: unmodulated intensity: solid (dotted) line: modulated intensity by means of the force field approximation with $\phi$=0.42 (0.62) GV. Fractional residual betweet PAMELA (middle) and AMS02 (bottom). VOYAGER1~\cite{stone2013}; PAMELA~\cite{Adriani:2011cu,Adriani:2013tif}; ACE-CRIS~\cite{acecris}; AMS01~\cite{ams01}; HEAO3-C2~\cite{heao3} (some data points are quoted from the Database of Charged Cosmic Rays~\cite{crdb}) }
\label{FigDragonFluka}
\end{figure*}

\section{Application to the Galactic emission}
\label{dragon}

In order to evaluate the emission from the Galaxy we use a customized version of the propagation code
{\tt DRAGON}~\cite{Evoli:2008dv,Gaggero:2013rya} (2D version)\footnote{A version of {\tt DRAGON} code is available for download at the link http://www.dragonproject.org/.} in which we introduced the present parameterizations in the secondary production routine.

We study the spectra of the following species: protons, Helium, Boron, Carbon, 
electrons, and positrons; for each one we evaluate the contribution of secondaries originating from nuclear interactions.
In our simulation we assume that the ISM is composed of Hydrogen and Helium with relative abundances $1:0.1$.

Therefore, in the case of leptons, we include the contributions from $p-p$, $p-^{4}He$, $^{4}He-p$ and $^{4}He - ^{4}He$ collisions. 
On the other hand, in the hadronic case, we include all the collisions of 
$p$, $D$, $T$, $^{3}He$, $^{4}He$, $^{6}Li$, $^{7}Li$, $^{9}Be$, $^{10}Be$, $^{10}B$, $^{11}B$, $^{12}C$, $^{13}C$, $^{14}C$, $^{14}N$, $^{15}N$, $^{16}O$, $^{17}O$, $^{18}O$, $^{20}Ne$, $^{24}Mg$ and $^{28}Si$ on proton and Helium targets.
The $\beta^{\pm}$ decays and the electron captures of unstable isotopes are taken into account according to their lifetimes. 
In the current version of the \texttt{DRAGON} code the daughter nucleus carries out all the energy of the parent nucleus, while the other products 
(i.e. leptons) are discarded.

The purpose of this section is to discuss the implications of the present parameterizations on some reference diffusion models.
The setup we are using is standard: CR transport properties are homogeneous and isotropic; the scalar diffusion coefficient depends on 
the particle rigidity $R$ and on the distance from the Galactic Plane $z$ according to the following parameterization:

\begin{equation}
D \,\, = \,\, D_0 \, \beta^{\eta} \, \left({\frac{R}{R_0}}\right)^{\delta} \, e^{|z|/z_t}.
\end{equation}
where:
\begin{itemize} 
\item $D_0$ is the diffusion coefficient normalization at the reference rigidity $R_0 = 4\units{GV}$;
\item $\eta$ effectively describes the complicated physical effects that may 
play a major role at low energy (below $1 \units{GeV}$), e.g. the dissipation of Alfv\'en waves 
due to the resonant interaction with CRs (here $\beta$ represents as usual the particle velocity 
in units of the speed of light);
\item $\delta$ is the spectral index: it is constrained by the data on the 
light nuclei ratios, in particular by the measurements of the Boron-over-Carbon (B/C) ratio;
\item $z_t$ is the scale height of the diffusive halo of the Galaxy. 
\end{itemize}

\begin{figure}[!ht]
\begin{center}
\includegraphics[width=1\columnwidth,height=0.25\textheight,clip]{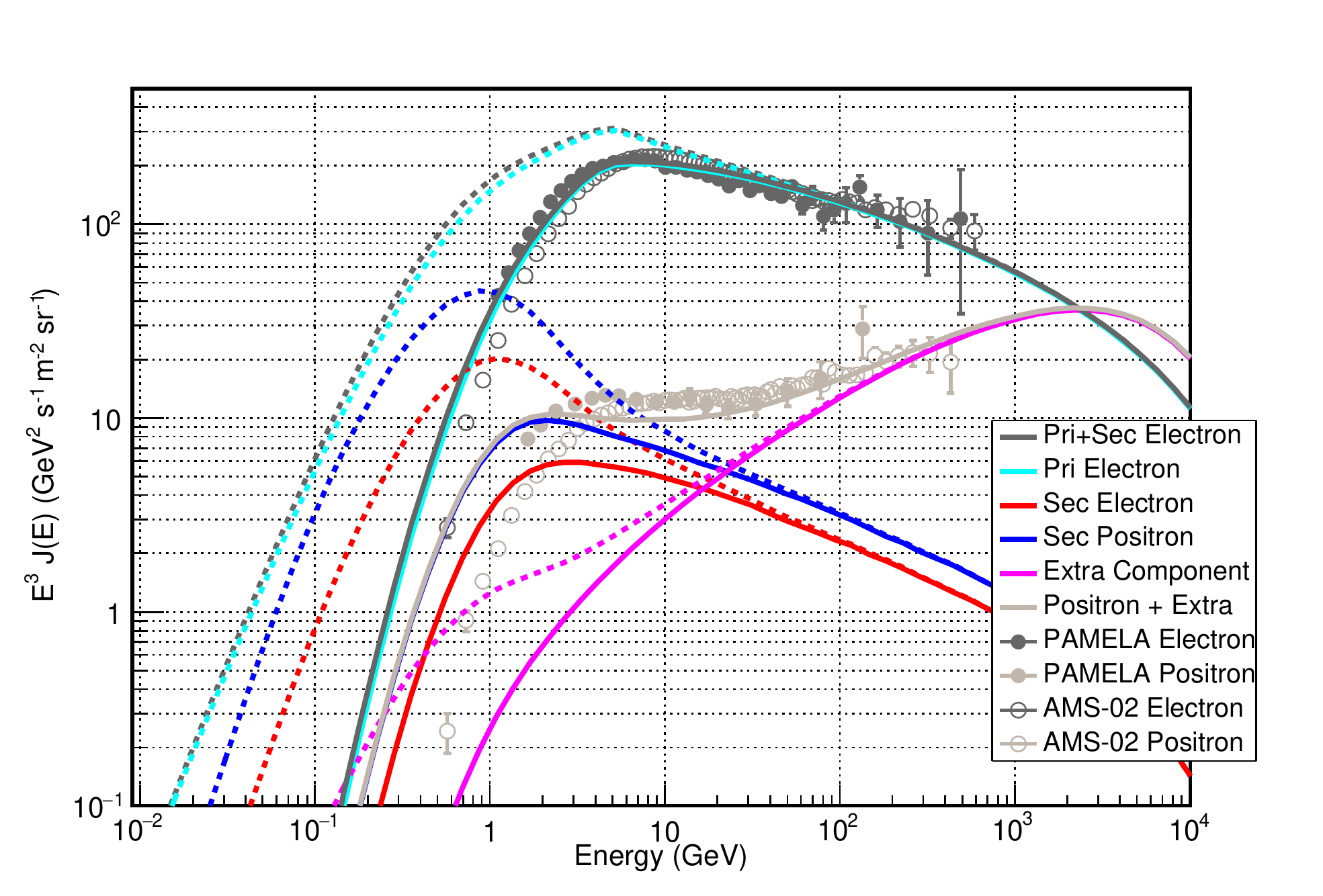} 
\includegraphics[width=1\columnwidth,height=0.25\textheight,clip]{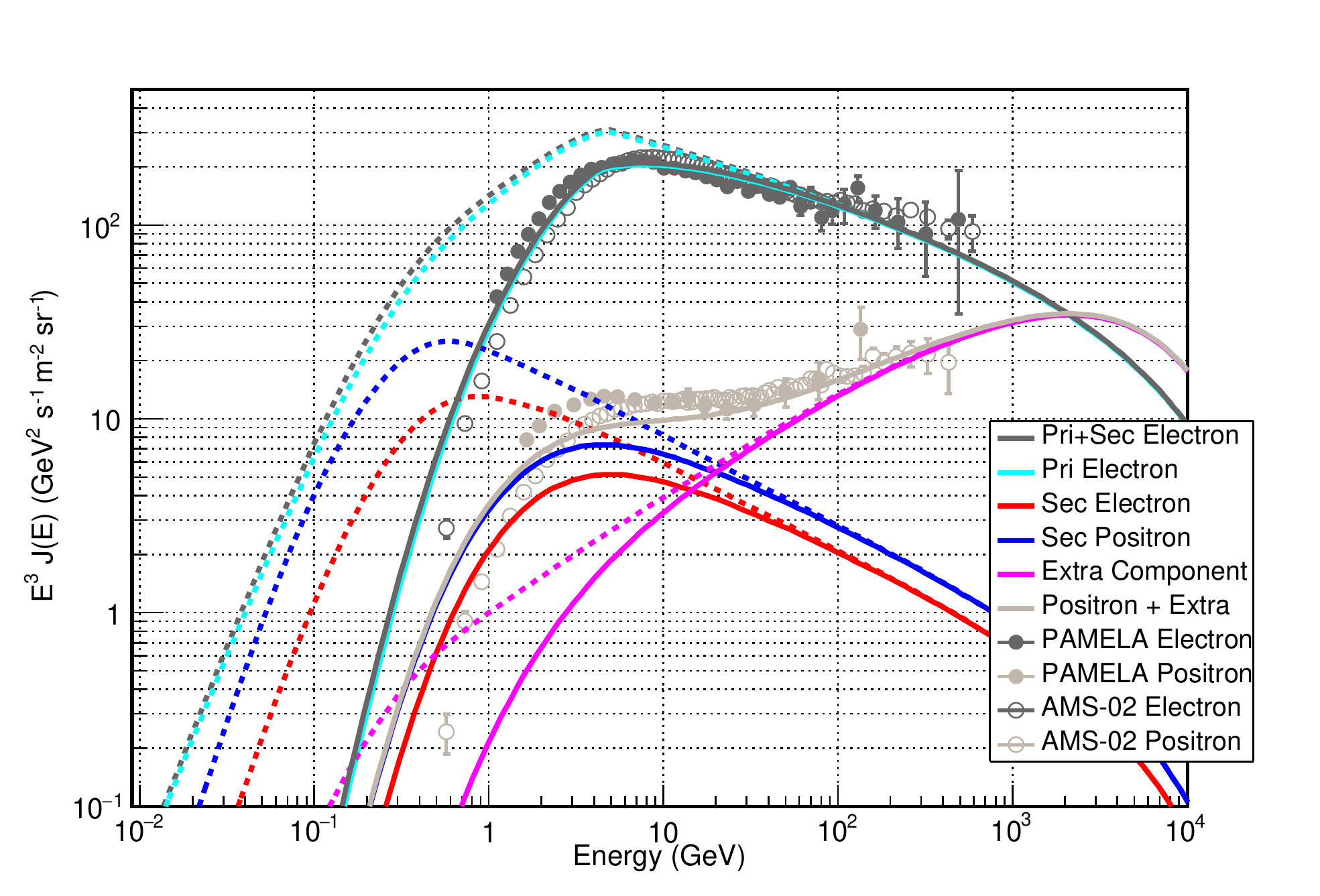}  
\includegraphics[width=1\columnwidth,height=0.25\textheight,clip]{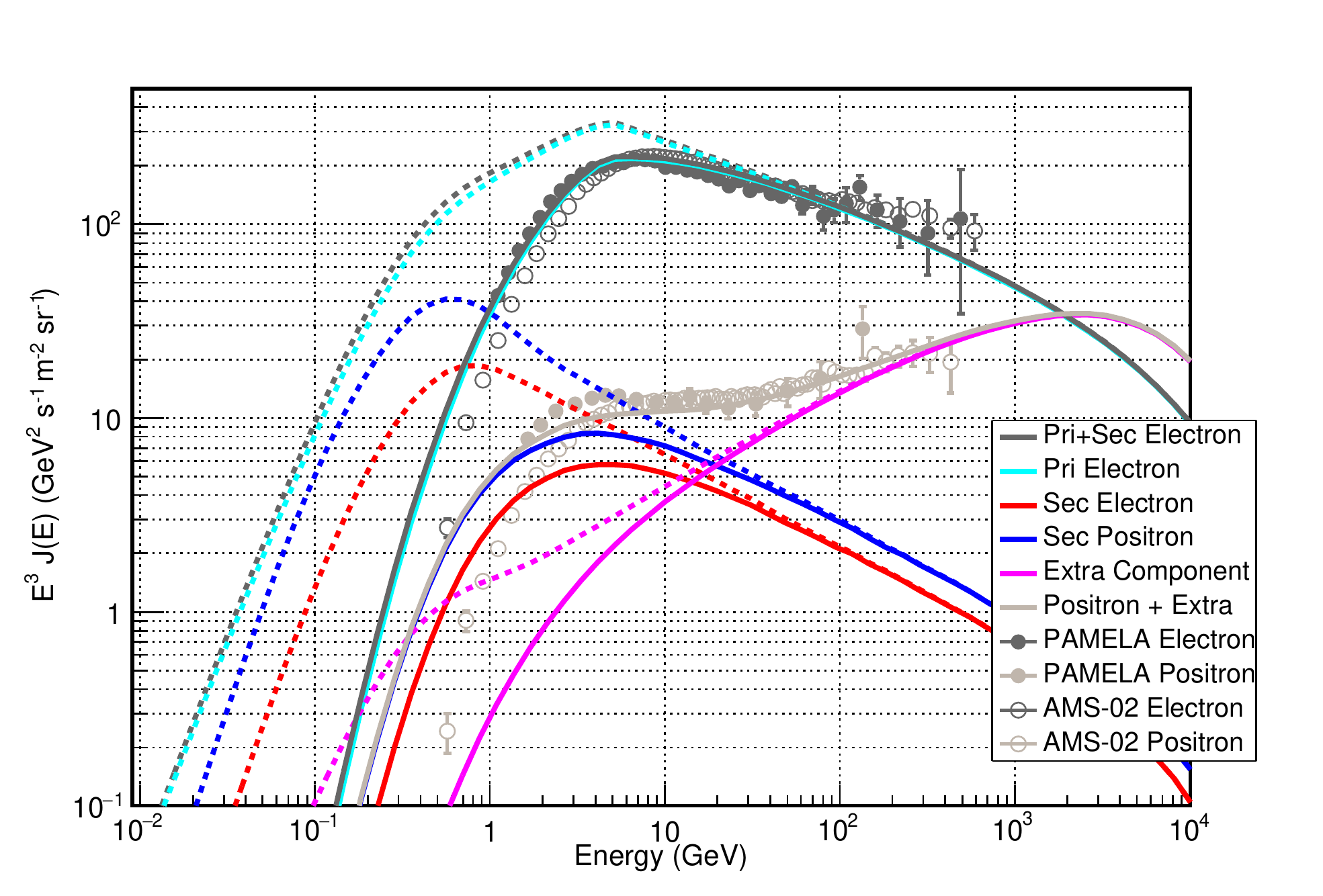} 
\end{center}
\caption{DRAGON results for electrons and positrons. The models are (from top to bottom): $\delta=0.33$ (KOL), $\delta=0.42$ and $\delta=0.50$ (KRA). 
Dashed line: unmodulated intensity: solid line: modulated intensity by means of the force field approximation with $\phi$=0.42 GV (some data point are quoted from the Database of Charged Cosmic Rays~\cite{crdb}) }
\label{FigDragonFlukaELE}
\end{figure}

\begin{table}[!hb]
\begin{center}
\begin{tabular}{|c|c|c|c|c|} \hline \hline 
Nuclei & \multicolumn{3}{ |c| }{Spectral indices} & Relative \\ \cline{2-4}
            &   KOL     & KRA       & AMS02     &  Fraction            \\ \hline 
$^{1}H$     & 2.05/2.48 & 2.05/2.33 & 2.05/2.43 & 1                     \\ \hline
$^{4}He$    & 2.18/2.40 & 2.18/2.24 & 2.18/2.32 & $6.13 \times 10^{-2}$ \\ \hline 
$^{12}C$    & 2.20/2.60 & 2.20/2.40 & 2.20/2.50 & $3.11 \times 10^{-3}$ \\ \hline
$^{14}N$    & 2.20/2.60 & 2.20/2.40 & 2.20/2.50 & $2.16 \times 10^{-4}$ \\ \hline
$^{16}O$    & 2.20/2.60 & 2.20/2.40 & 2.20/2.50 & $3.61 \times 10^{-3}$ \\ \hline
$^{20}Ne$   & 2.20/2.60 & 2.20/2.40 & 2.20/2.30 & $2.94 \times 10^{-4}$ \\ \hline
$^{24}Mg$   & 2.20/2.60 & 2.20/2.40 & 2.20/2.50 & $6.21 \times 10^{-4}$ \\ \hline
$^{28}Si$   & 2.20/2.60 & 2.20/2.40 & 2.20/2.50 & $6.85 \times 10^{-4}$ \\ \hline
$e^-$       & 1.60/2.50 & 1.60/2.50 & 1.60/2.50  &                       \\ \hline
Extra $e^+$ & 1.78      & 1.78      & 1.78      &                       \\ \hline \hline
\end{tabular}
\caption{Main parameters of the injection spectra of the primary species.}
\label{tab1}
\end{center}
\end{table}

Our starting points consist of two diffusion models 
considered as a reference in several previous works (see e.g. \cite{Evoli:2012ha}) labelled as KRA and KOL: 
they are mainly tuned on PAMELA B/C data. We also present a new model tuned on the recent preliminary AMS02 
data \cite{ams_preliminary}. In the following we summarize the main features of these models:

\begin{itemize}
\item
KRA features $\delta = 0.5$ (compatible with a Kraichnan turbulence in the
framework of Quasi-Linear Theory), a moderate level of reacceleration (The Alfv\'en velocity is $\simeq 15\units{km/s}$), and an altered diffusion coefficient at low energy parametrized by $\eta \simeq - 0.4$. 
\item
KOL is based on $\delta = 0.33$ (compatible with a Kolmogorov turbulence in the
framework of Quasi-Linear Theory), a high level of reacceleration (The Alfv\'en velocity is $\simeq 30\units{km/s}$).
\item
The model tuned on recent preliminary AMS02 data has $\delta = 0.42$ and the same  Alfv\'en velocity as the KRA model. This model is based on the one presented in \cite{Evoli:2015}
\end{itemize}

In all cases the injected spectra up to Silicon nuclei are described by a broken power law model; the main parameters describing the injection spectra are summarized in Table~\ref{tab1}. The source term distrubution is taken from \cite{Ferriere:2001rg}. The gas density distribution is taken from the public {\tt Galprop} version \cite{galprop_website,galprop1998,galprop2002}.

We also consider a primary electron+positron extra component with harder injection spectrum in order to explain the anomalous rise of the positron fraction above $10 \units{GeV}$ reported by PAMELA, Fermi-LAT and AMS02. This behaviour cannot be reproduced in 
the standard scenario in which positrons only originate from spallation of protons, helium and heavier nuclei on interstellar gas. 

\begin{figure}[!h]
\includegraphics[width=1.\columnwidth,height=0.25\textheight,clip]{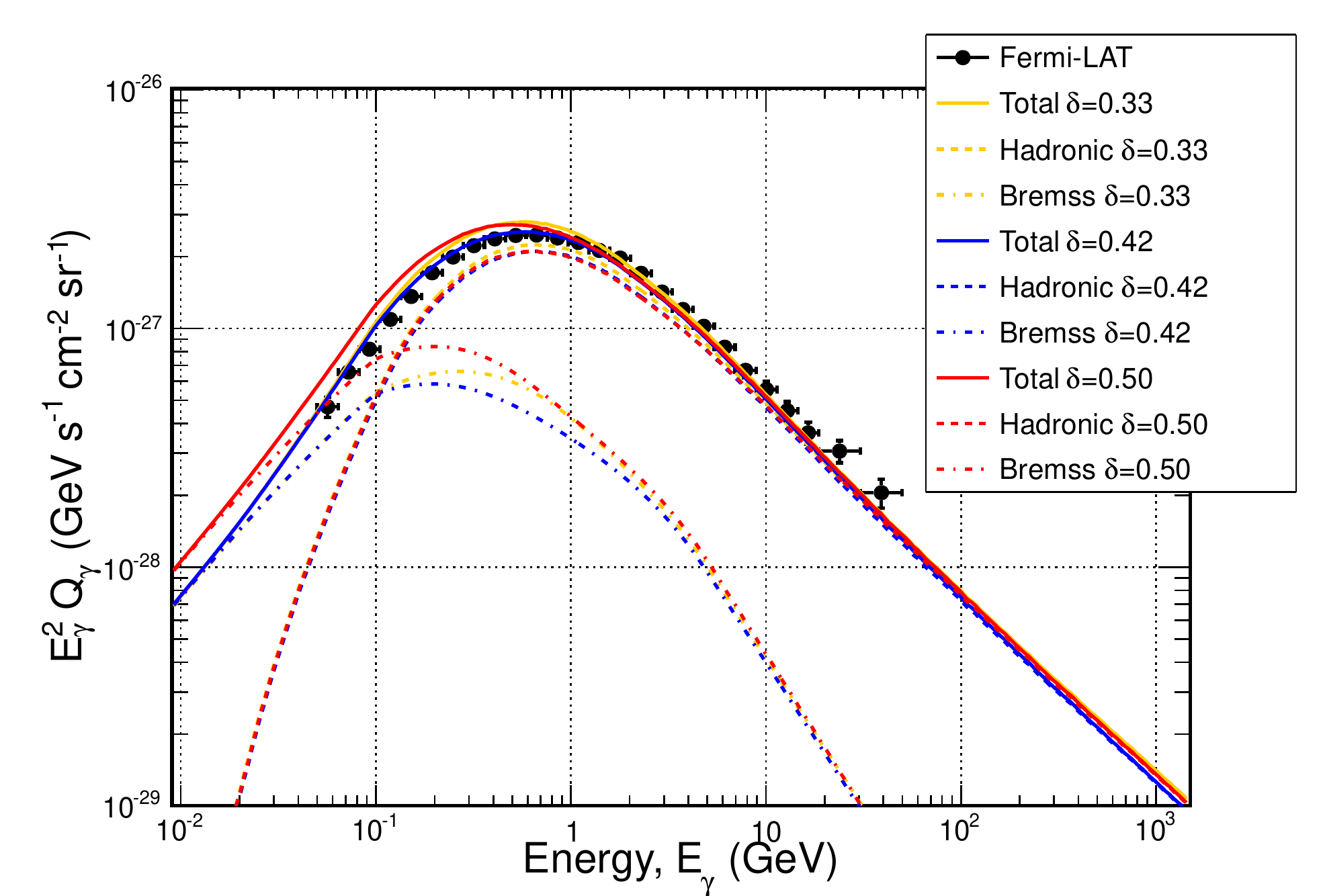}
\caption{Differential gamma-ray emissivity as a function of the gamma-ray energy. We calculated the contribution associated with the hadronic collisions and bremsstrahlung by folding the proton, the Helium and electron LIS with $\gamma$-ray production cross sections. Orange: KOL model; Red: KRA model; Blue: AMS02 model (see text). The data point are quoted from~\cite{jm2015}. }
\label{FigDragonFlukaEmis}
\end{figure}

In order to compare our computations with the data, we have to consider that, in the final stage of the propagation, 
charged particles enter the sphere of influence of the Sun. Here they diffuse in the Heliospheric 
magnetic field, and suffer adiabatic energy losses and convection due to the presence of the solar wind: 
this process is called {\it solar modulation} 
and is relevant for low energy ($< 10 \units{GeV}$) particles. 
In this work we 
adopt the so-called \textit{``force field approximation''}~\cite{GA}, a simplified 
description in which a single free parameter is involved: the 
modulation potential $\phi$.

Given all these ingredients, our method is the following: we compare the current data to 
the computations performed with our modified version of {\tt DRAGON} with the new cross section parametrizations and we retune the model accordingly.
Our results are summarized in Fig.s~\ref{FigDragonFluka} and ~\ref{FigDragonFlukaELE}.
Very remarkably, we find that it is possible to reproduce correctly all the relevant observables (namely, proton, Helium, light nuclei spectra, and B/C ratio) with the reference models considered, after a rescaling of the diffusion coefficient normalization ($D_0$) and a minor fine-tuning of the other parameters.

\begin{itemize}
\item For the KOL model we use $D_{0}$ = $4.25\times 10^{28}\units{cm^2~s^{-1}}$, $\eta = 0$ and $v_{\rm A} = 33\units{km~s^{-1}}$ (Alfv\'en velocity): while
\item For the KRA model we use $D_{0}$ = $2.8\times 10^{28}\units{cm^2~s^{-1}}$, $\eta = -0.4$ and $v_{\rm A} = 17.5\units{km~s^{-1}}$. 
\item For the AMS02 model we use $D_{0}$ = $3.35\times 10^{28}\units{cm^2~s^{-1}}$, $\eta = -0.4$ and $v_{\rm A} = 17.5\units{km~s^{-1}}$. 
\end{itemize}

For the three models we use the solar modulation parameter $\phi = 0.42$ and 0.62 $\units{GV}$ for all the species to reproduce the PAMELA and AMS02 data respectively. The comparison of these results with PAMELA \footnote{We lowered the proton spectrum by 3.2\% according to the prescription in Ref.~\cite{Adriani:2013as}.} and AMS02 proton data is within $\pm 10\%$ or better, as shown in Fig.~\ref{FigDragonFluka}.
Clearly, the model with $\delta=0.42$ is designed to have the best compatibility with AMS02 preliminary B/C data.

We remind that in the low energy region it is useful 
in many cases to go beyond the very simplified picture of the force field approximation, 
since the modulation process depends on the charge of the particle (including its sign). 
A more accurate treatment consists in using a dedicated numerical transport tool 
to describe the propagation process in the Solar System, e.g. {\tt HelioProp}~\cite{helioprop}, 
but that is far beyond the scope of this paper.

It is worth to point out that in these three models we do not consider the presence of the break around 
$300 \units{GeV}$ in the proton and nuclei spectra, since it is beyond the aim of this paper. However, with this complete sample of inclusive cross sections
we do not need to introduce the concept of nuclear enhancement factor to describe secondary production in interaction 
among heavier nuclei, extrapolating the results of $p-p$ interaction~\cite{mori2009}. 
In particular, this factor depends on the shape of the spectrum (see for instance \cite{Kachelriess:2014mga}) and it does not work in case of abrupt change in the CR spectra (as in case of breaks).

Here we want to stress that with this new evaluation we have a much better insight on the nuclear processes, since 
the cross-sections can be computed accurately in a wide energy range. 
For this reason, our approach opens the possibility
of more detailed studies on the CR spectra, especially in the low-energy range (below $10 \units{GeV}$).

\begin{figure*}[!ht]
\begin{center}
\begin{tabular}{cc}
\includegraphics[width=1\columnwidth,height=0.25\textheight,clip]{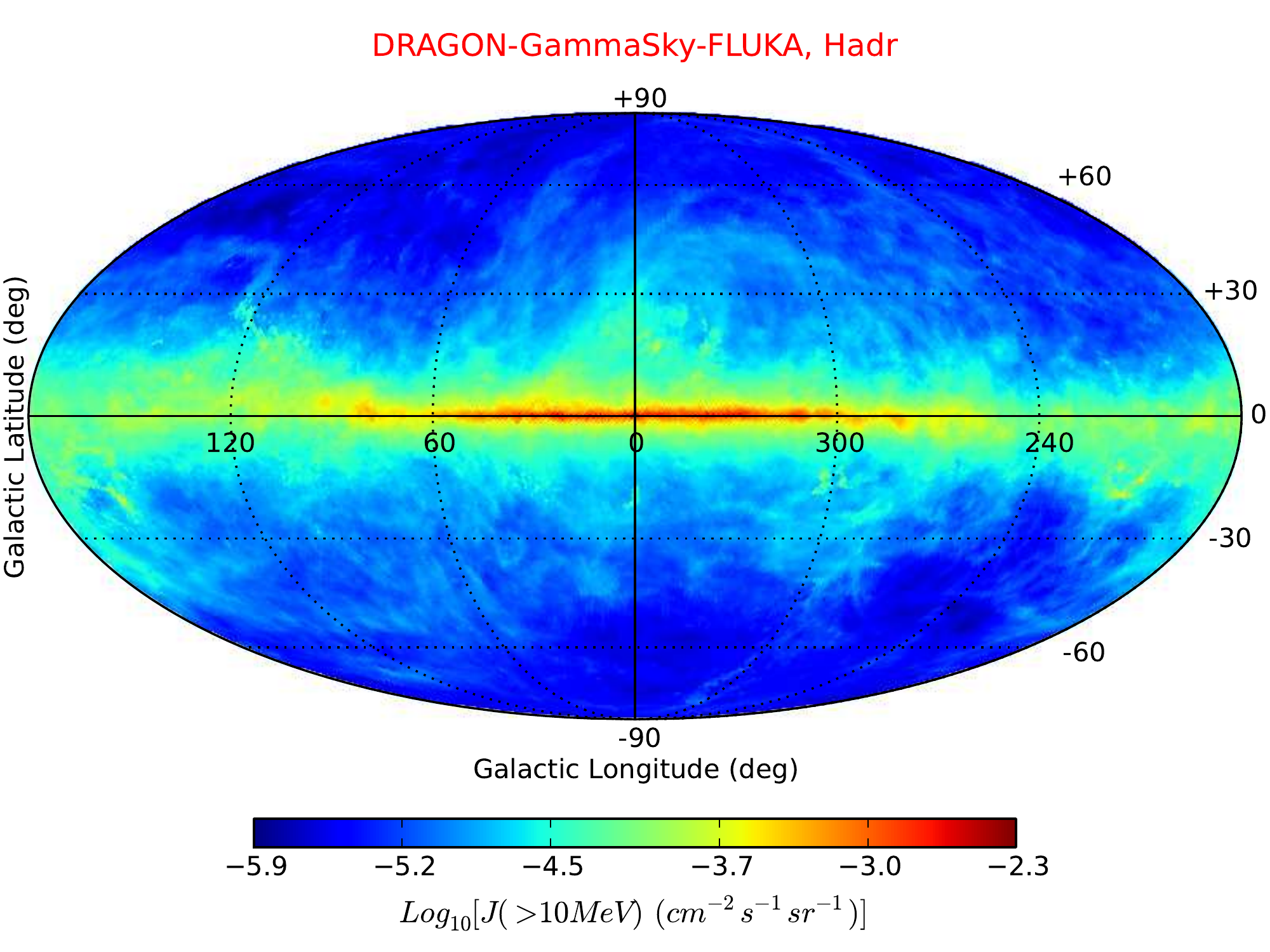} &
\includegraphics[width=1\columnwidth,height=0.25\textheight,clip]{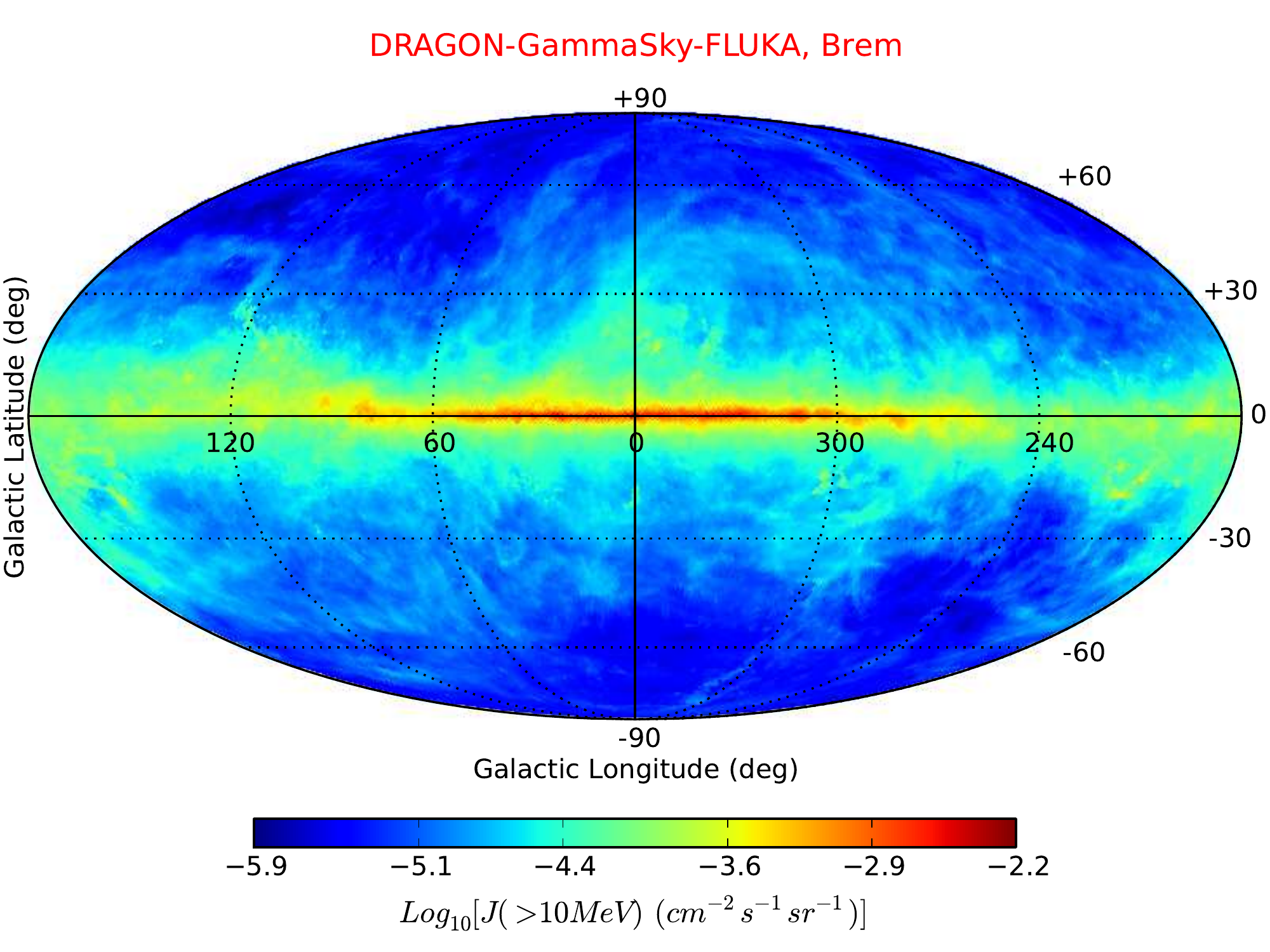} \\
\includegraphics[width=1\columnwidth,height=0.25\textheight,clip]{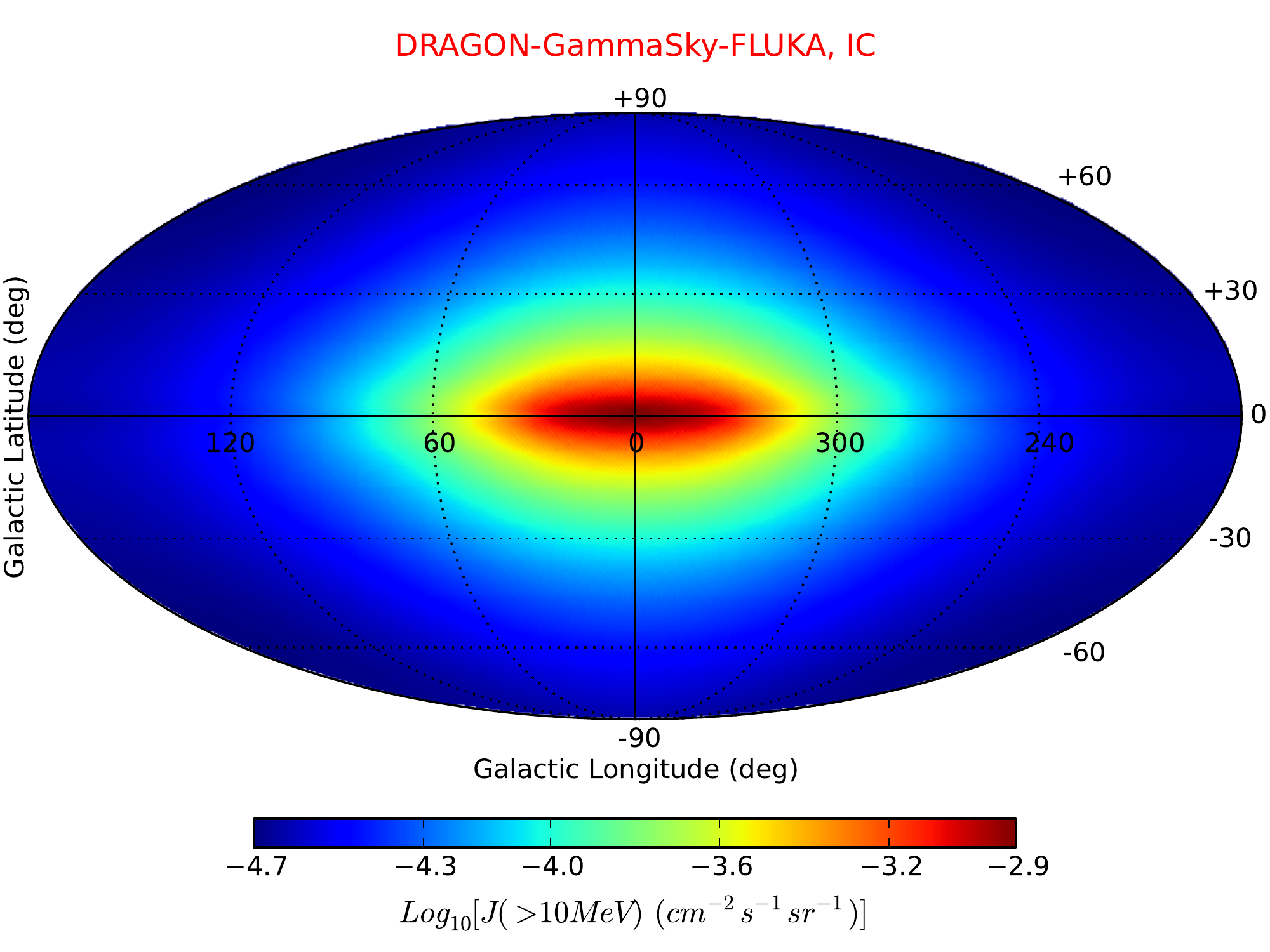} &
\includegraphics[width=1\columnwidth,height=0.22\textheight,clip]{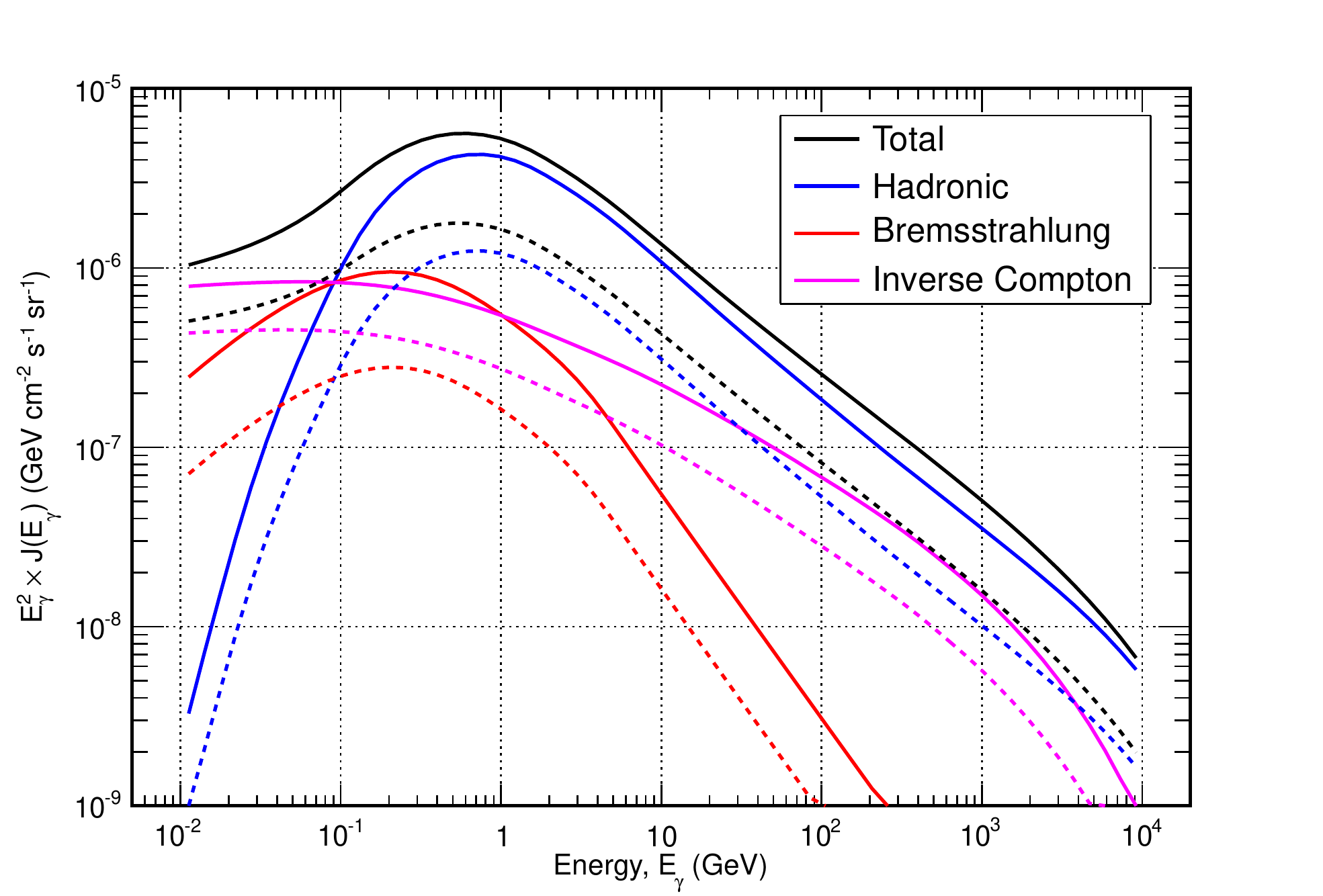} \\
\end{tabular}
\end{center}
\caption{Gamma-ray results obtained with {\tt GammaSky}. The top left, top right and bottom left plots show the sky maps
of the integral gamma-ray intensities above $10\units{MeV}$ for the hadronic, bremsstrahlung and inverse Compton (IC) emission. 
The maps have been built using the HEALPix~\cite{healpix} pixelization in the Galactic reference frame (Mollweide projection). The bottom right plot
shows the differential average intensities multiplied by $E_{\gamma}^{2}$ as a function of the gamma-ray energy. The solid lines
have been obtained summing the intensities over the whole sky, the dashed lines refer to the region $|b|>10^{\circ}$.}
\label{FigGamSkyFluka}
\end{figure*}

\begin{figure*}[!ht]
\begin{center}
\begin{tabular}{cc}
\includegraphics[width=1\columnwidth,height=0.25\textheight,clip]{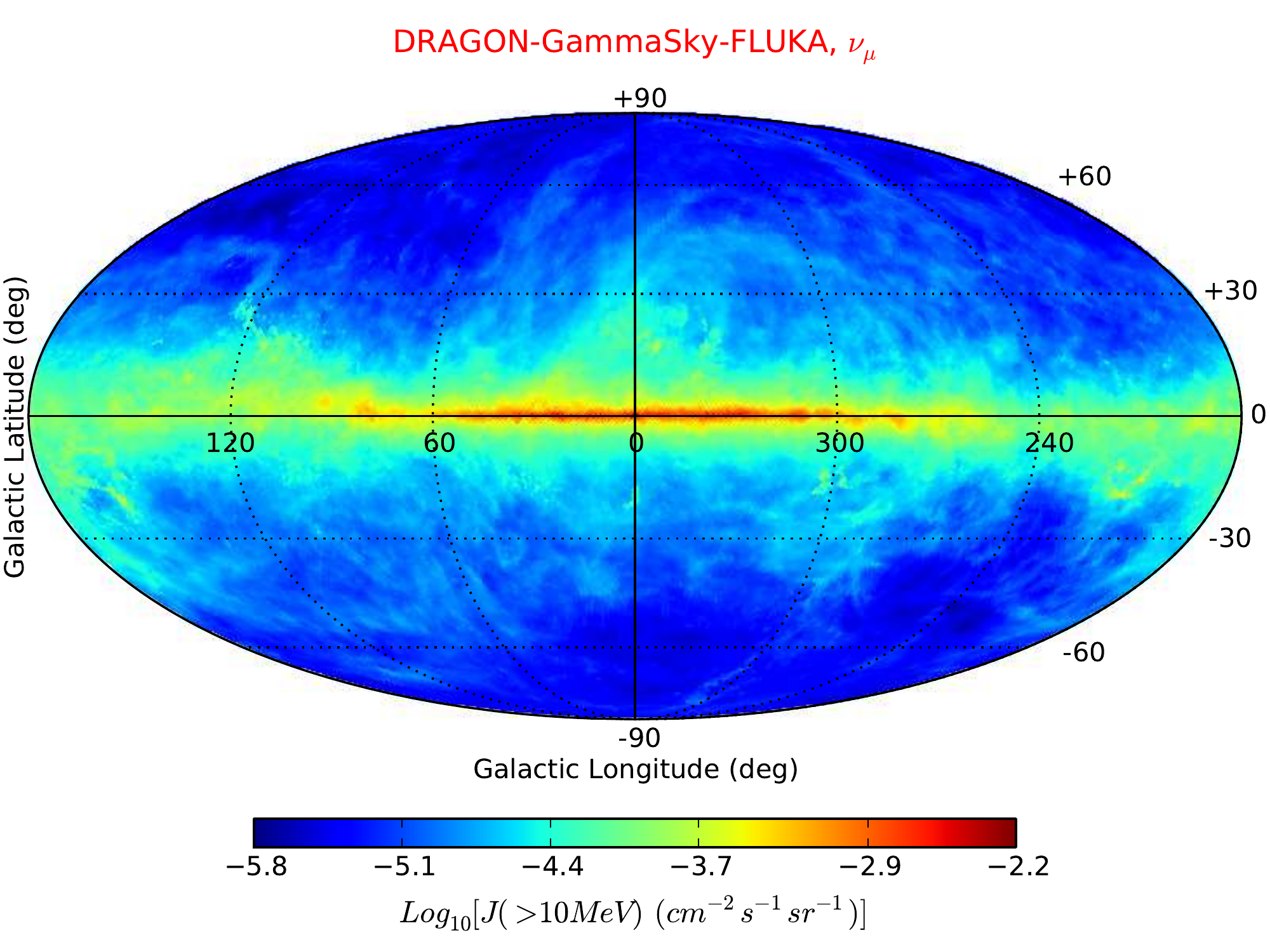} &
\includegraphics[width=1\columnwidth,height=0.25\textheight,clip]{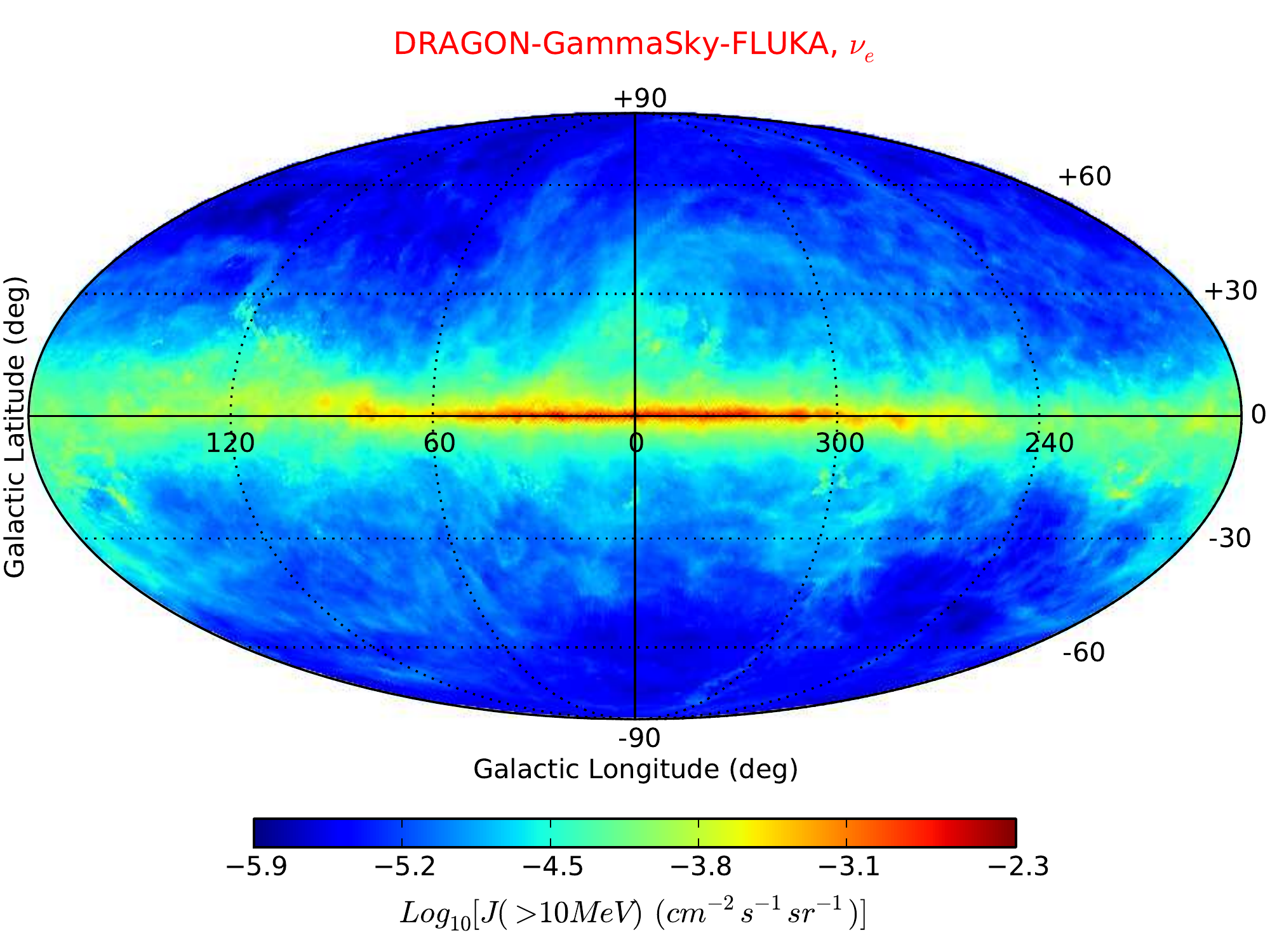} \\
\includegraphics[width=1\columnwidth,height=0.25\textheight,clip]{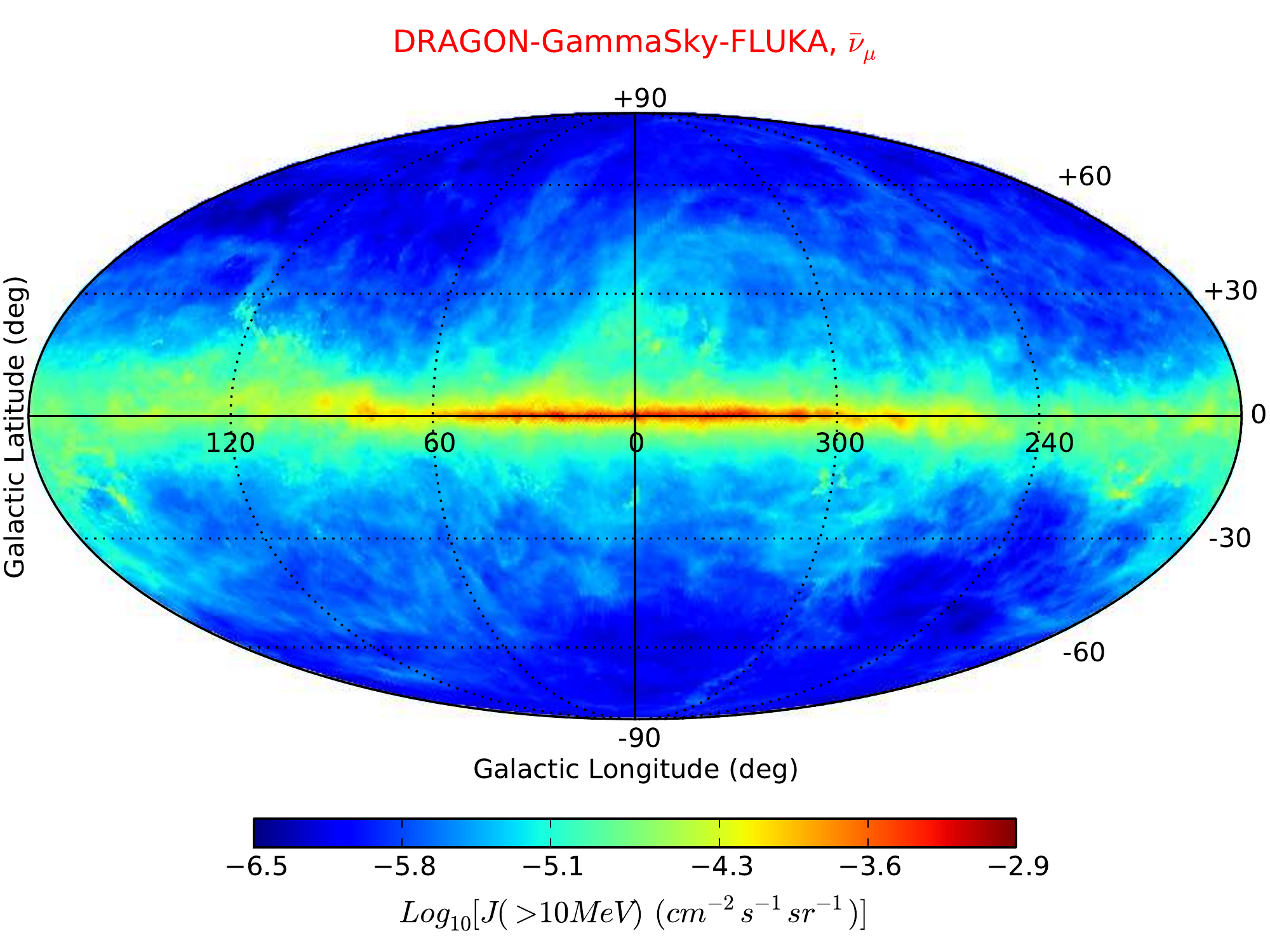} &
\includegraphics[width=1\columnwidth,height=0.25\textheight,clip]{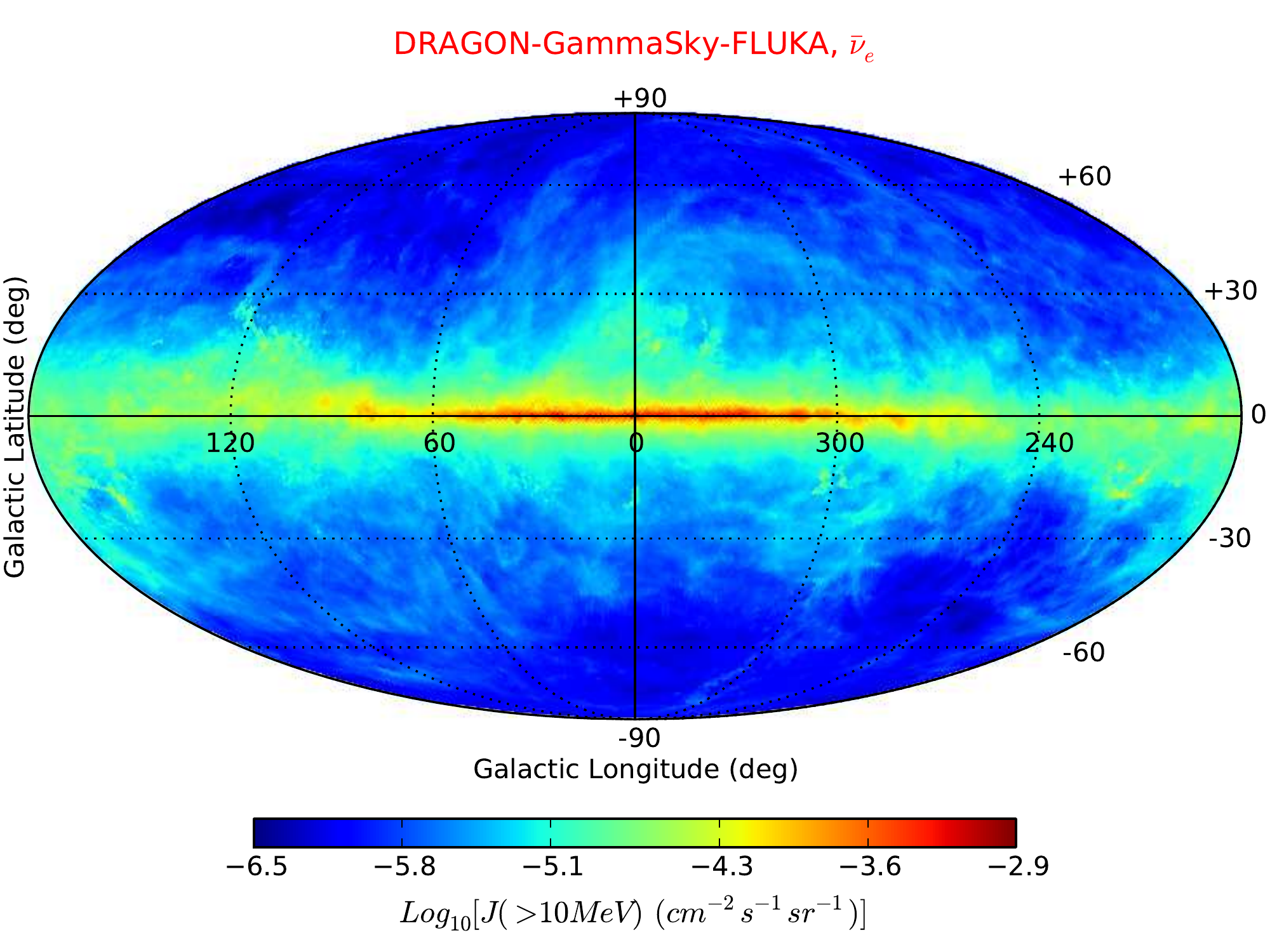} \\
\includegraphics[width=1\columnwidth,height=0.22\textheight,clip]{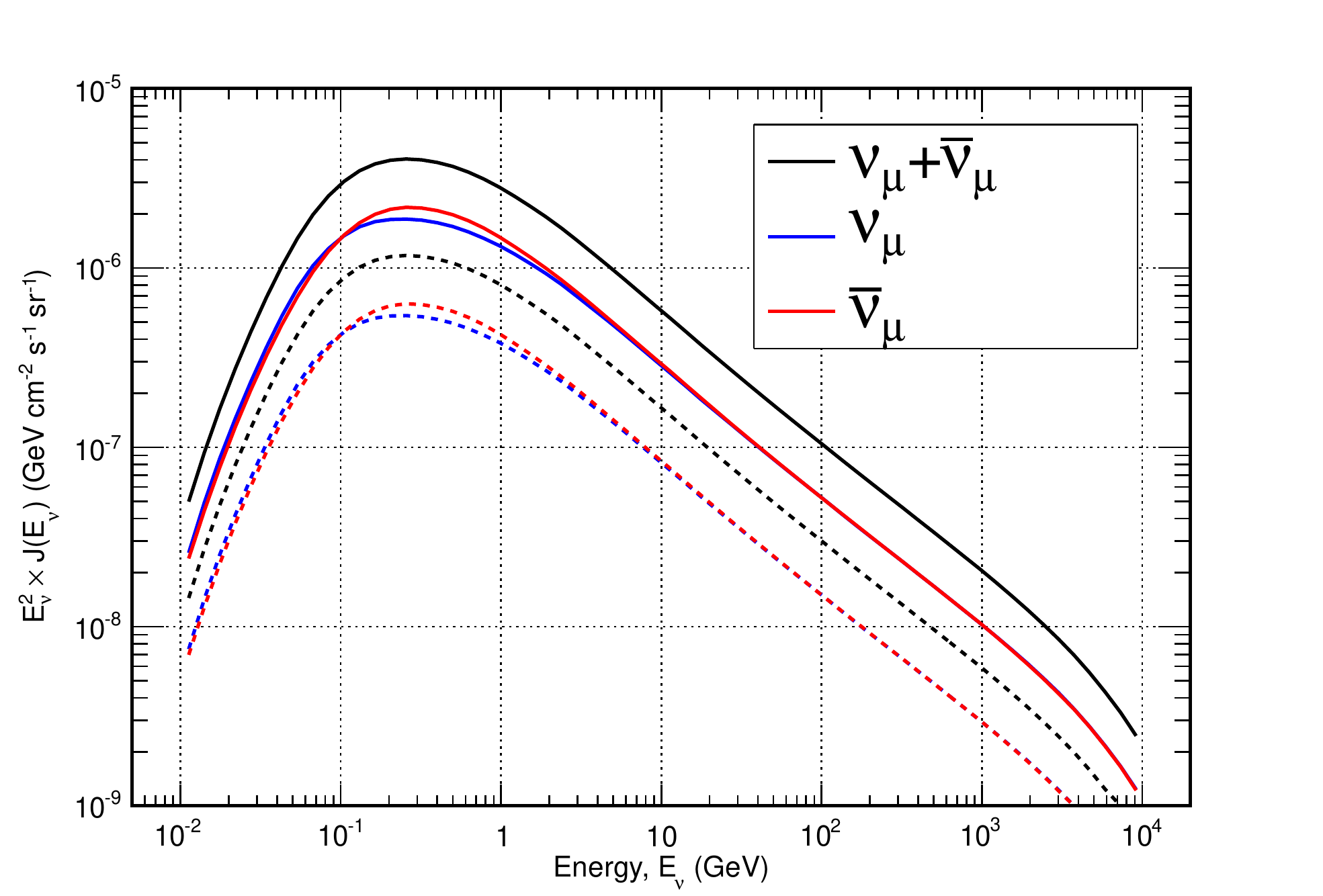} &
\includegraphics[width=1\columnwidth,height=0.22\textheight,clip]{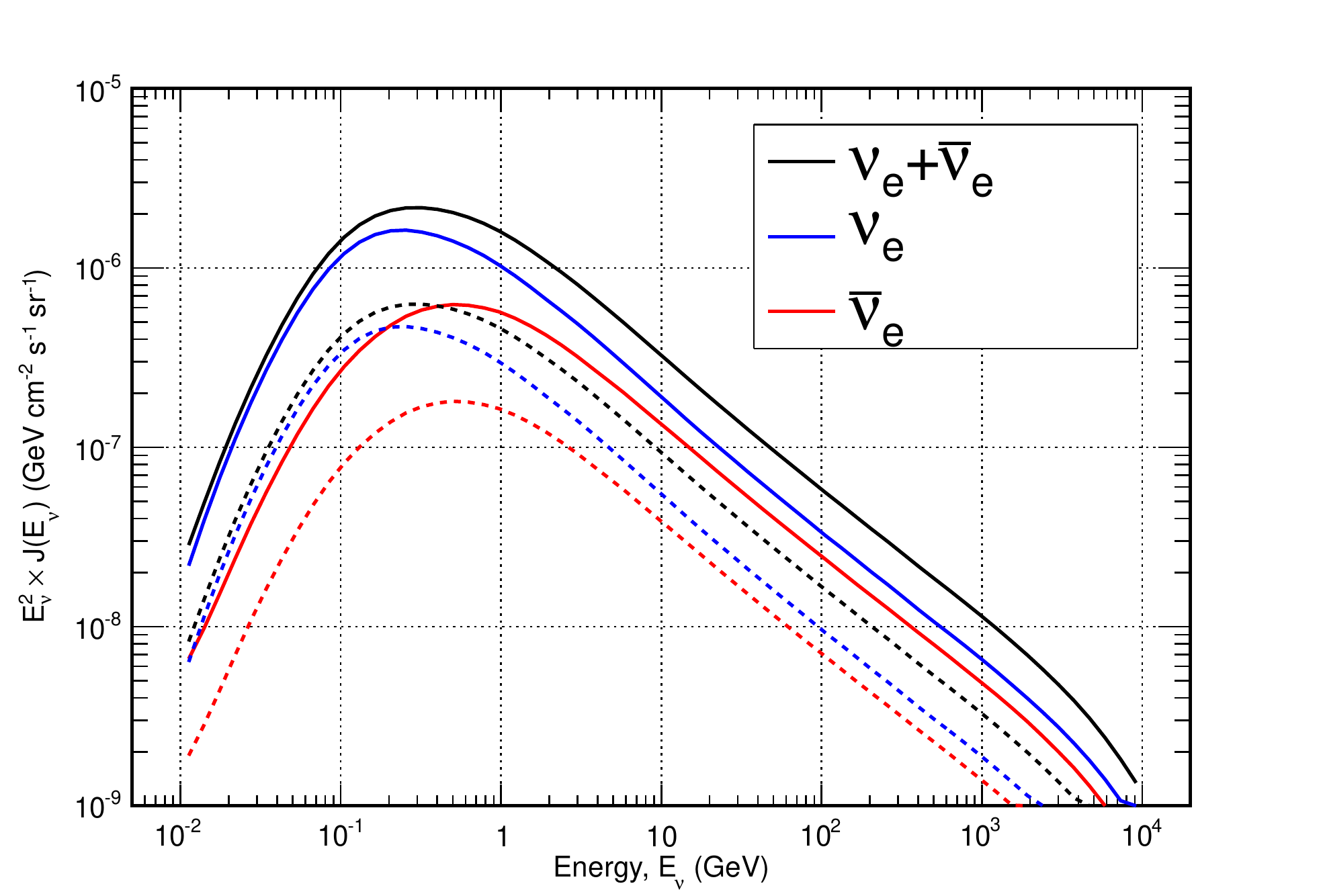} \\
\end{tabular}
\end{center}
\caption{Neutrino results obtained with {\tt GammaSky}. The plots in the left column refer to muon neutrinos and antineutrinos; 
the plots in the right column refer to electron neutrinos and antineutrinos. The four top panels show the sky maps of the integral intensities 
above $10\units{MeV}$. The maps are in Galactic coordinates and have been built using the HEALPix pixelization scheme (Mollweide projection).
The bottom plots show the differential average intensities multiplied by $E_{\nu}^{2}$ as a function of the neutrino energy.
The solid lines have been obtained summing the intensities over the whole sky, the dashed lines refer to the region $|b|>10^{\circ}$.}
\label{FigNuSkyFluka}
\end{figure*}

For example, given the more accurate determination of the gamma-ray emissivity, we can turn our attention
to the Local Interstellar Spectra (LIS) of protons and Helium nuclei, not affected by solar modulation.
This quantity can be inferred from the Fermi-LAT gamma-ray data~\cite{jm2015} and its determination is 
useful in order to constrain the propagation models described above, since each of them implies a different shape and normalization for the LIS. Fig.~\ref{FigDragonFlukaEmis} shows the gamma-ray emissivity calculated by means of the proton, Helium and electron LIS.
We calculate the contribution associated with the hadronic collisions and bremsstrahlung by folding the proton, Helium and electron LIS 
with the corresponding $\gamma$-ray production cross sections (according to Eq.~\ref{Eq:emiss}) on the ISM with relative abundance of $H:He:C:N:O:Ne:Mg:Si=
1:0.096:4.65\times 10^{-4}:8.3\times 10^{-5}:8.3\times 10^{-4}:1.3\times 10^{-4}:3.9\times 10^{-5}:3.69\times 10^{-5}$.
An inspection of Fig.~\ref{FigDragonFlukaEmis} shows that at low energies the contribution from electron bremsstrahlung is not negligible. 
This contribution mainly depends on the electron spectrum that should also be constrained taking synchrotron radiation into account (see for instance \cite{DiBernardo:2012zu,Orlando:2013ysa}).

In order to test our model even further, we evaluate the Galactic diffuse gamma-ray and neutrino emission adopting the numerical package {\tt GammaSky} as post-processor of the {\tt DRAGON} output.
{\tt GammaSky} is a dedicated code recently used e.g. in~\cite{Evoli:2012ha,Tavakoli:2011wz,Cirelli:2014lwa} and other works to compute gamma-ray, synchrotron and neutrino maps and spectra. This package includes, among the different options, the gas model taken from the public version of {\tt Galprop} (see e.g. \cite{galprop2007,galprop2012} and references therein).

We implement in {\tt GammaSky} the present 
parameterizations for the gamma-ray and neutrino secondary production in the hadronic interactions 
(taking into account the $p-p$, $p-^{4}He$, $^{4}He-p$ and $^{4}He - ^{4}He$ collisions with ISM and assuming an ISM $H:He=1:0.1$).

Fig.~\ref{FigGamSkyFluka} shows the sky maps of the integral gamma-ray intensity above $10\units{MeV}$ for 
the various components. The hadronic contribution is evaluated using the present parameterizations, 
while the bremsstrahlung and inverse Compton contributions are evaluated using the built-in models in {\tt GammaSky}.
The last panel of Fig.~\ref{FigGamSkyFluka} shows the average gamma-ray intensities as a function of 
the energy for the different components. The calculation is performed considering 
either the whole sky or only the region outside the galactic plane ($|b|>10^{\circ}$).
In both cases the contribution from the hadronic component becomes dominant in the energy range
above $100\units{MeV}$.   

Fig.~\ref{FigNuSkyFluka} shows the maps of the integral intensities above $10\units{MeV}$ of
the various species of neutrinos and antineutrinos. The bottom panels of Fig.~\ref{FigNuSkyFluka}
show the average differential intensity of electron and muon neutrinos and antineutrinos,
evaluated considering either the whole sky or only the region outside the galactic plane ($|b|>10^{\circ}$).

\section{Conclusions}

We have used the {\tt FLUKA} simulation code to evaluate the inclusive cross sections for the production of
stable secondary particles and as well as nuclei from spallation in the interactions of several CR projectiles 
with different target nuclei. 

The results are provided in tables, that can be used for the computation of the production
spectra of secondary CRs starting from any given spectra of the primaries.
The propagation and the interactions of these particles in astronomical
environments can then be studied using dedicated tools, as {\tt DRAGON, GammaSky} and \texttt{Galprop}.

As an example of application we implemented the values of the cross sections in a custom version of the
{\tt DRAGON} CR propagation code, showing that it is possible to reproduce with accuracy the measured 
CR spectra and light nuclei ratios as well as the gamma-ray emissivity.
In the present work we have propagated only CR projectiles up to $Z=14$ negleting heavier nuclei that could affect the present results at the level of a few percent.
The production and the propagation of nuclei up to iron will be investigated in a future work.

Finally, we evaluated the gamma-ray and neutrino sky-maps and energy distributions using a custom version of
the {\tt GammaSky} code.



\appendix
\section{Additional plots}
In this section we present additional results of inclusive secondary cross sections.

\begin{figure*}[!ht]
\begin{tabular}{cc}
\includegraphics[width=0.99\columnwidth,height=0.19\textheight,clip]{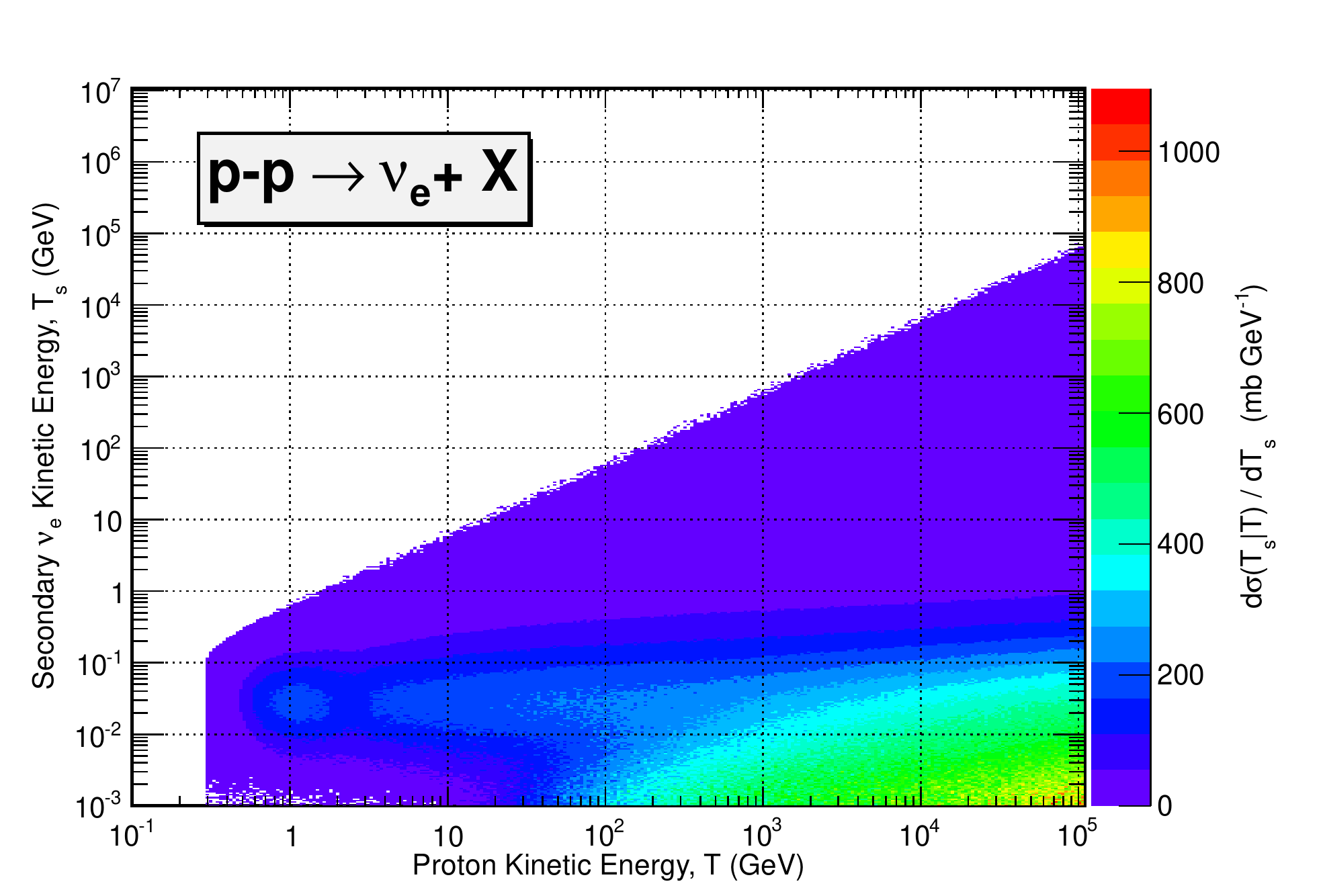} &
\includegraphics[width=0.99\columnwidth,height=0.19\textheight,clip]{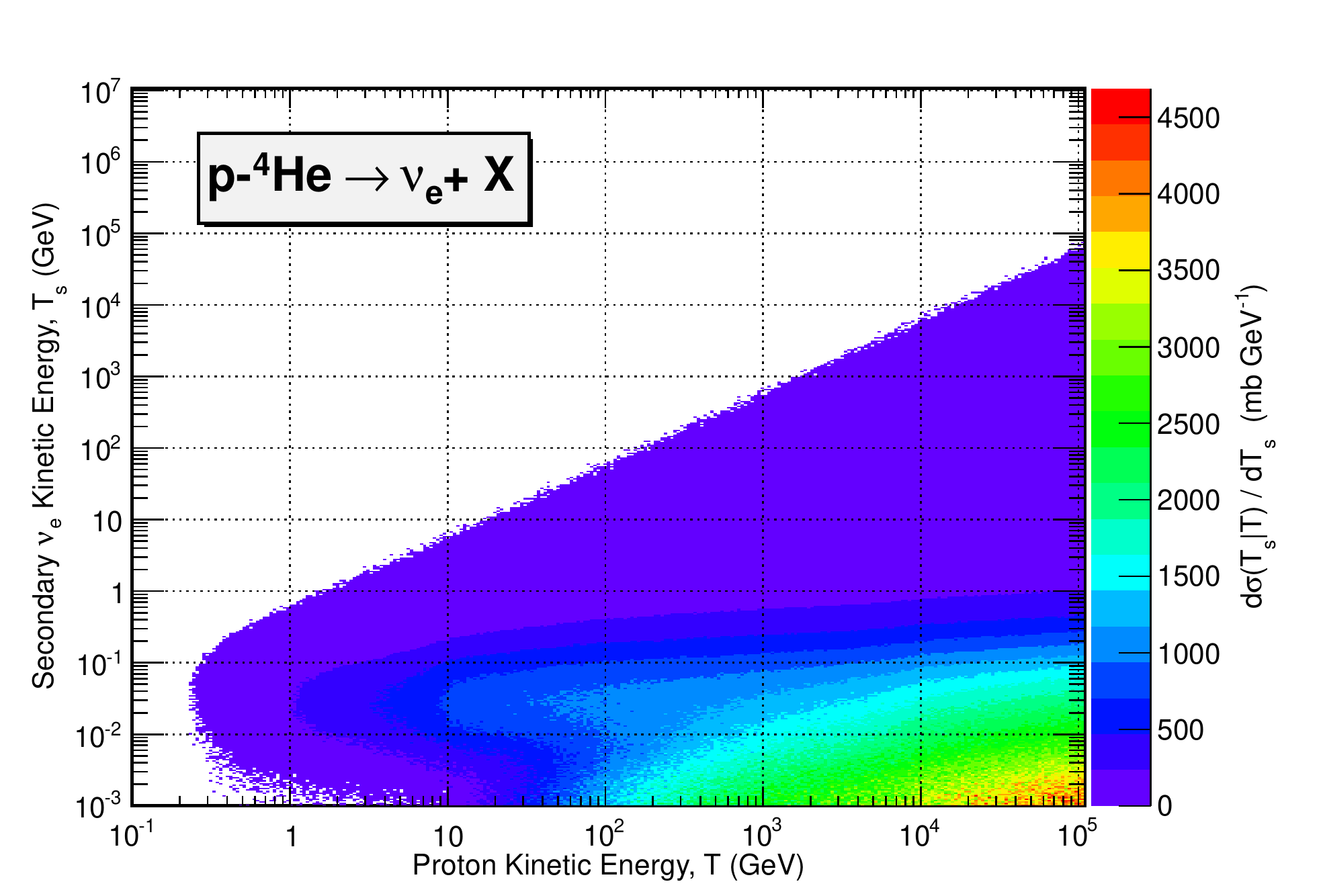} \\
\includegraphics[width=0.99\columnwidth,height=0.19\textheight,clip]{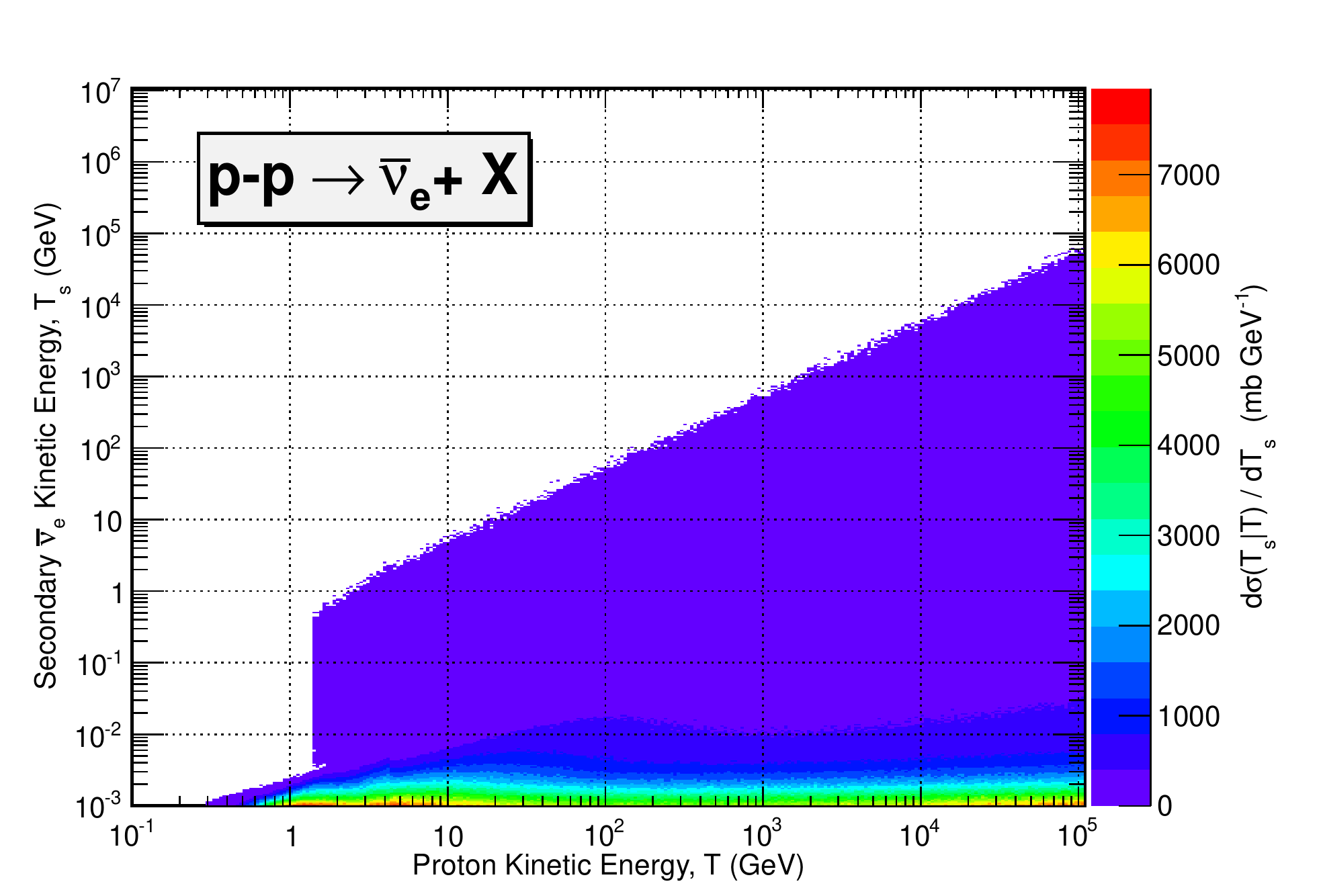} &
\includegraphics[width=0.99\columnwidth,height=0.19\textheight,clip]{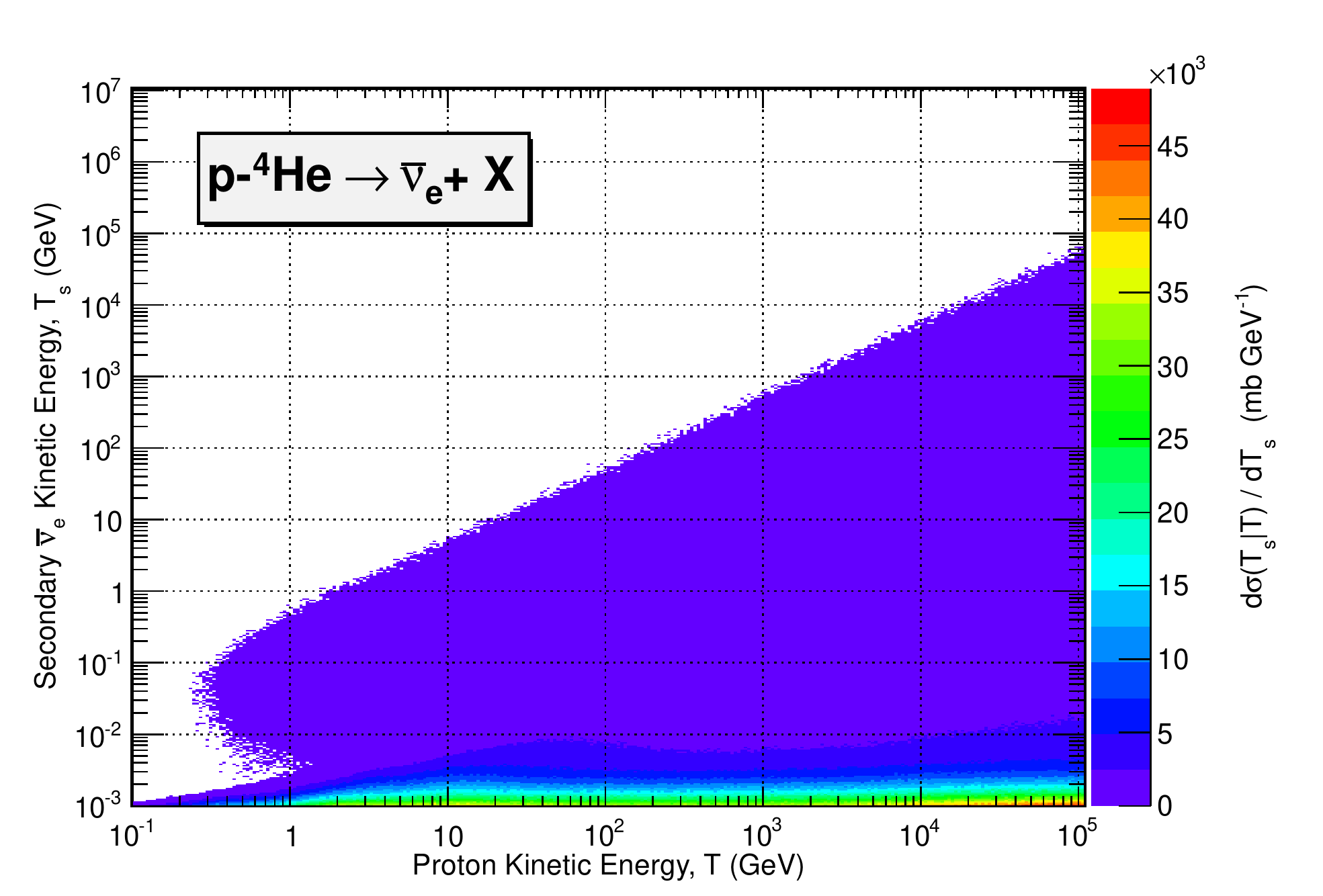} \\
\includegraphics[width=0.99\columnwidth,height=0.19\textheight,clip]{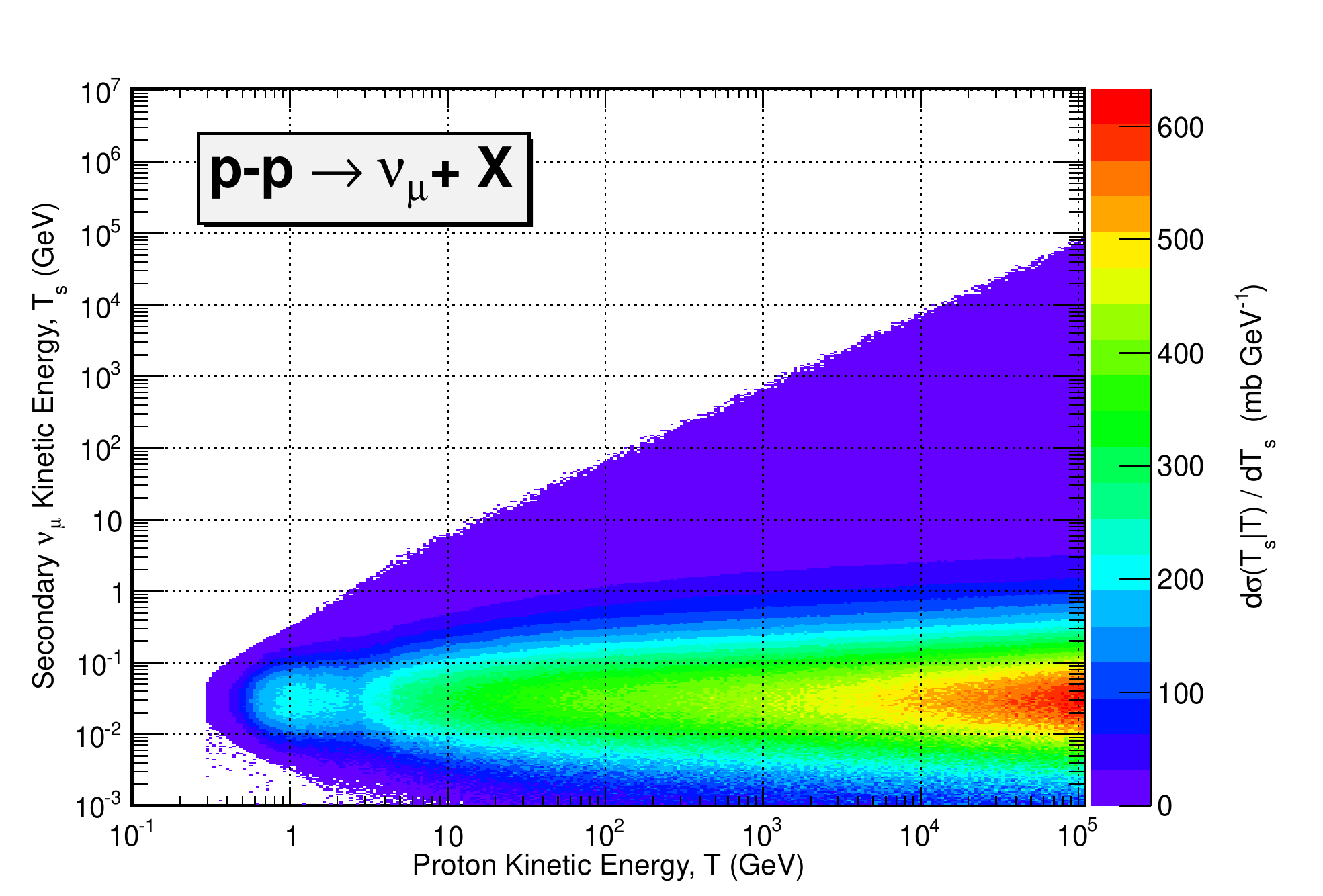} &
\includegraphics[width=0.99\columnwidth,height=0.19\textheight,clip]{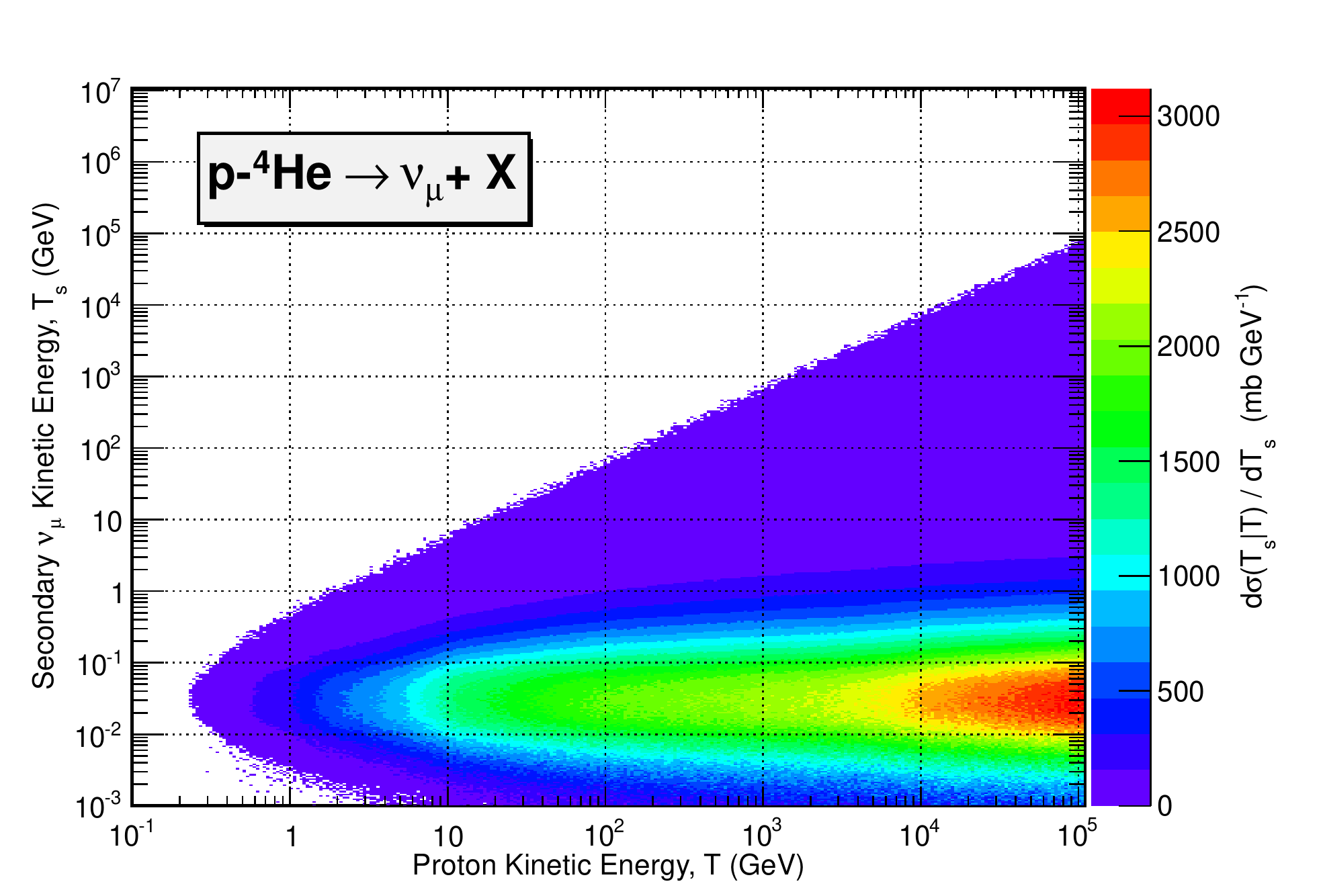} \\
\includegraphics[width=0.99\columnwidth,height=0.19\textheight,clip]{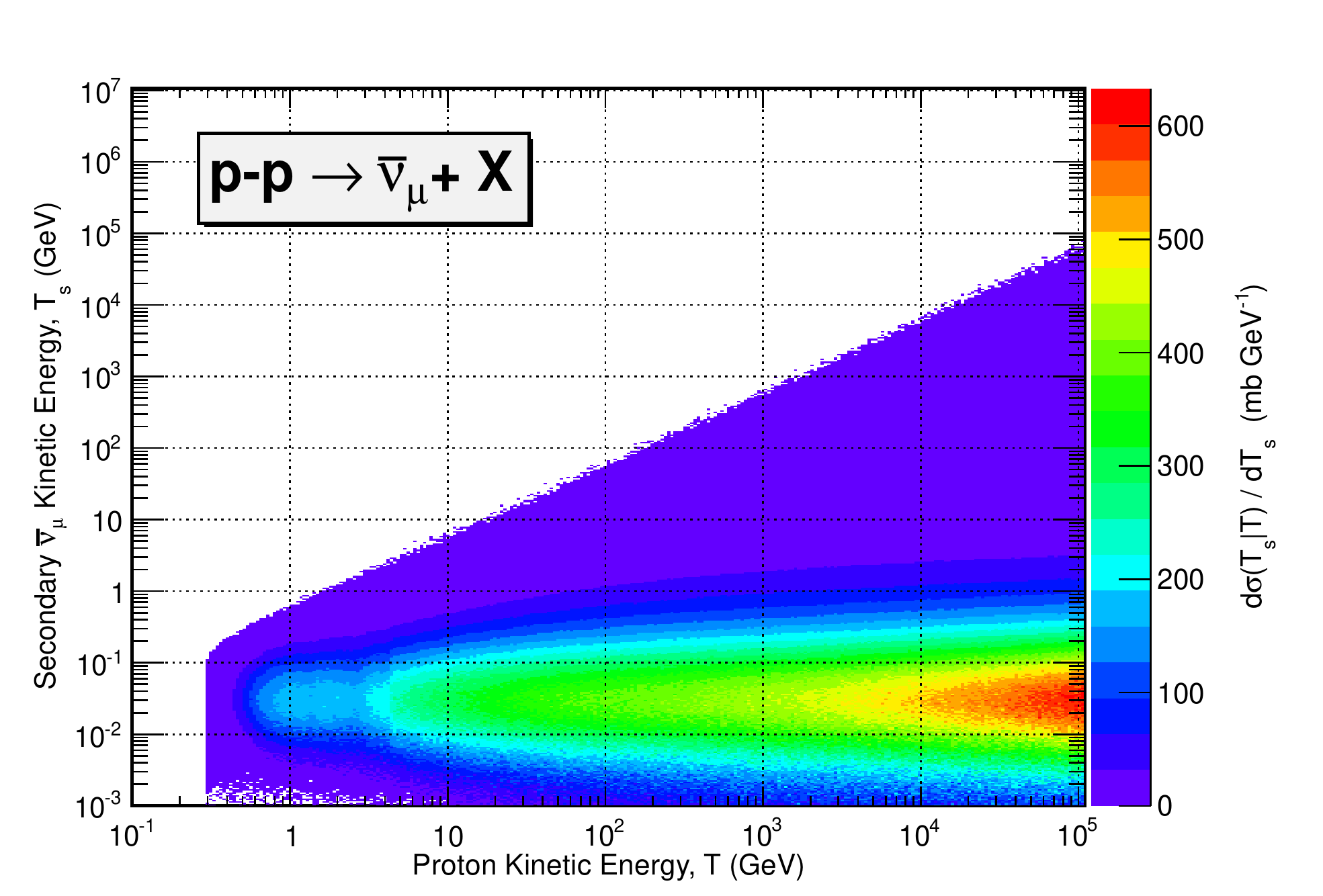} &
\includegraphics[width=0.99\columnwidth,height=0.19\textheight,clip]{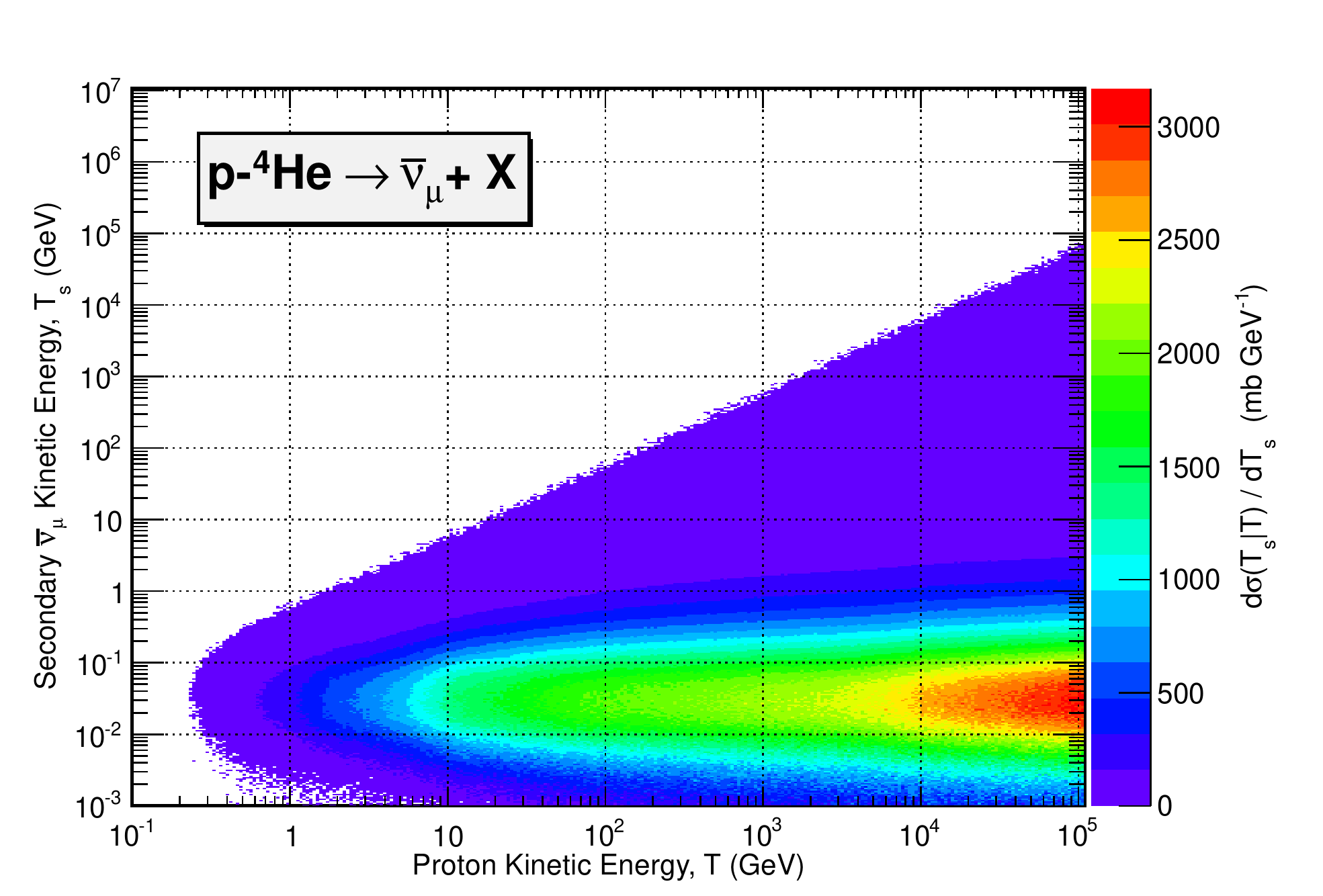} \\
\includegraphics[width=0.99\columnwidth,height=0.19\textheight,clip]{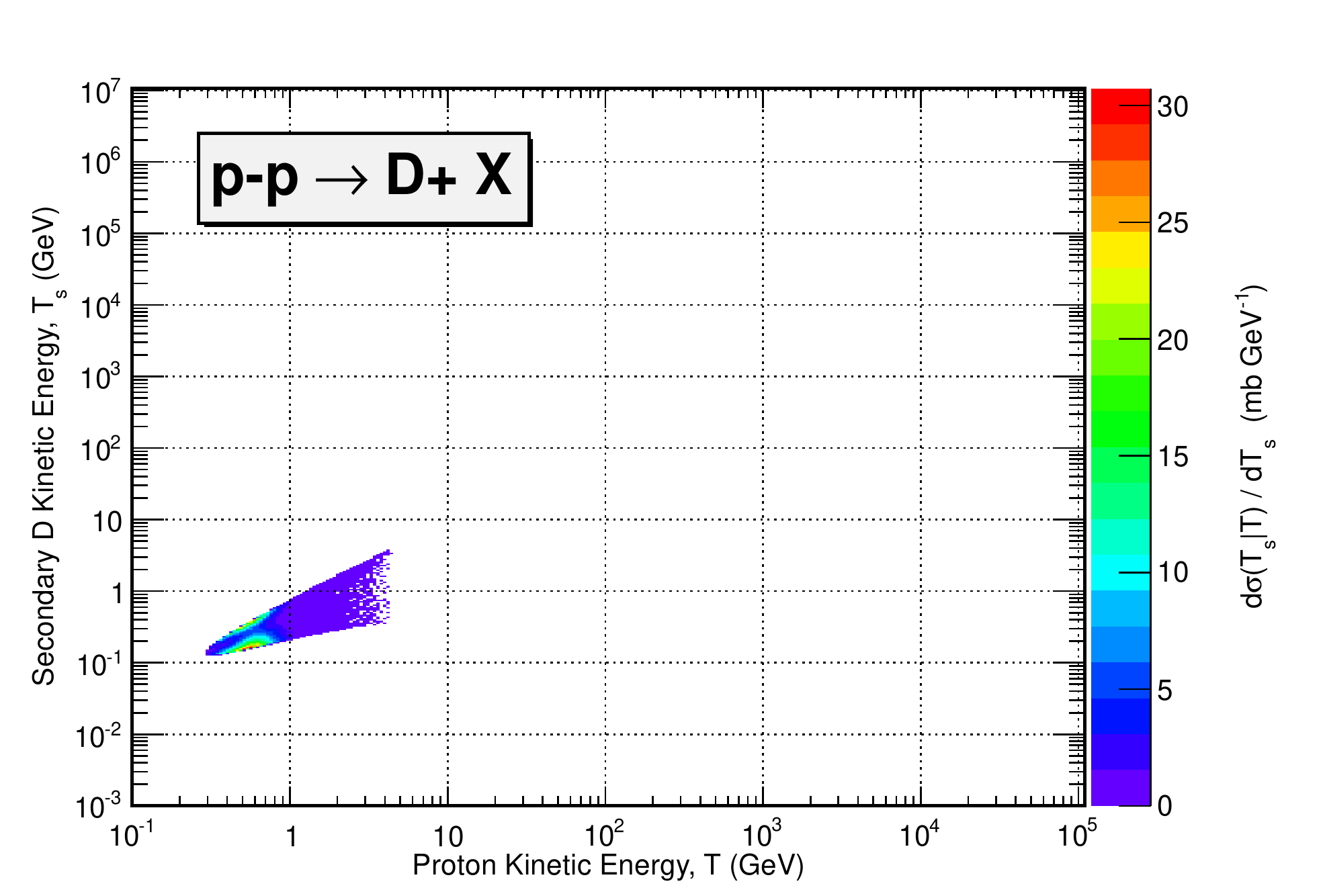} &
\includegraphics[width=0.99\columnwidth,height=0.19\textheight,clip]{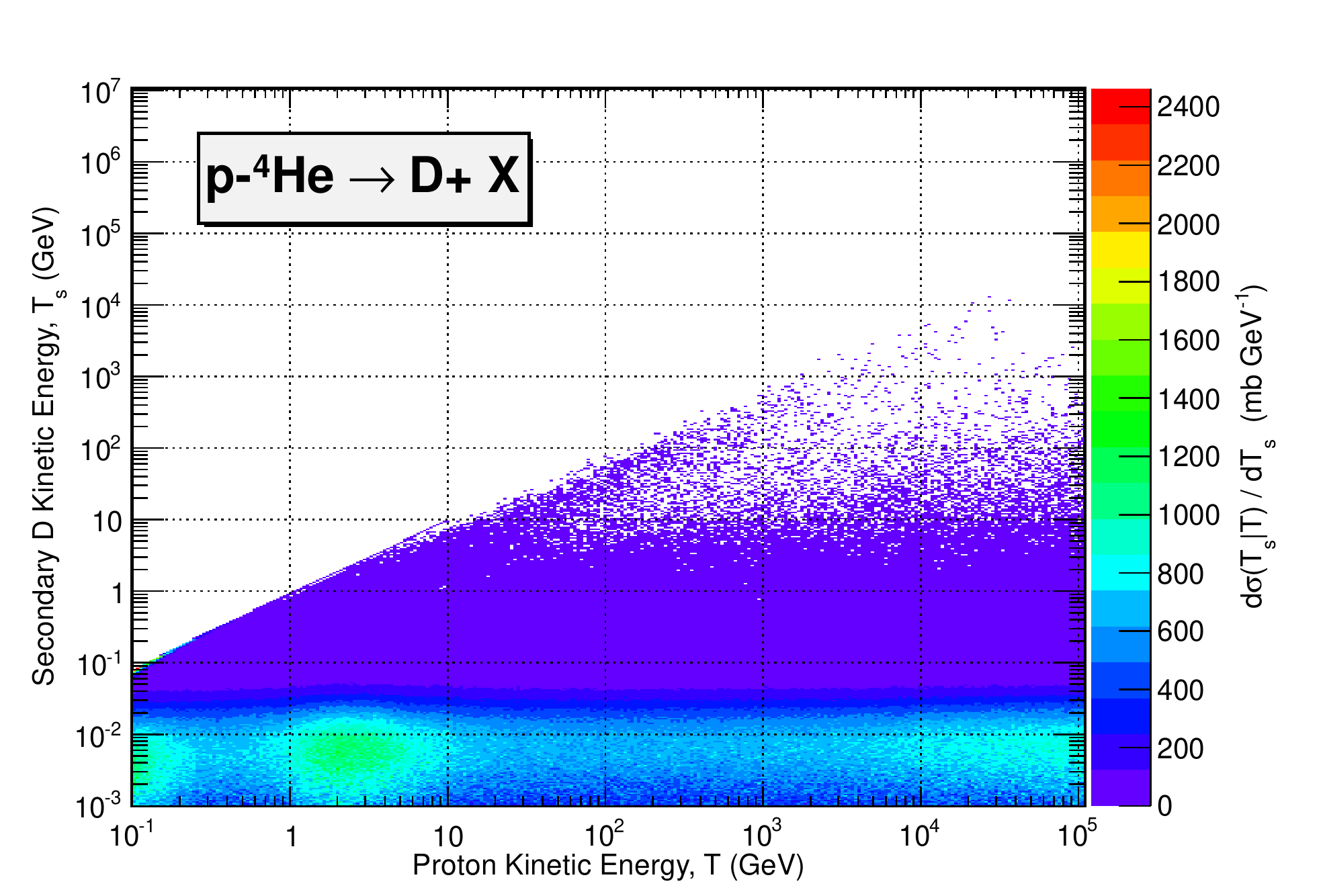} \\
\end{tabular}
\caption{Differential inclusive secondary cross sections for the production of $\nu$, $\bar{\nu}$ and deuteron  
in $p-p$ (left) and $p-^{4}He$ (right) interactions. The values of the cross sections (color scales) are in $\units{mb~GeV^{-1}}$.}
\label{FigpHe2DXsecApx}
\end{figure*}

\begin{figure*}[!ht]
\begin{tabular}{cc}
\includegraphics[width=0.99\columnwidth,height=0.19\textheight,clip]{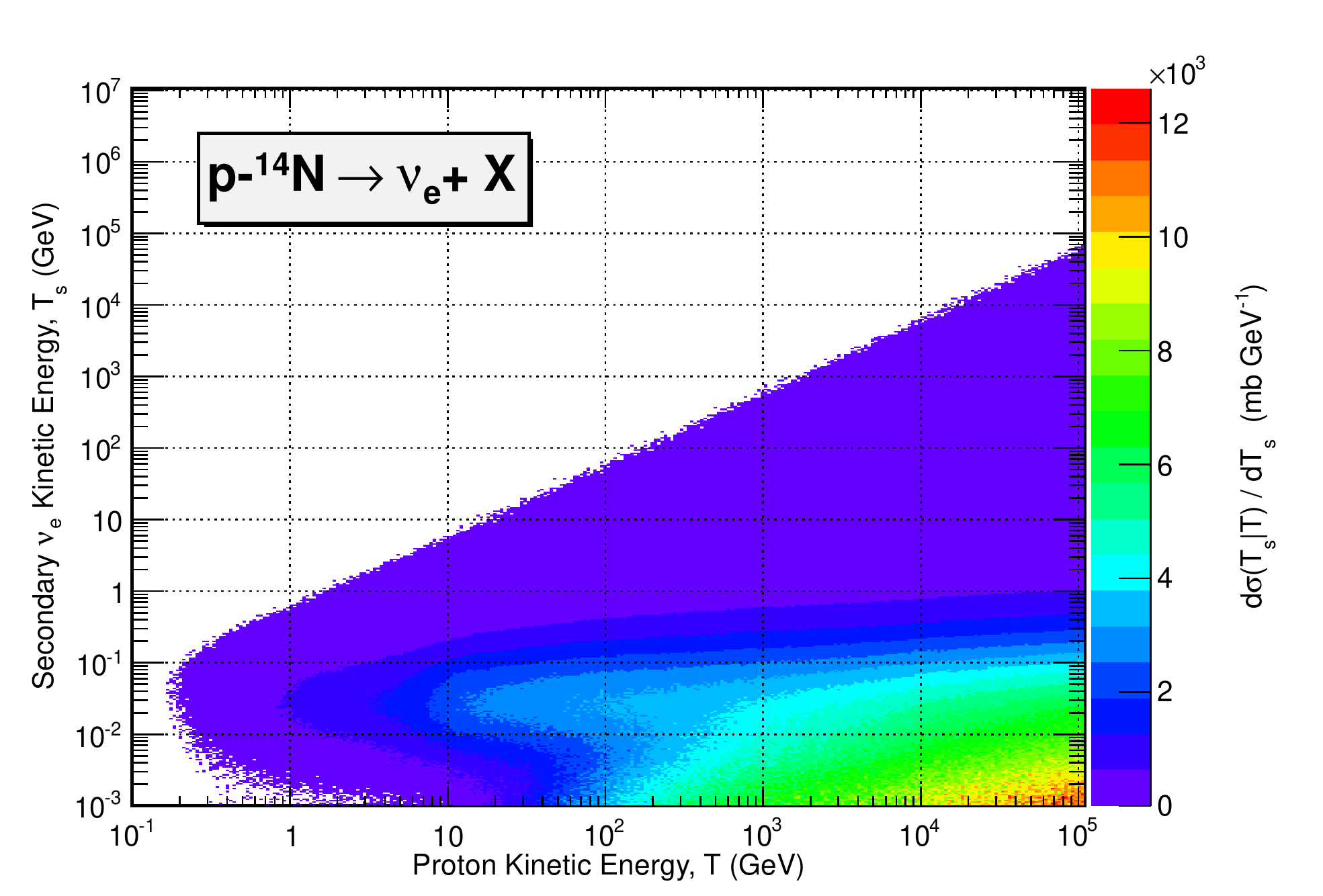} &
\includegraphics[width=0.99\columnwidth,height=0.19\textheight,clip]{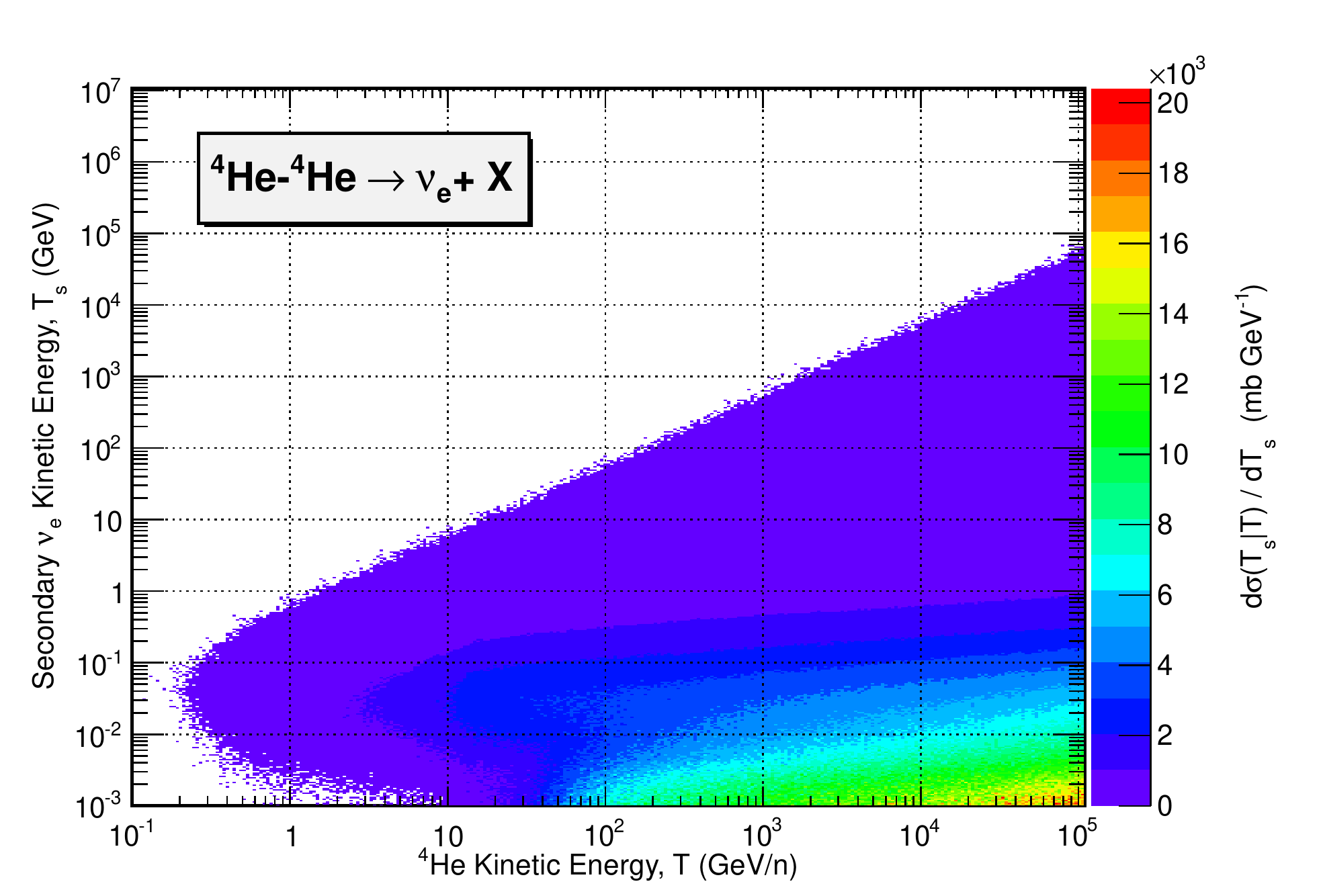} \\
\includegraphics[width=0.99\columnwidth,height=0.19\textheight,clip]{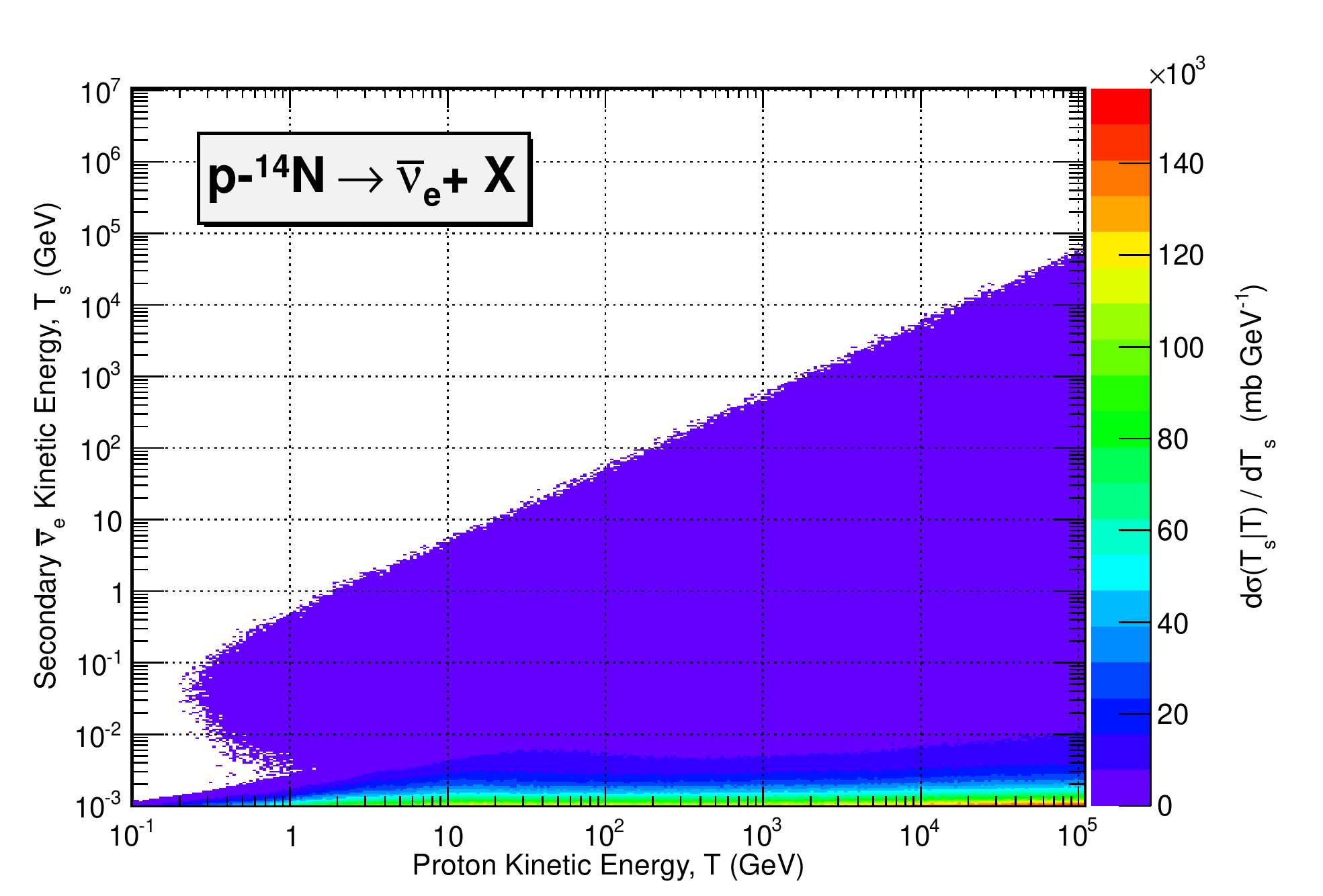} &
\includegraphics[width=0.99\columnwidth,height=0.19\textheight,clip]{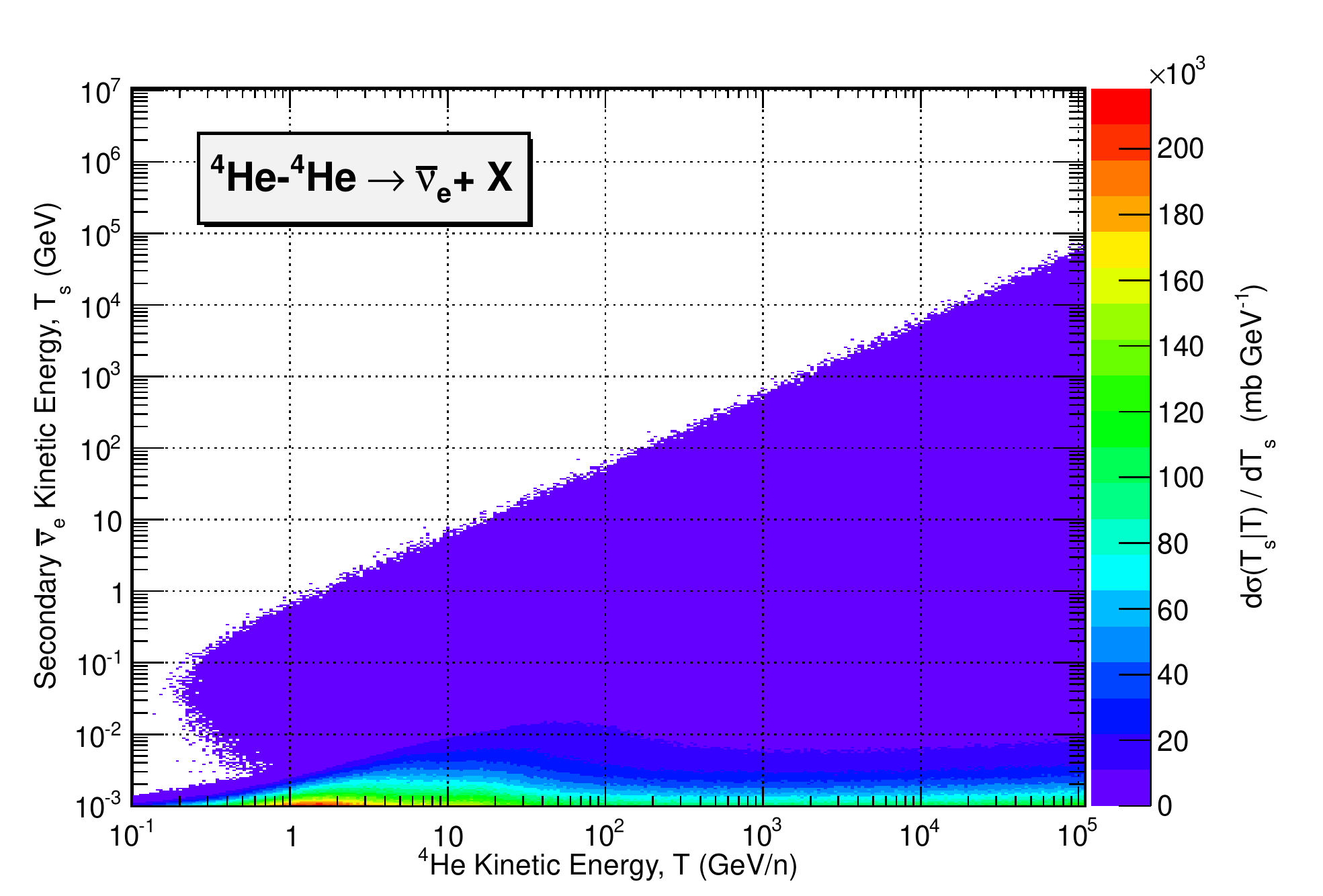} \\
\includegraphics[width=0.99\columnwidth,height=0.19\textheight,clip]{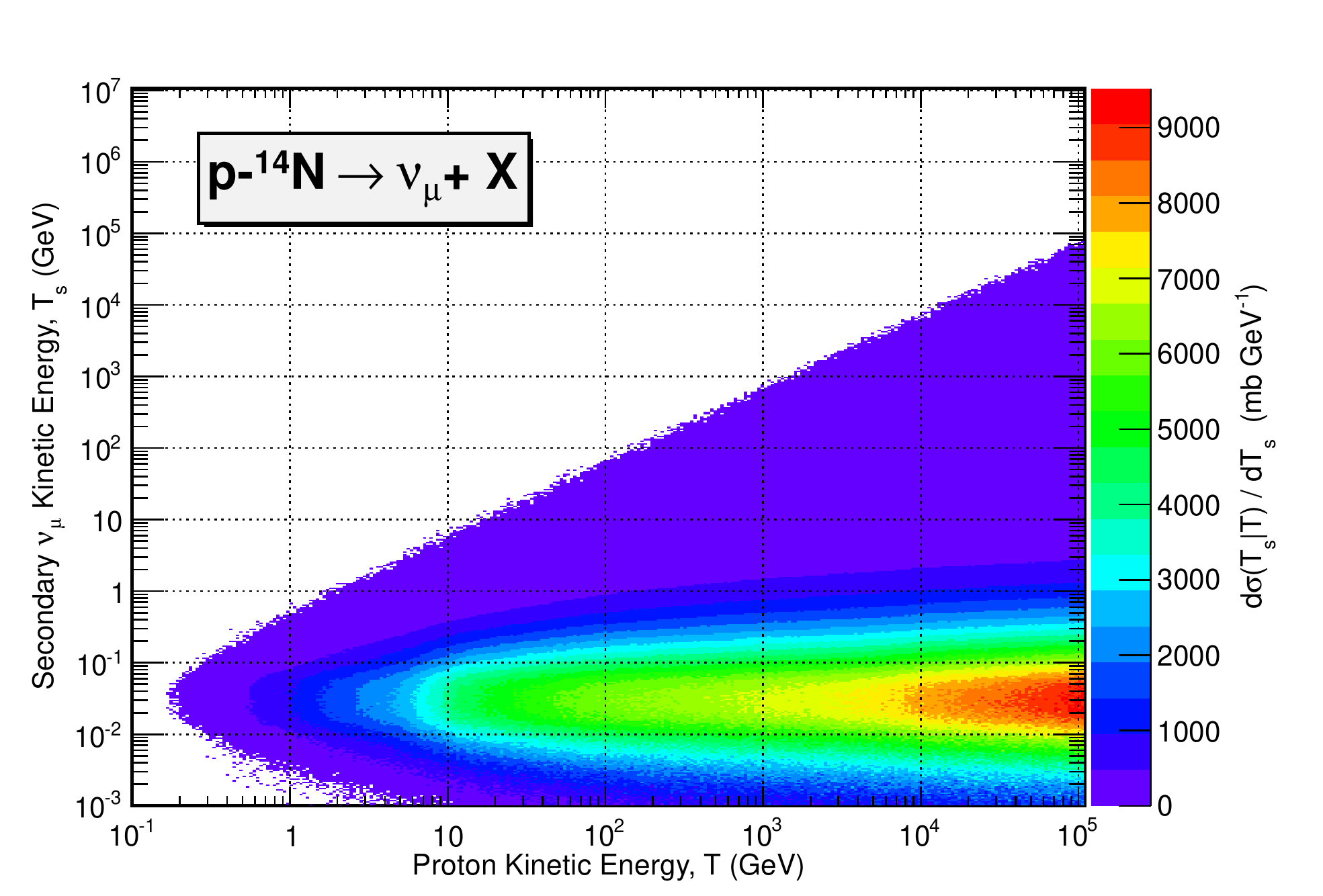} &
\includegraphics[width=0.99\columnwidth,height=0.19\textheight,clip]{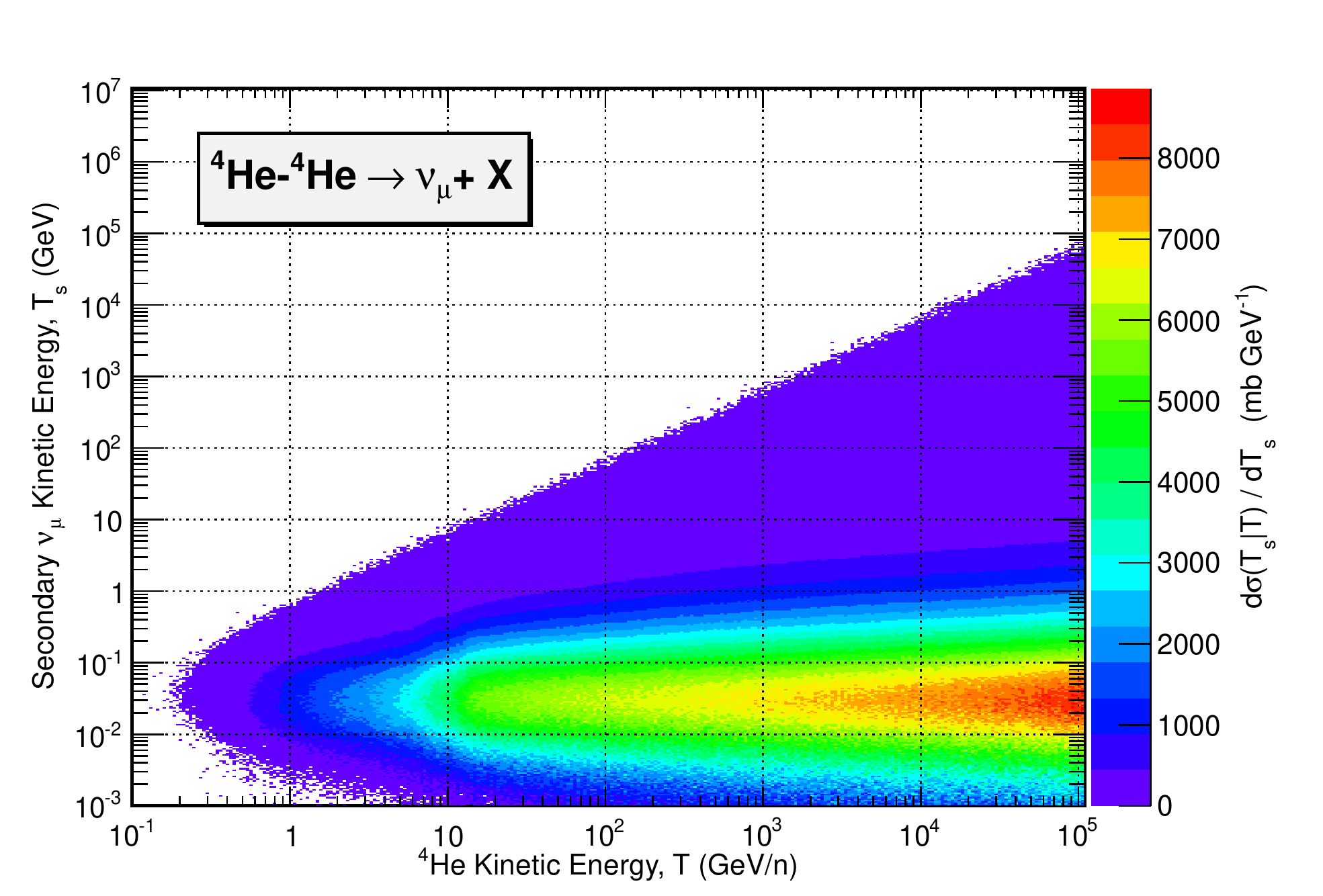} \\
\includegraphics[width=0.99\columnwidth,height=0.19\textheight,clip]{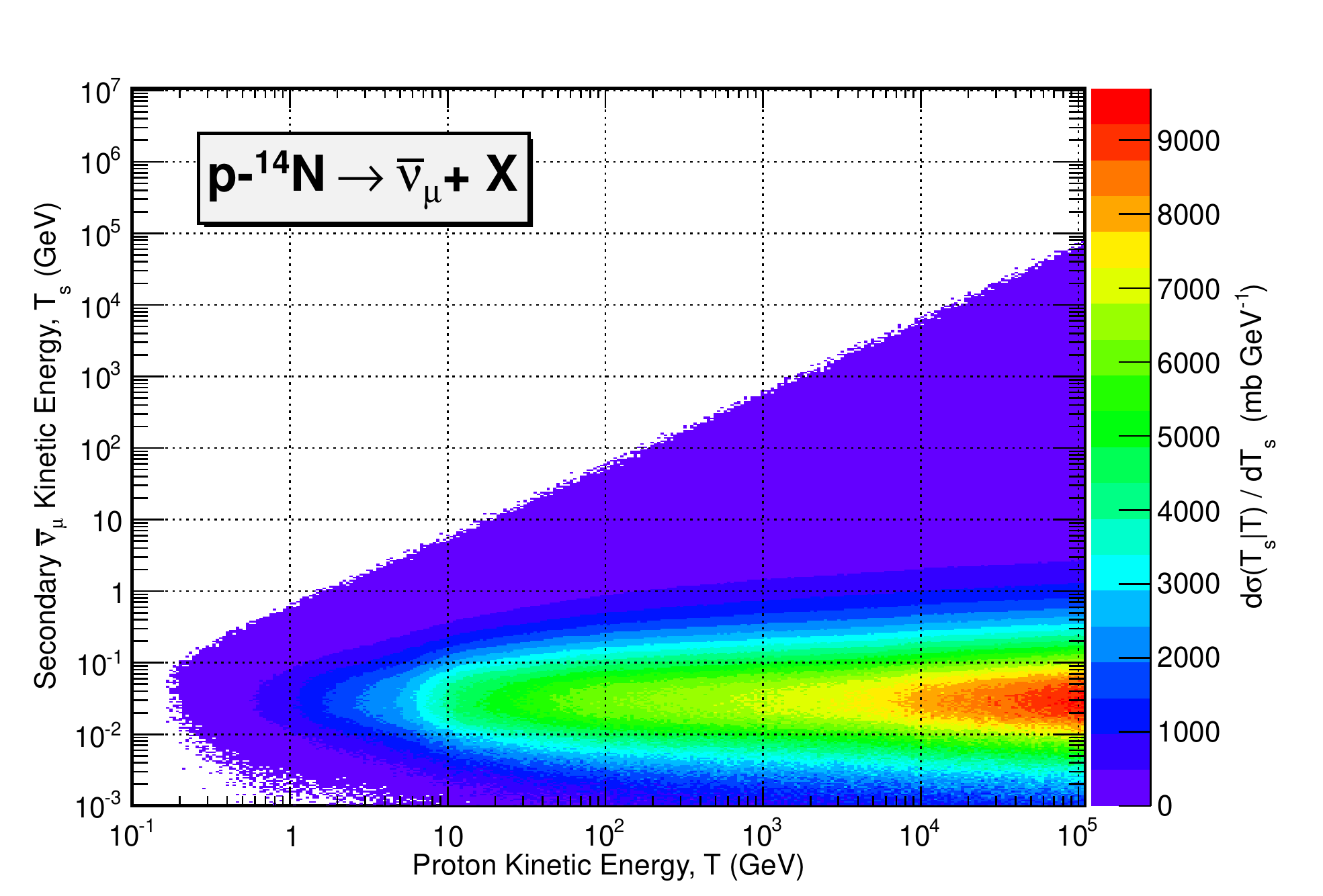} &
\includegraphics[width=0.99\columnwidth,height=0.19\textheight,clip]{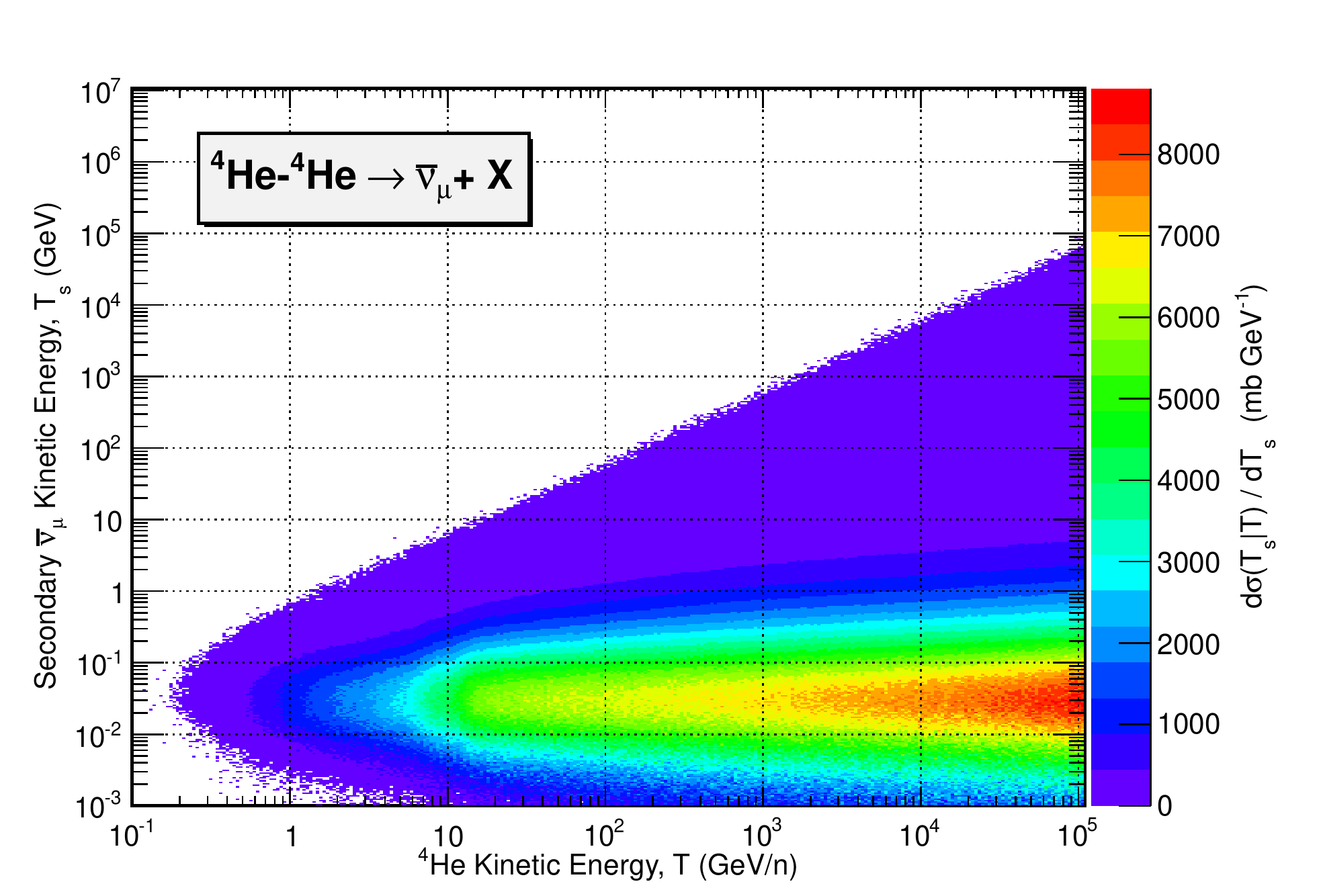} \\
\includegraphics[width=0.99\columnwidth,height=0.19\textheight,clip]{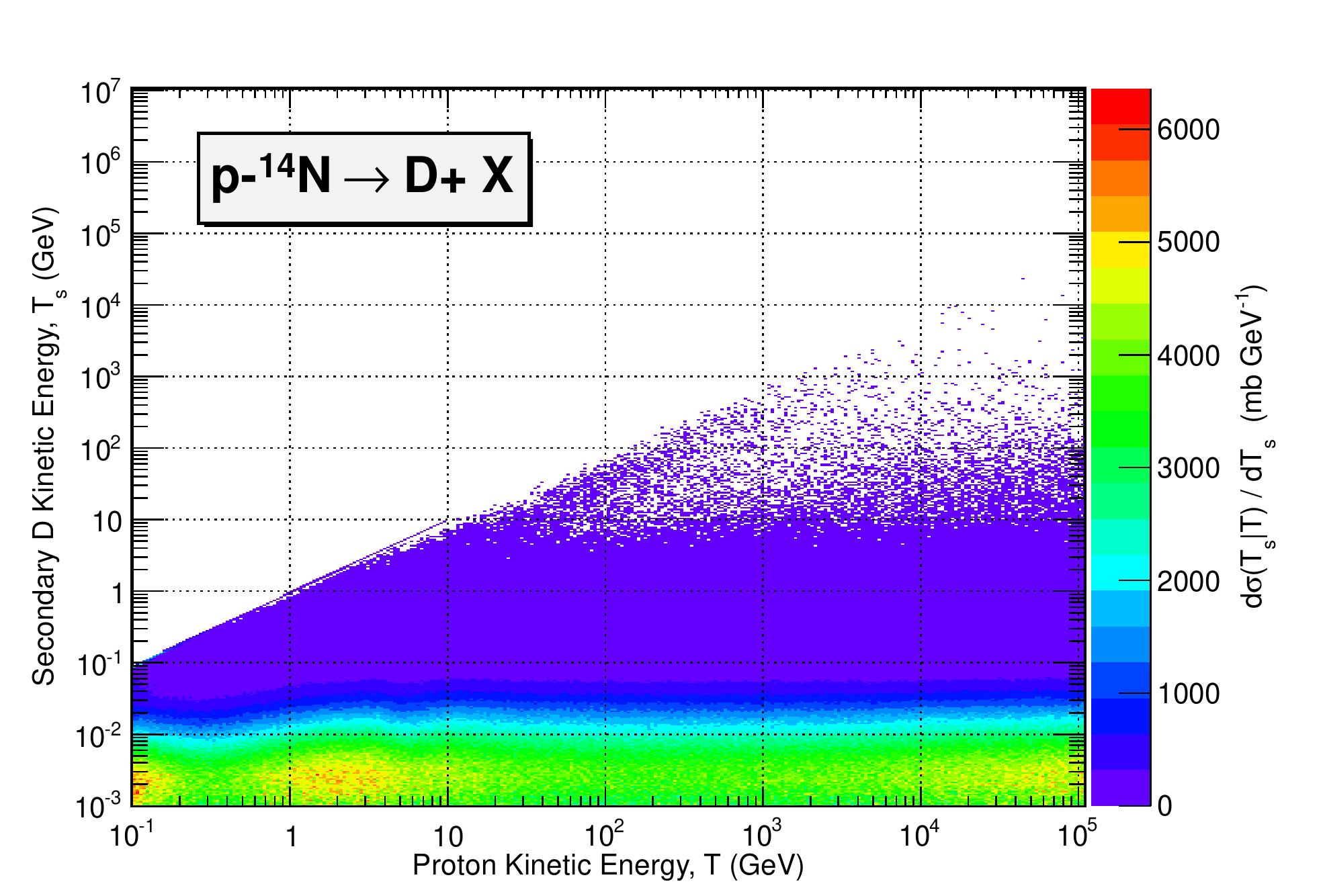} &
\includegraphics[width=0.99\columnwidth,height=0.19\textheight,clip]{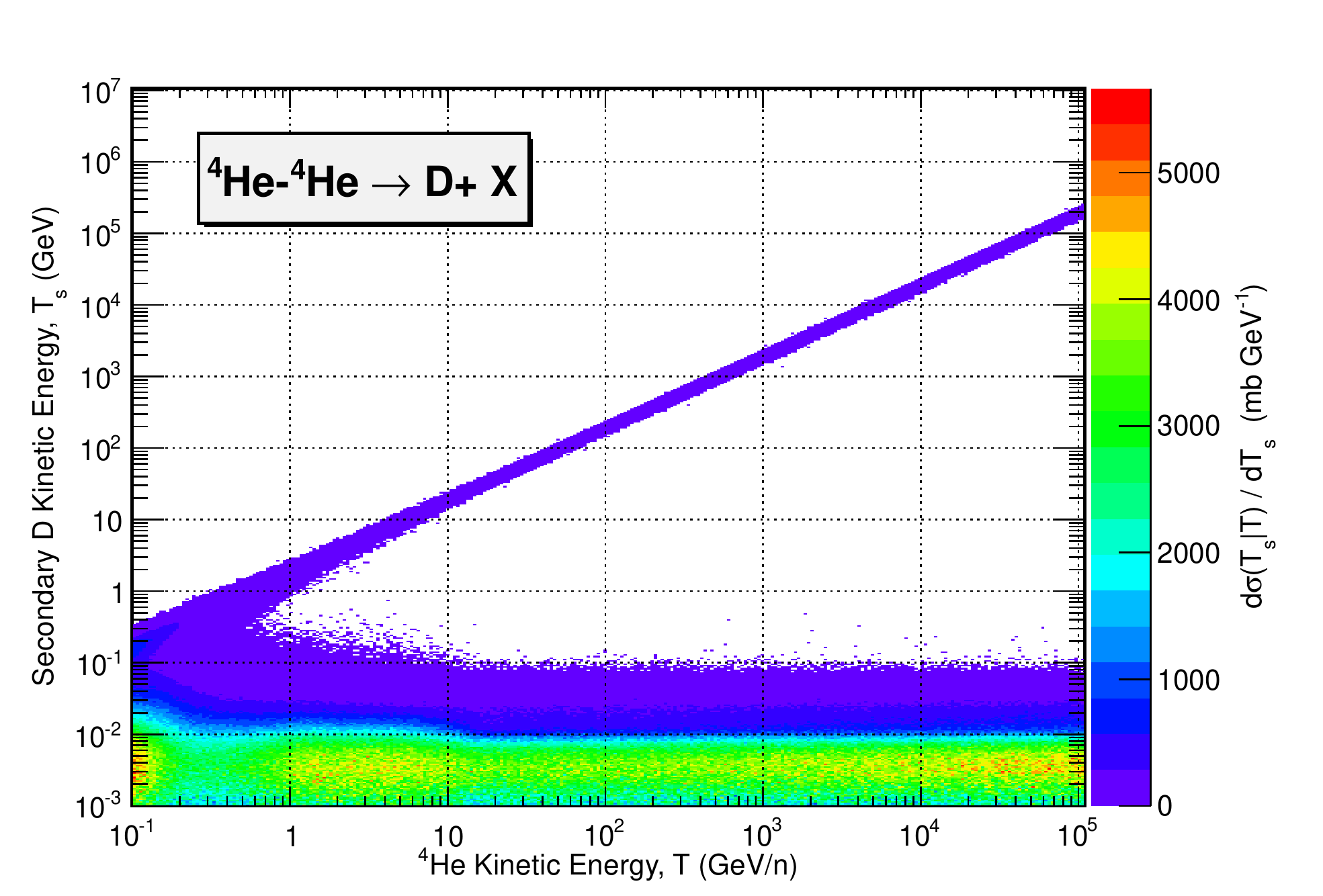} \\
\end{tabular}
\caption{Differential inclusive secondary cross sections for the production of $\nu$, $\bar{\nu}$ and deuteron 
in $p-^{14}N$ (left) and  $^{4}He-^{4}He$ (right) interactions. The values of the cross sections (color scales) are in $\units{mb~GeV^{-1}}$.}
\label{FigpNHeHe2DXsecDPMApx}
\end{figure*}

\begin{figure*}[!ht]
\begin{tabular}{cc}
\includegraphics[width=1\columnwidth,height=0.18\textheight,clip]{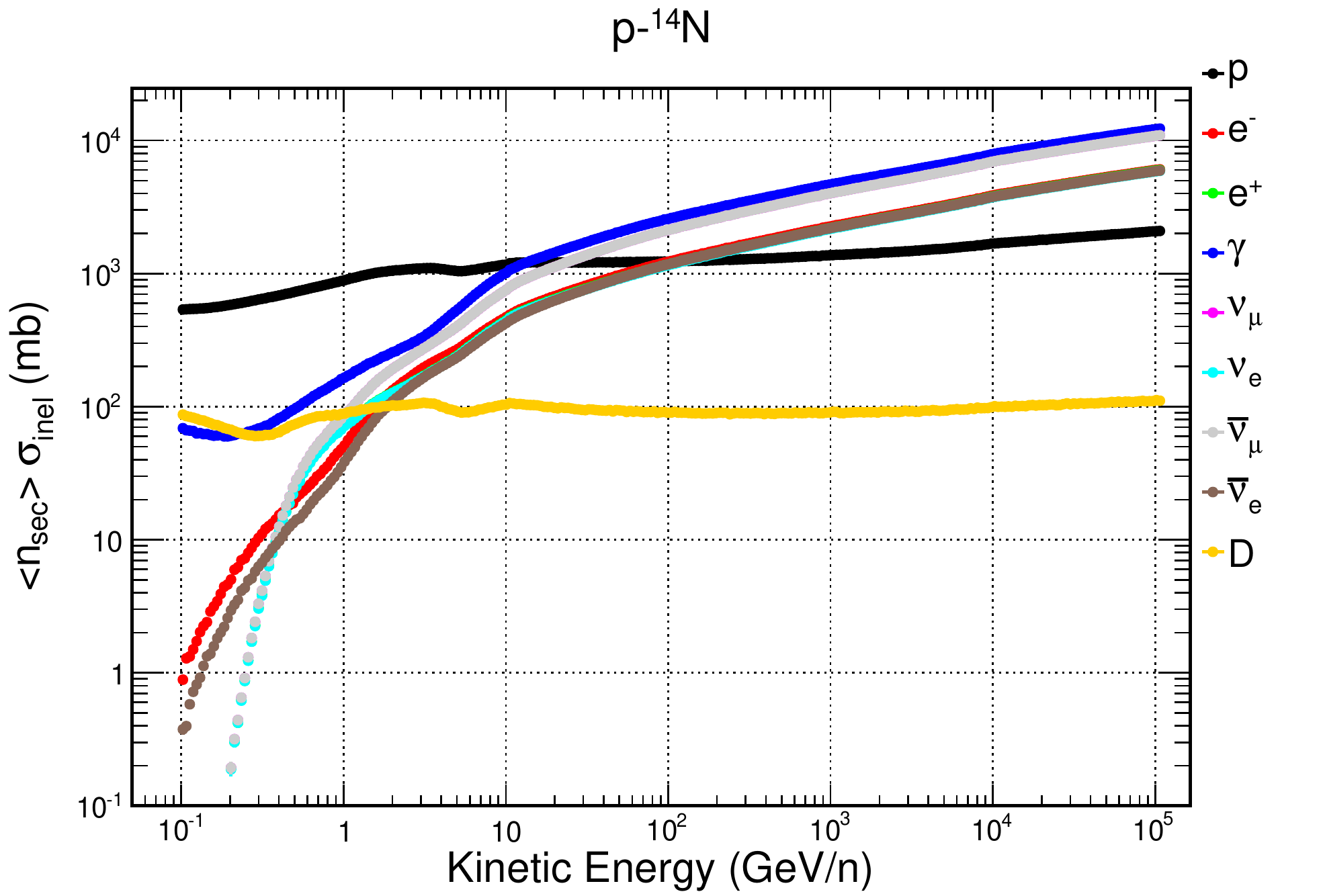} &
\includegraphics[width=1\columnwidth,height=0.18\textheight,clip]{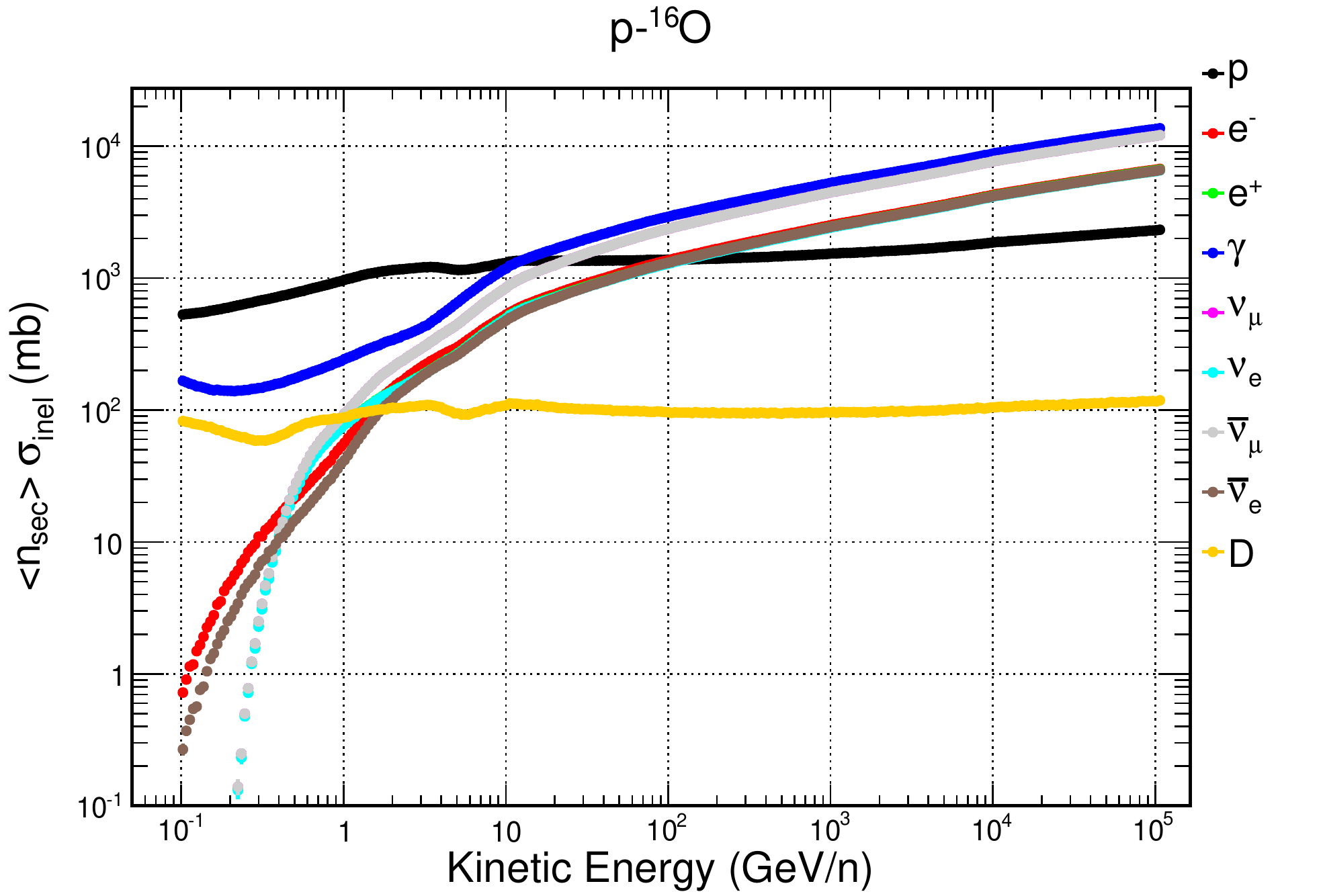} \\
\includegraphics[width=1\columnwidth,height=0.18\textheight,clip]{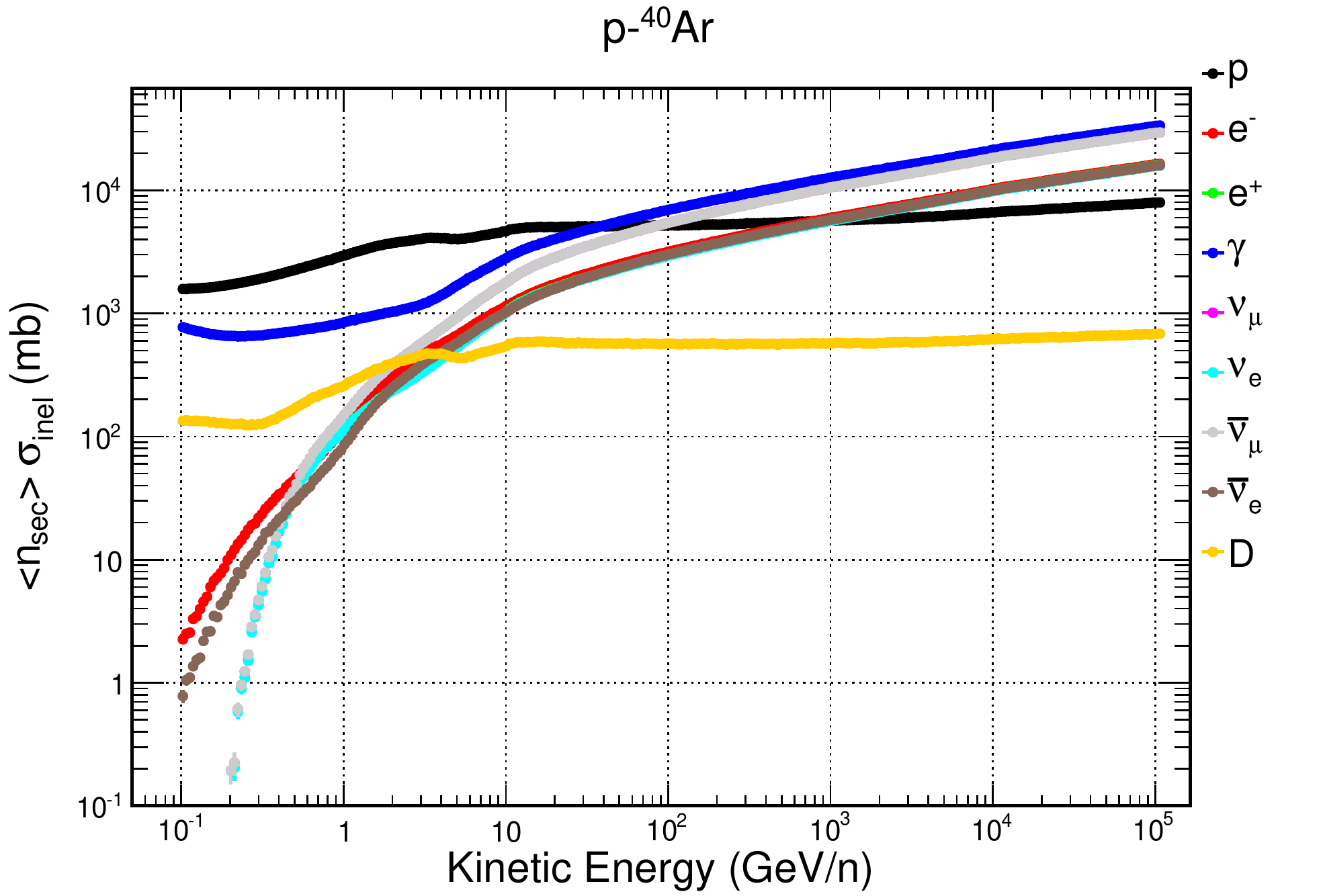} &
\includegraphics[width=1\columnwidth,height=0.18\textheight,clip]{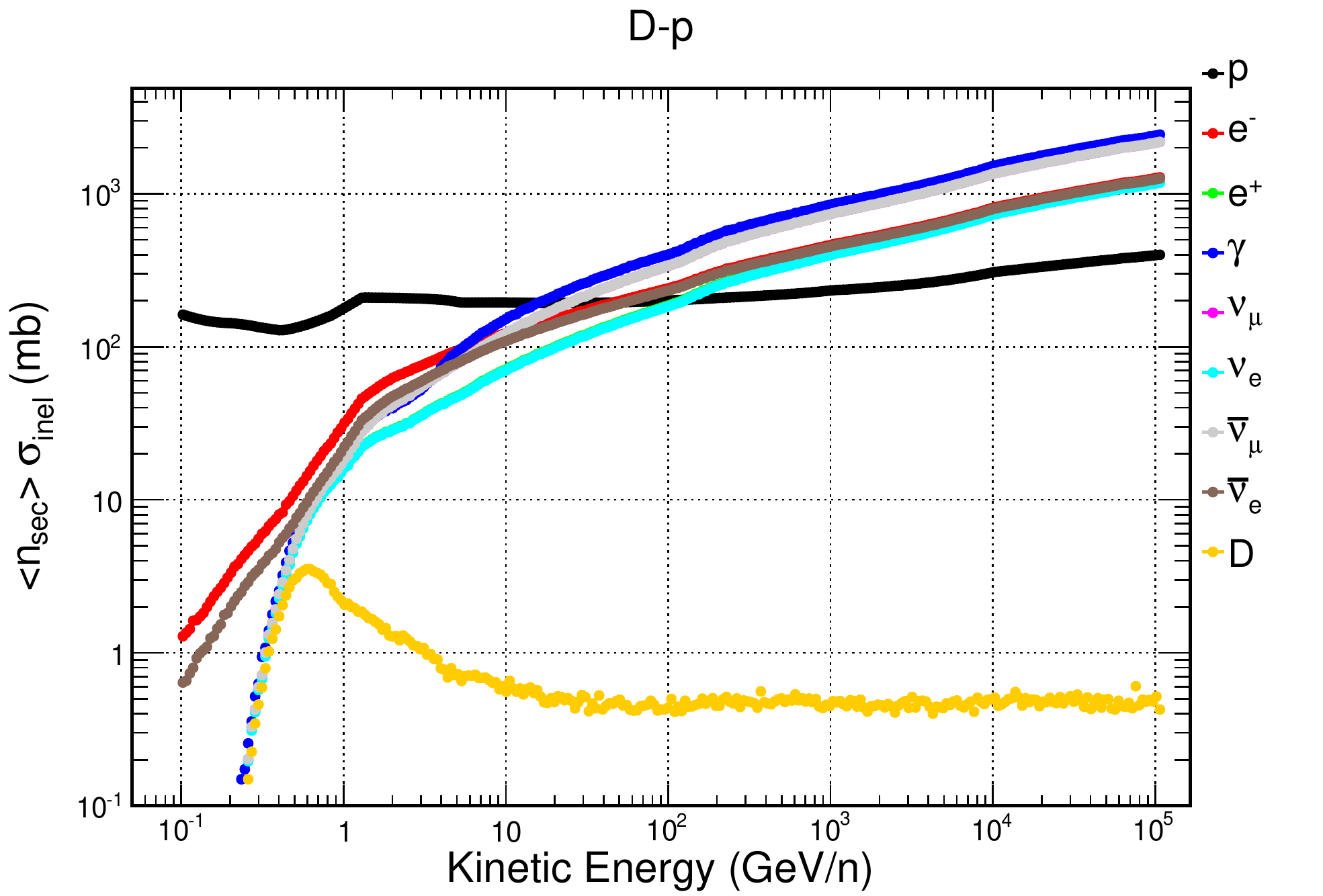} \\
\includegraphics[width=1\columnwidth,height=0.18\textheight,clip]{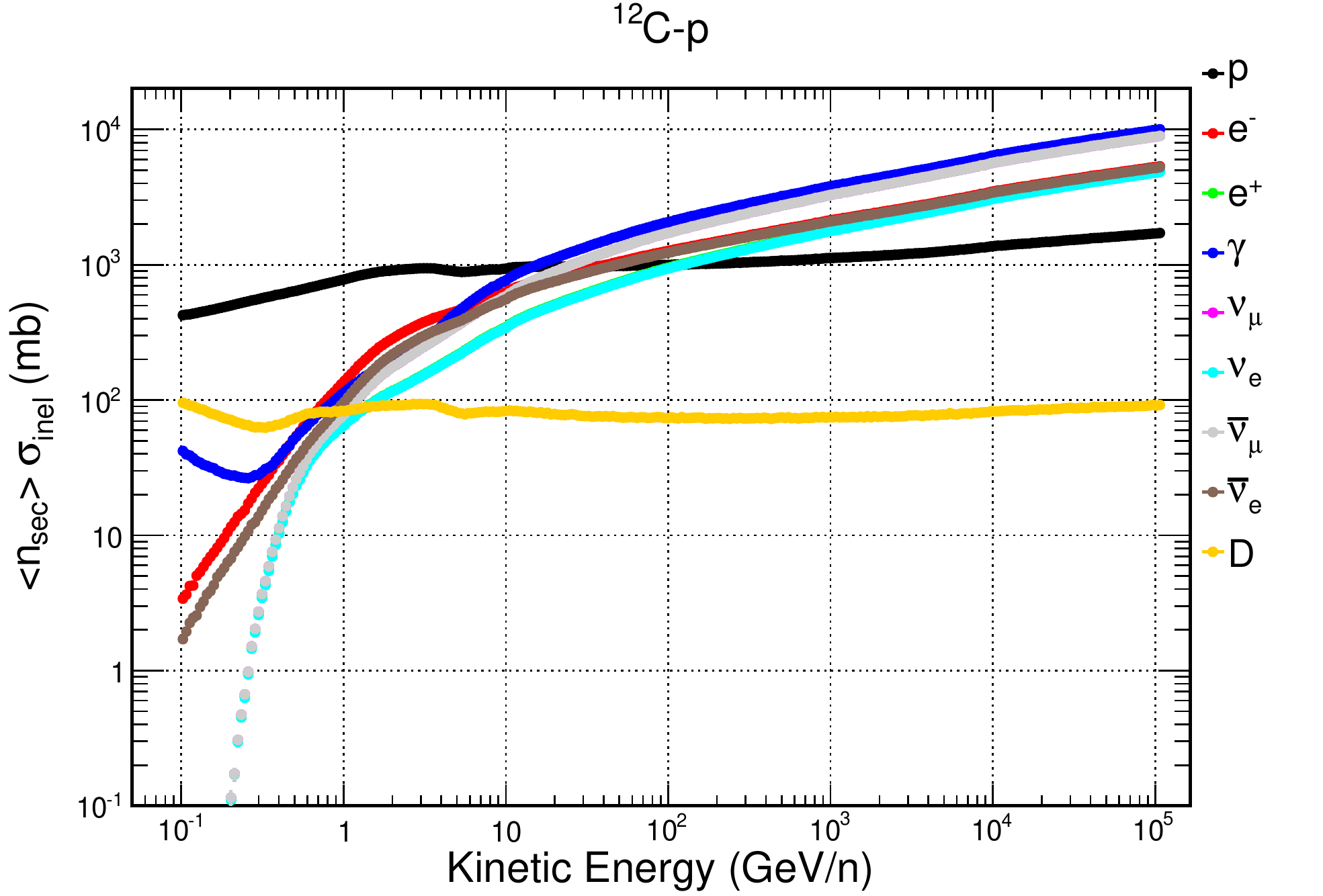}  &
\includegraphics[width=1\columnwidth,height=0.18\textheight,clip]{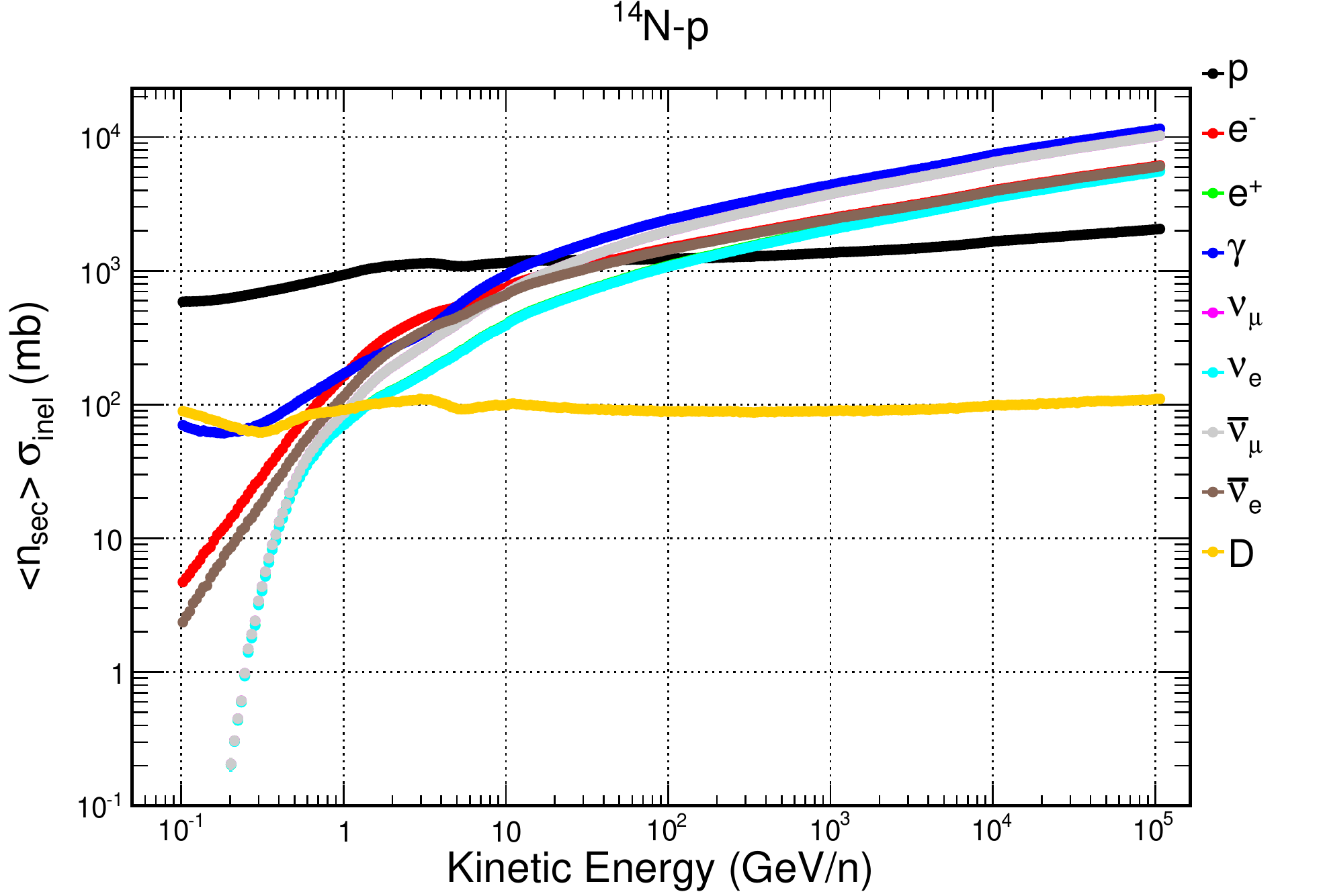}  \\
\includegraphics[width=1\columnwidth,height=0.18\textheight,clip]{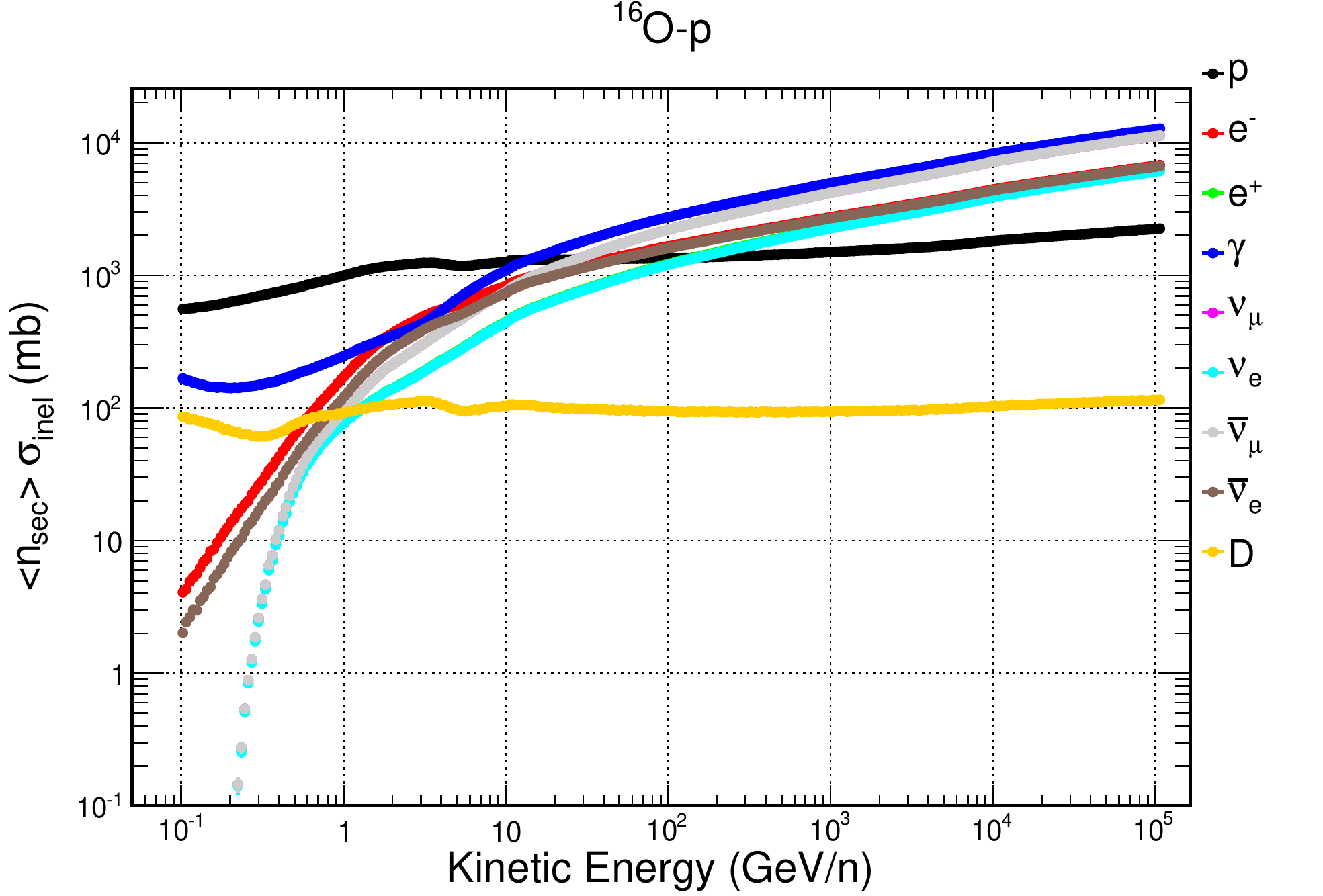} &
\includegraphics[width=1\columnwidth,height=0.18\textheight,clip]{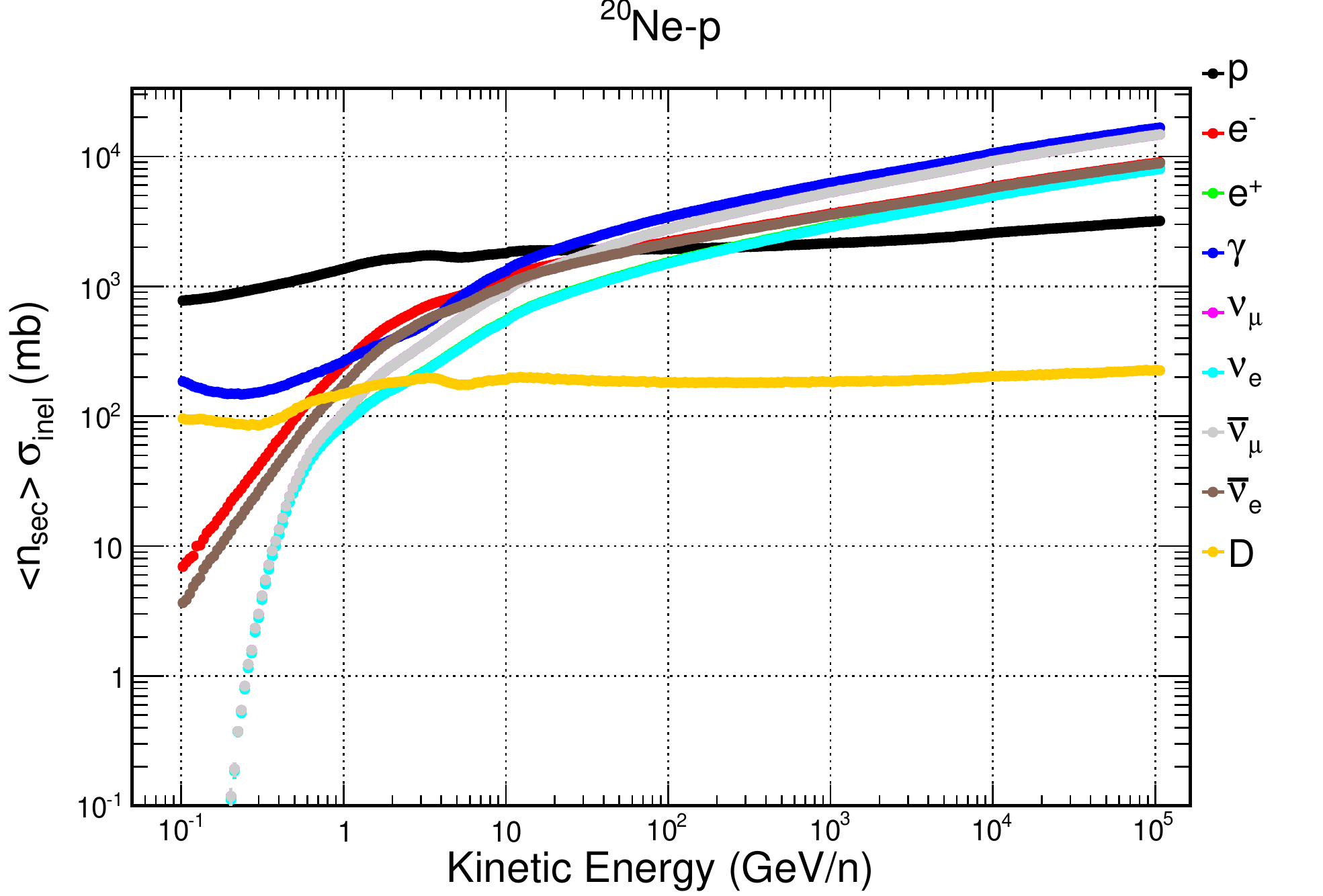} \\
\includegraphics[width=1\columnwidth,height=0.18\textheight,clip]{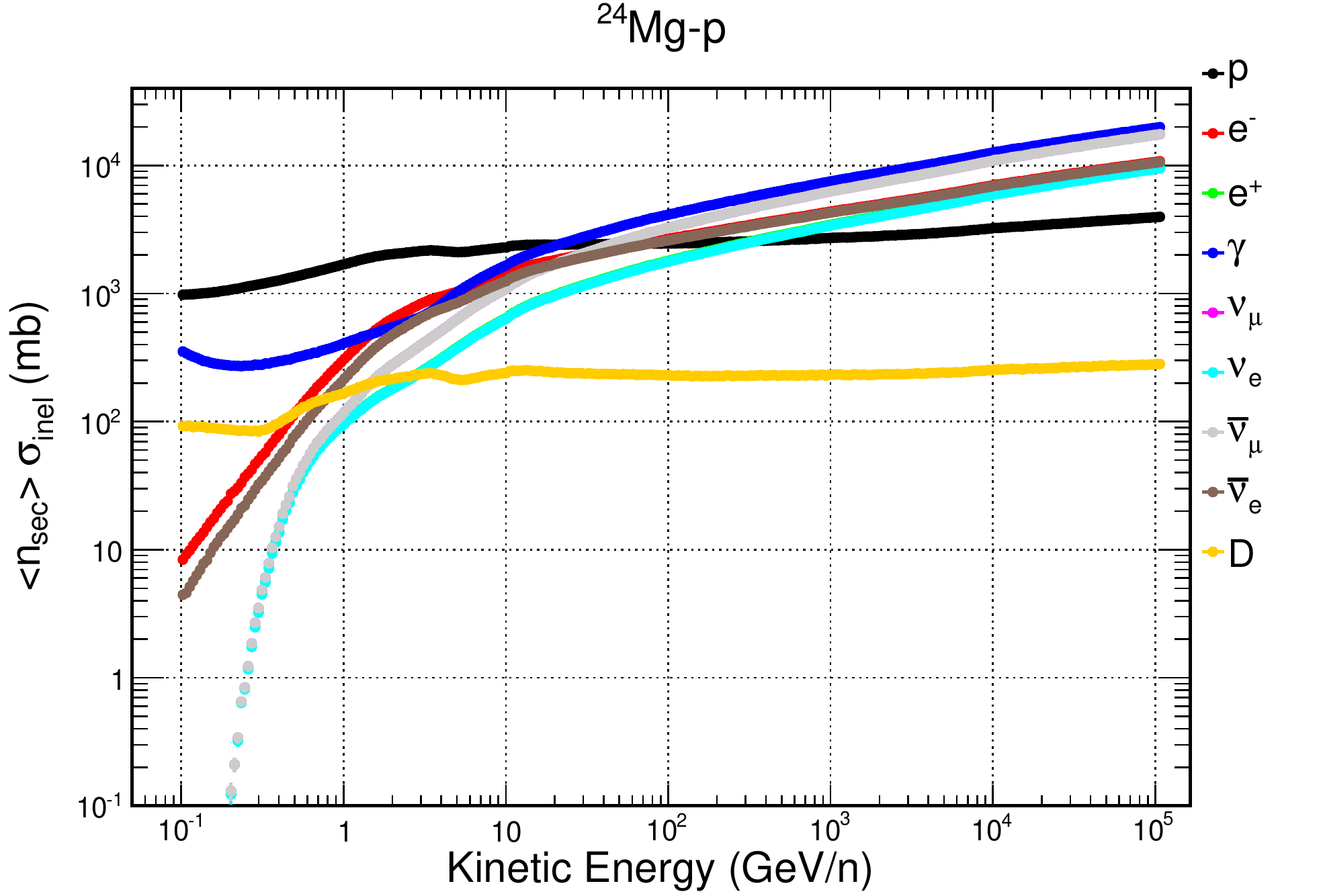} &
\includegraphics[width=1\columnwidth,height=0.18\textheight,clip]{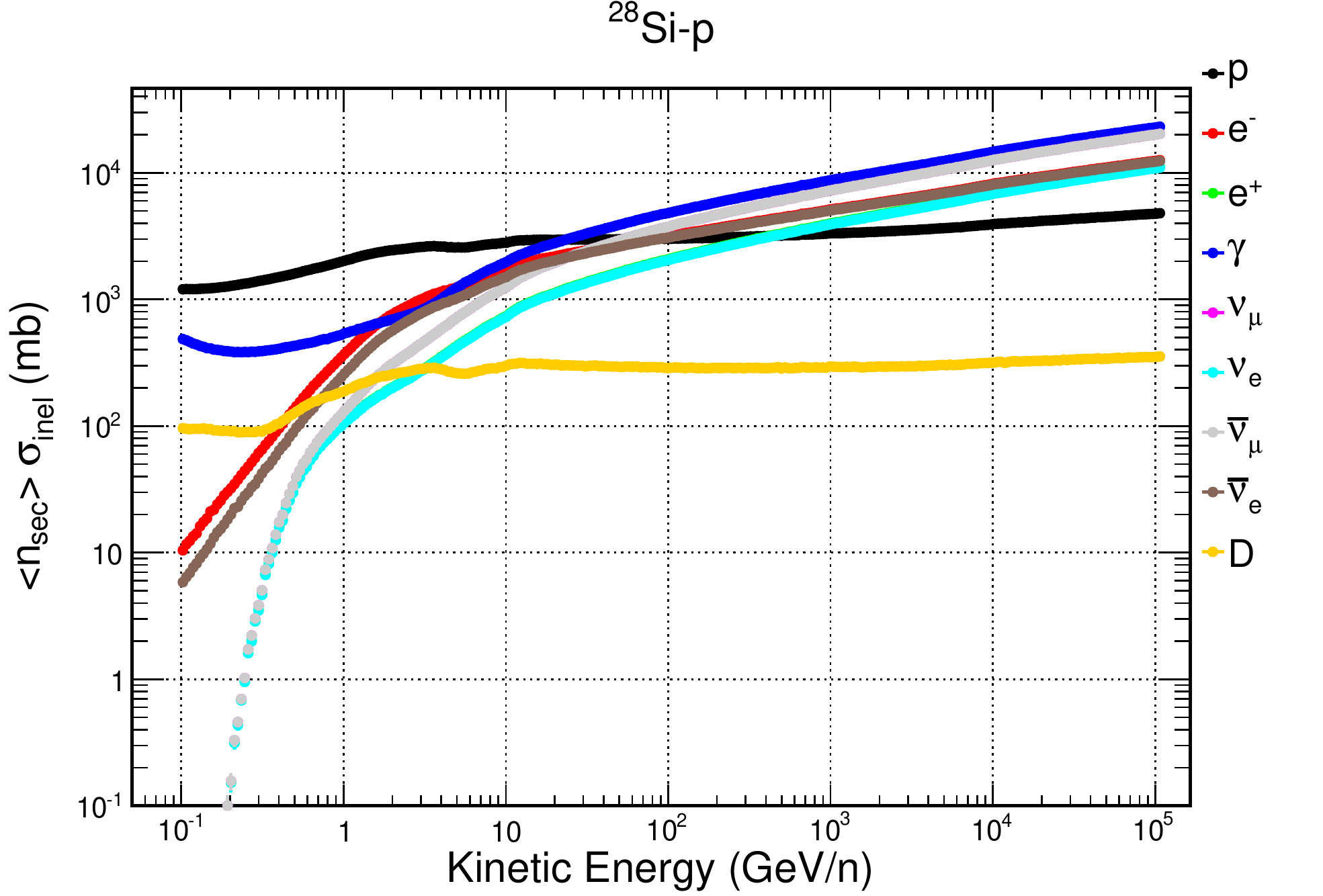} \\
\end{tabular}
\caption{Inclusive cross sections for the production of protons (black), electrons (red), 
positrons (green), gamma rays (blue), electron neutrinos (cyan), 
electron antineutrinos (grey), muon neutrinos (magenta), muon antineutrinos (brown)
and Deuterons (orange) in the collisions of several CR projectiles with different target nuclei.}
\label{FigHeOthXsecApx1}
\end{figure*}

\begin{figure*}[!ht]
\begin{tabular}{cc}
\includegraphics[width=1\columnwidth,height=0.18\textheight,clip]{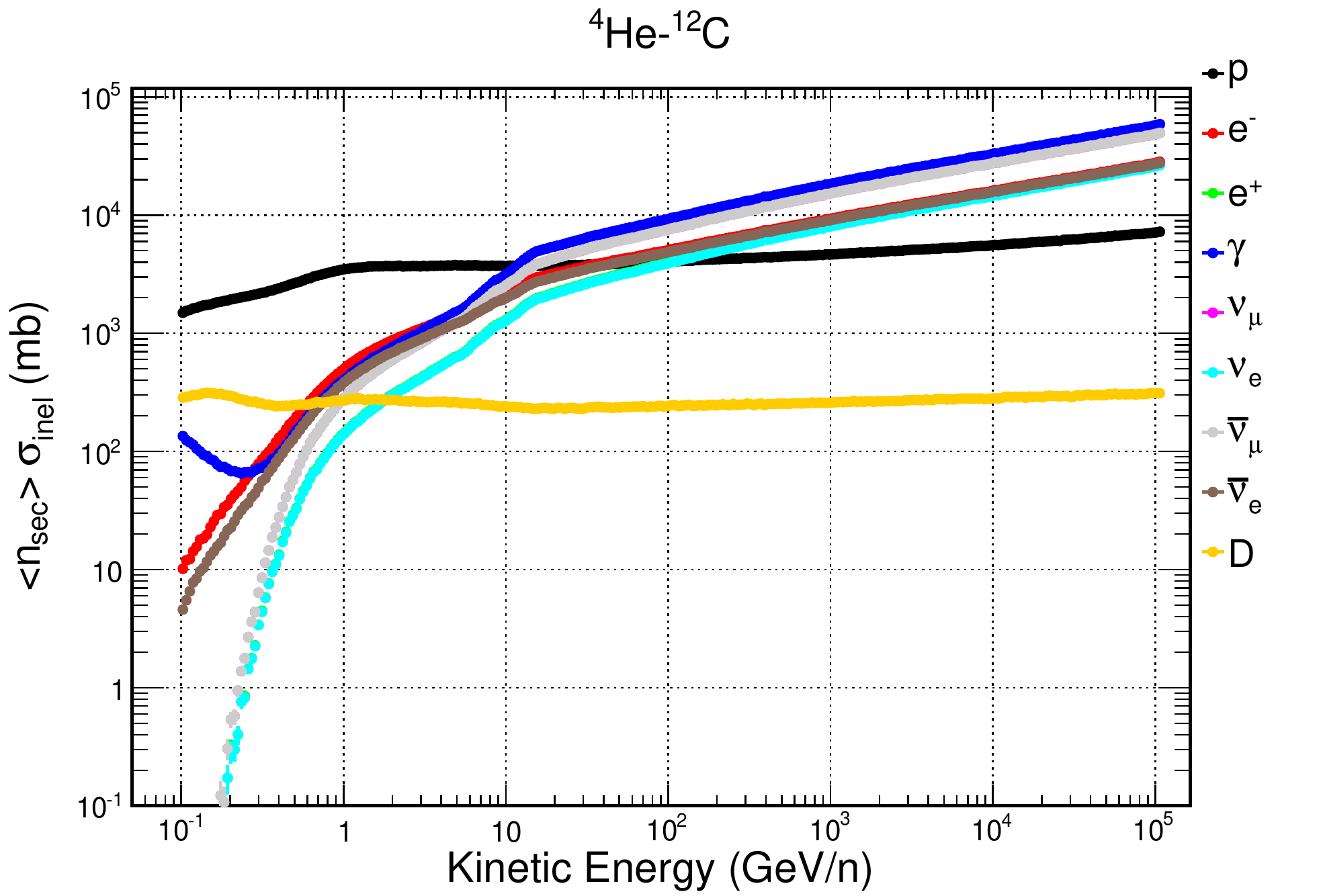} &
\includegraphics[width=1\columnwidth,height=0.18\textheight,clip]{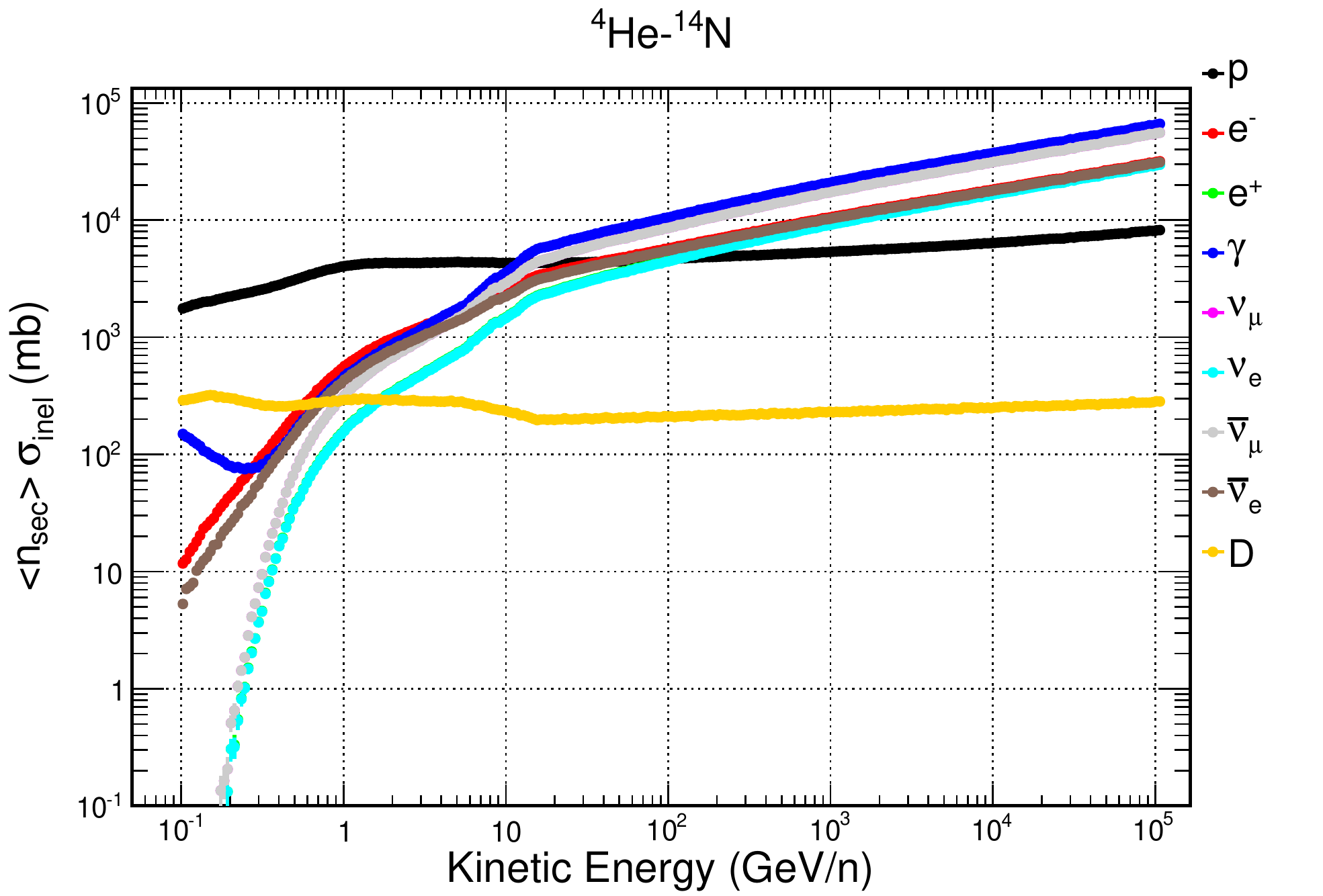} \\
\includegraphics[width=1\columnwidth,height=0.18\textheight,clip]{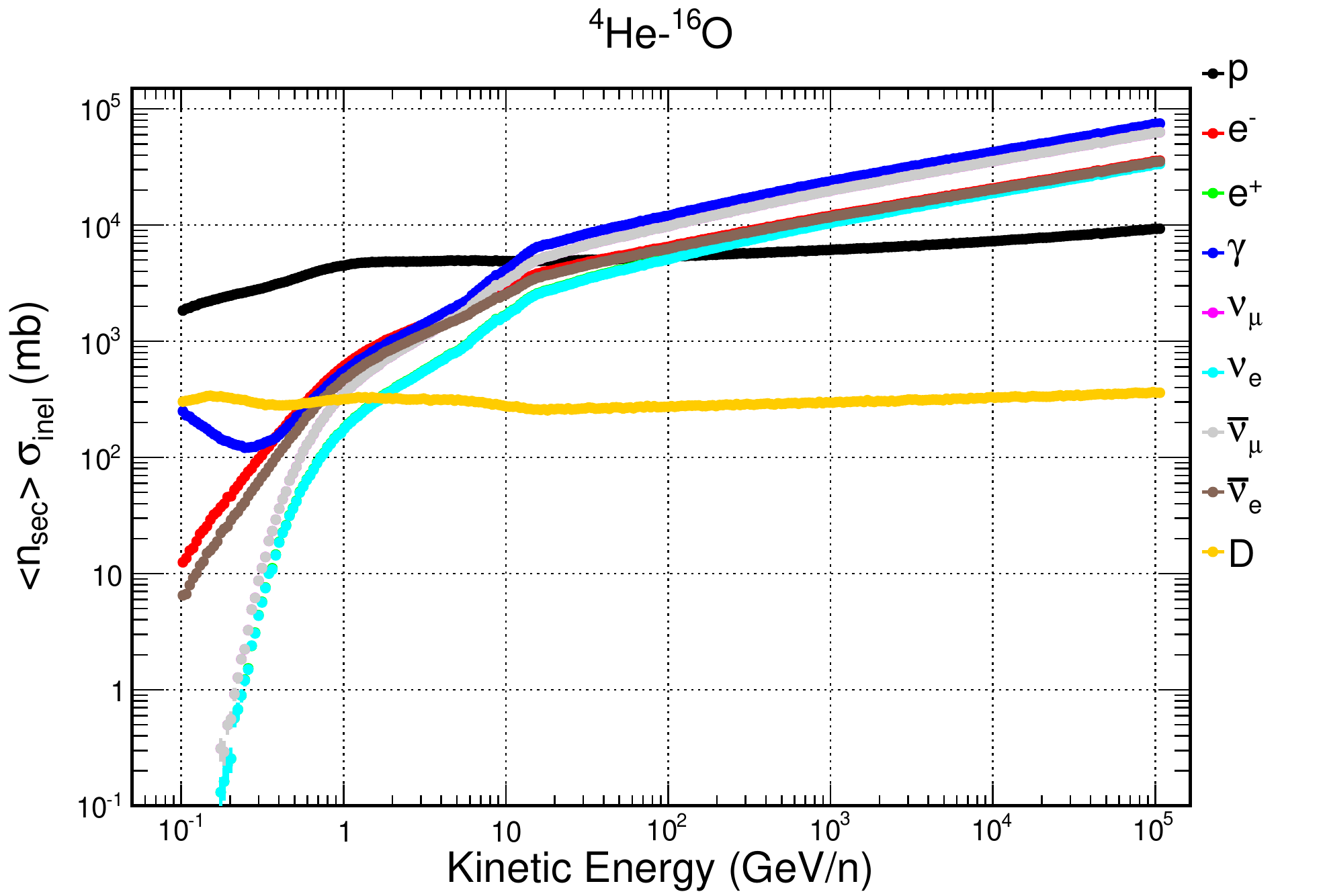} &
\includegraphics[width=1\columnwidth,height=0.18\textheight,clip]{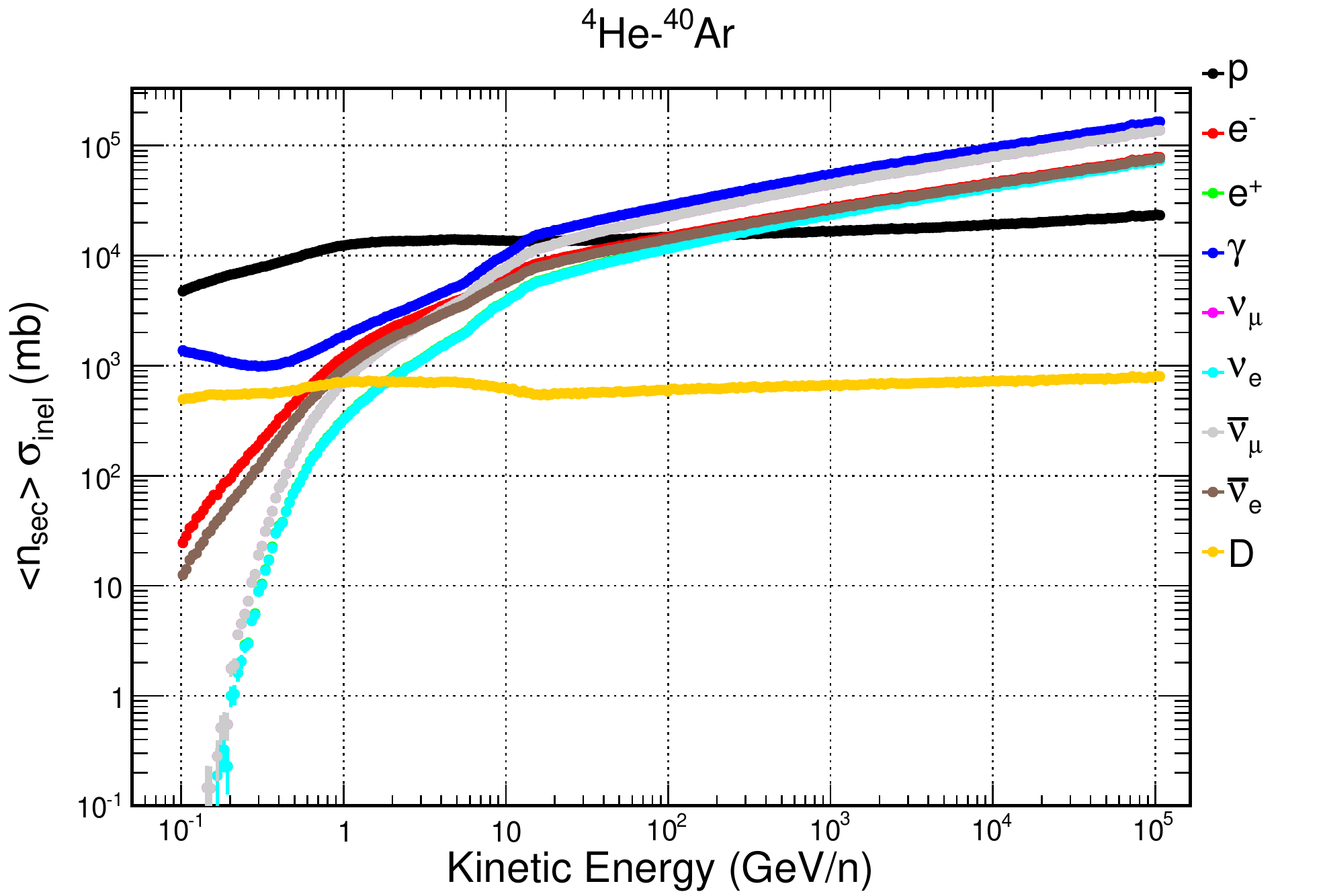} \\
\includegraphics[width=1\columnwidth,height=0.18\textheight,clip]{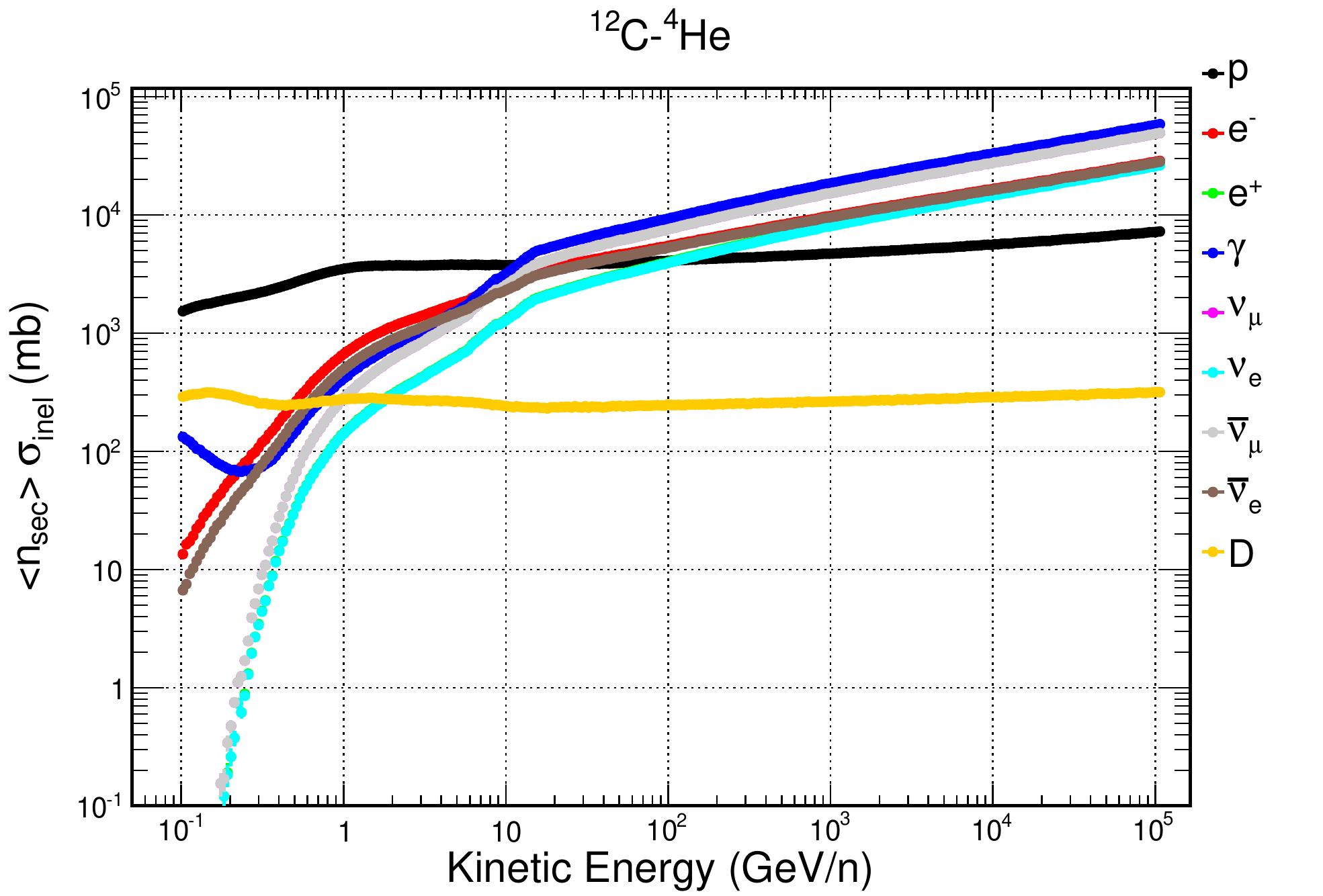} &
\includegraphics[width=1\columnwidth,height=0.18\textheight,clip]{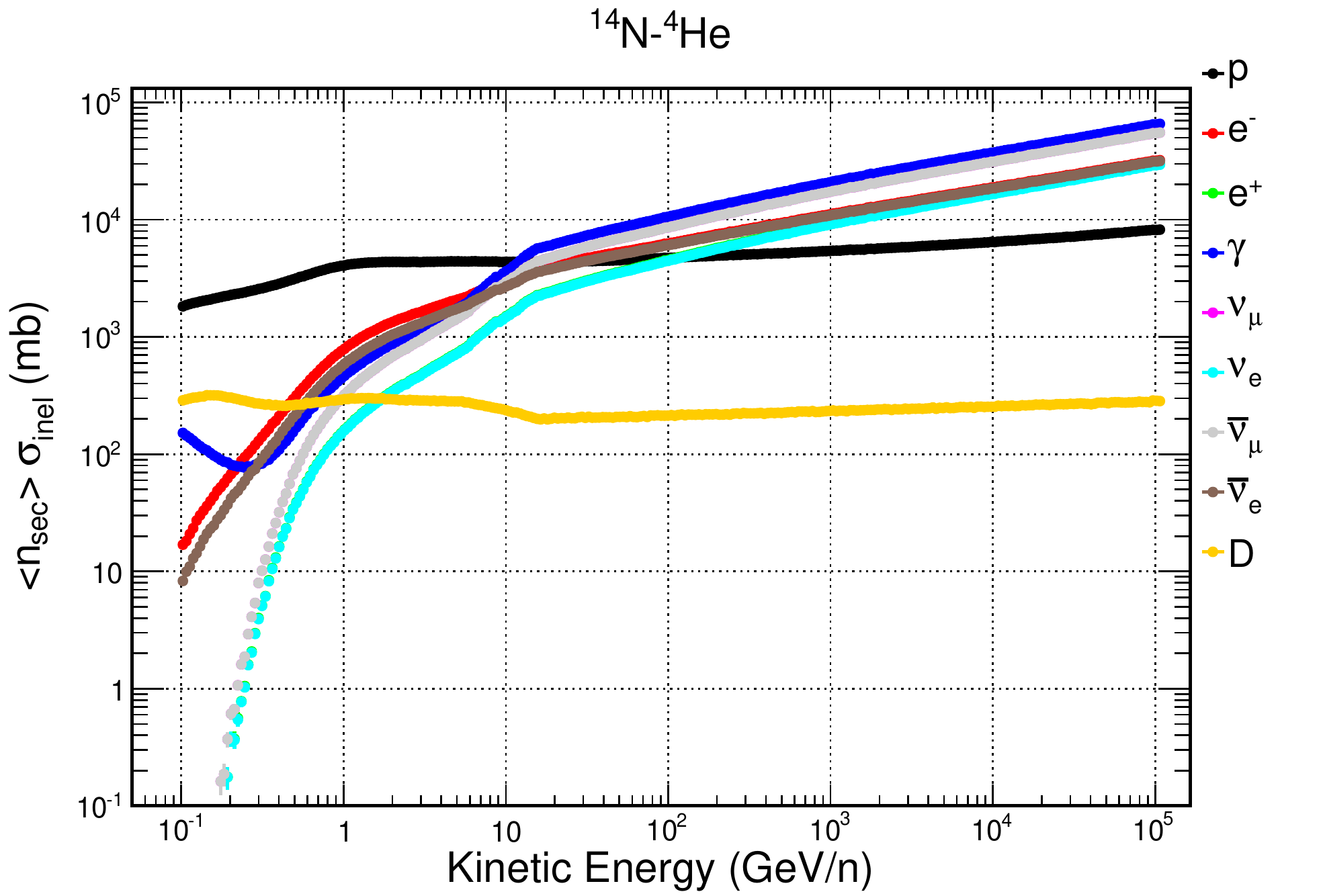} \\
\includegraphics[width=1\columnwidth,height=0.18\textheight,clip]{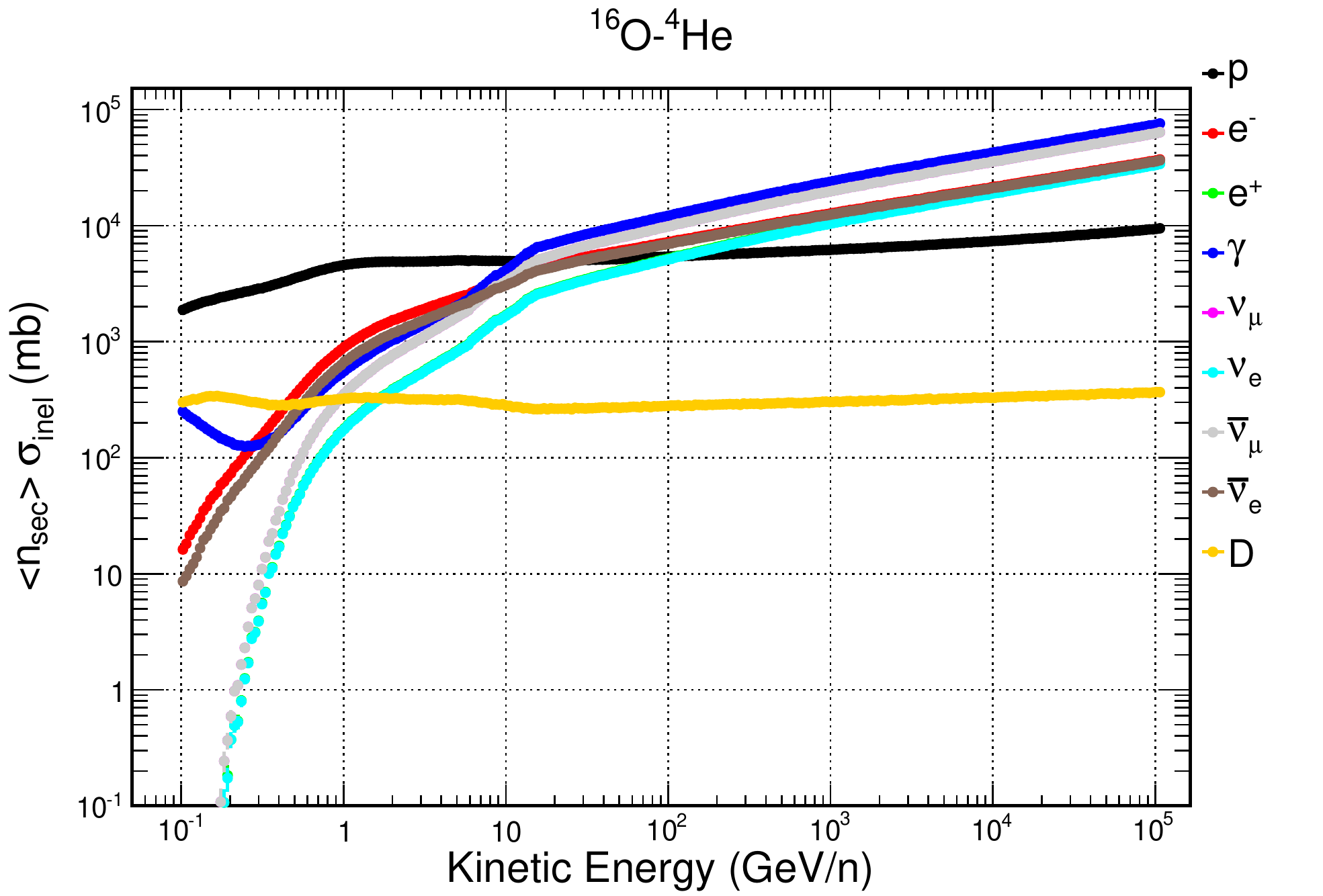} &
\includegraphics[width=1\columnwidth,height=0.18\textheight,clip]{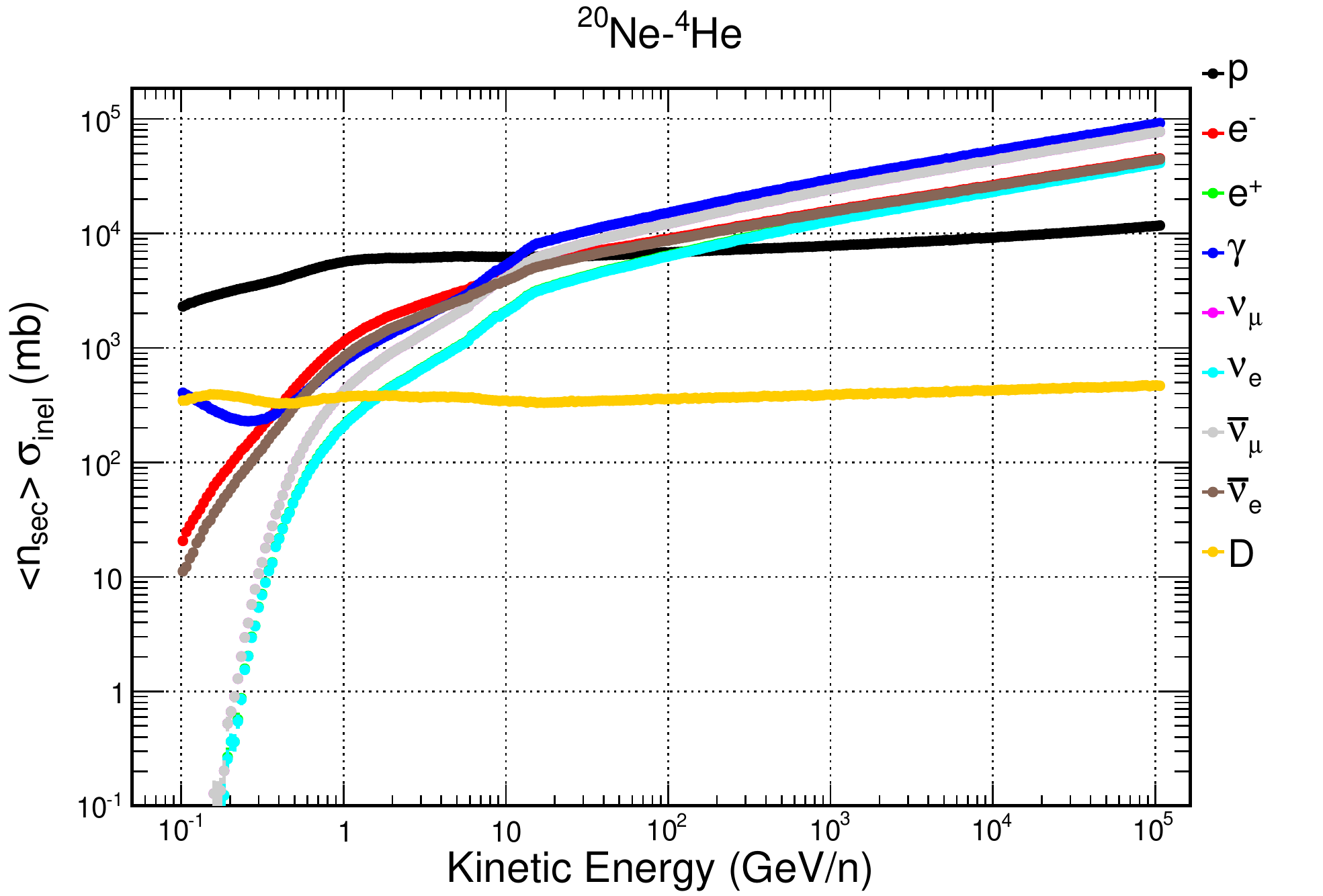} \\
\includegraphics[width=1\columnwidth,height=0.18\textheight,clip]{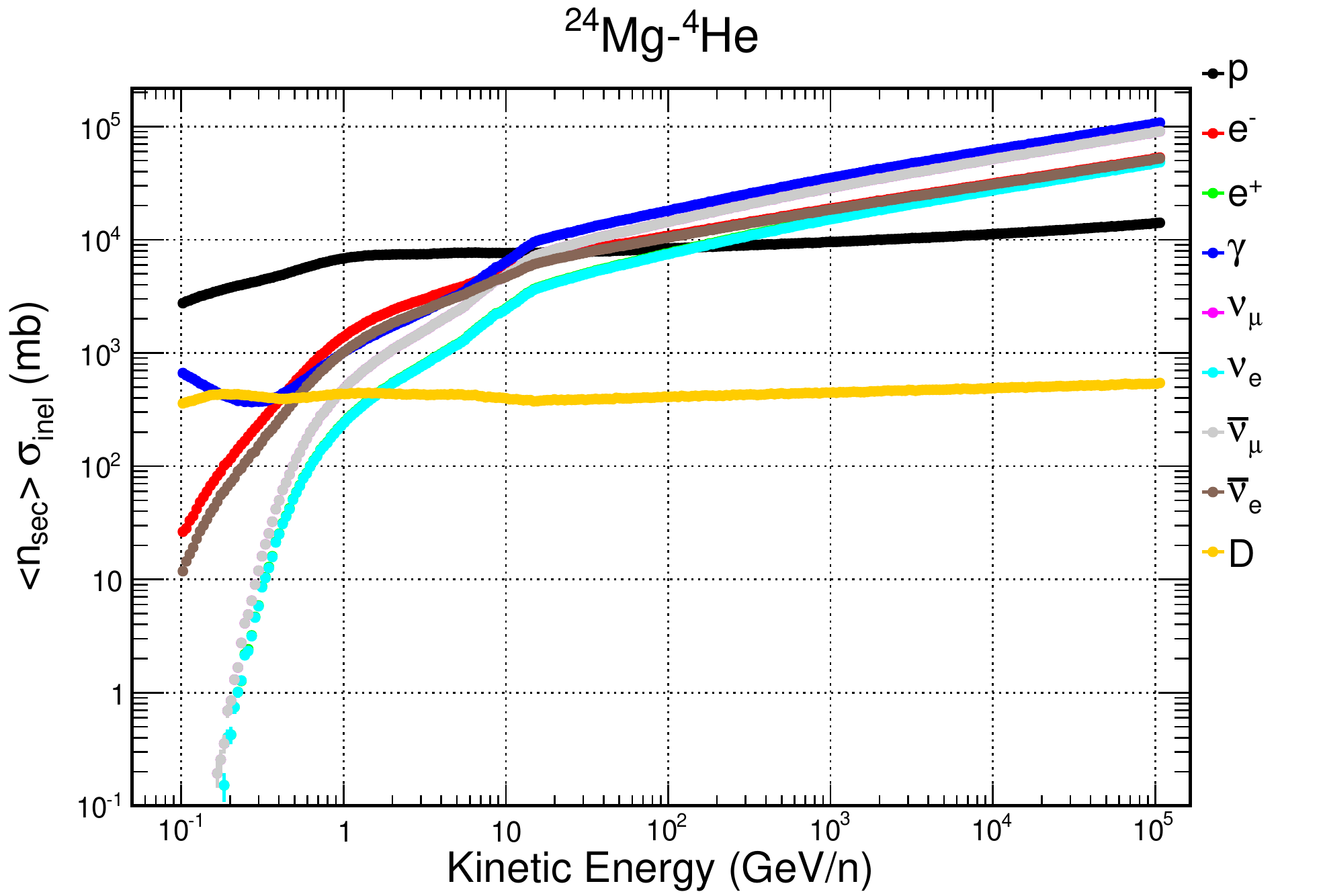} &
\includegraphics[width=1\columnwidth,height=0.18\textheight,clip]{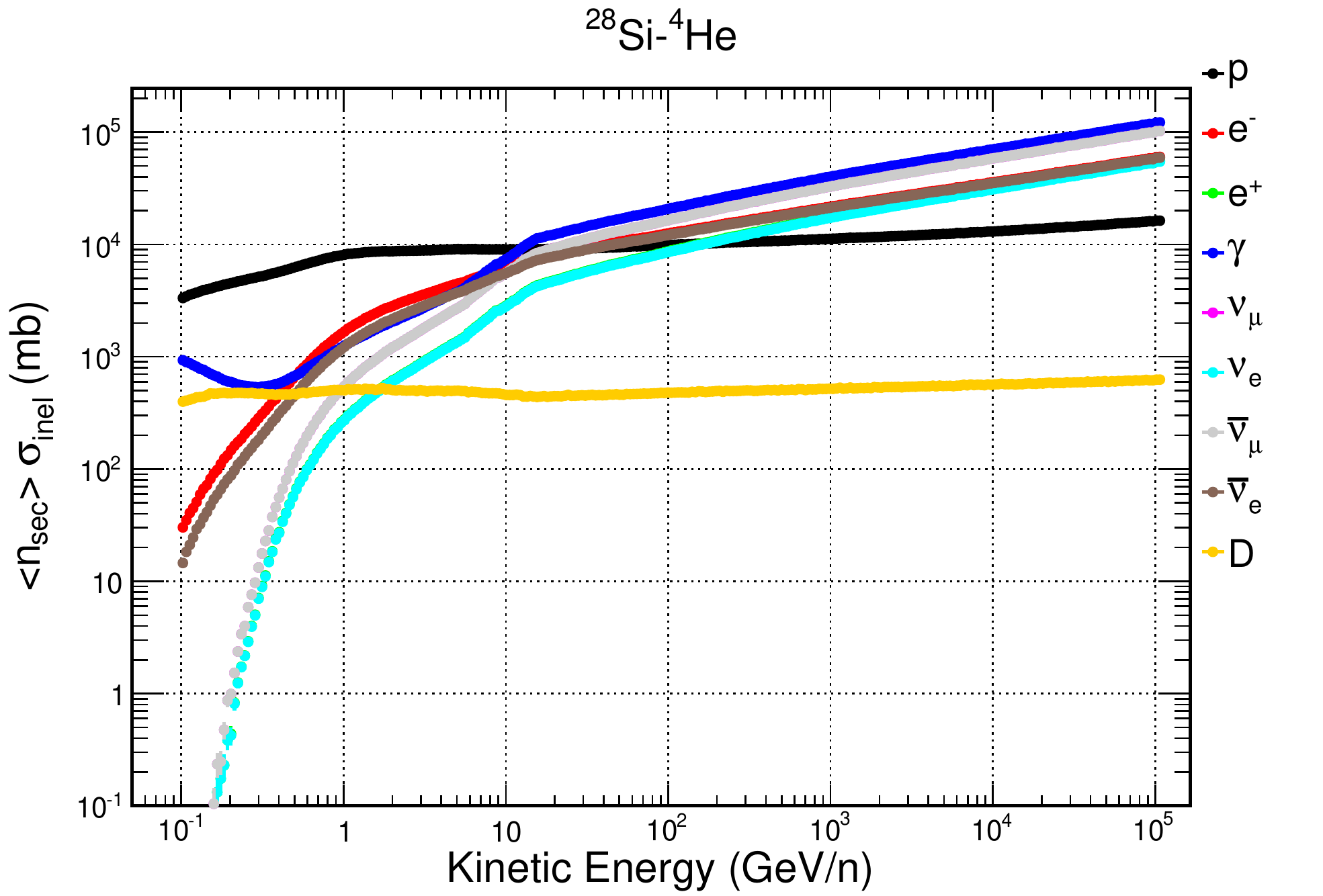} \\
\end{tabular}
\caption{Inclusive cross sections for the production of protons (black), electrons (red), 
positrons (green), gamma rays (blue), electron neutrinos (cyan), 
electron antineutrinos (grey), muon neutrinos (magenta), muon antineutrinos (brown)
and Deuterons (orange) in the collisions of several CR projectiles with different target nuclei.}
\label{FigHeOthXsecApx2}
\end{figure*}

\begin{figure*}[!ht]
\begin{tabular}{cc}
\includegraphics[width=0.99\columnwidth,height=0.225\textheight,clip]{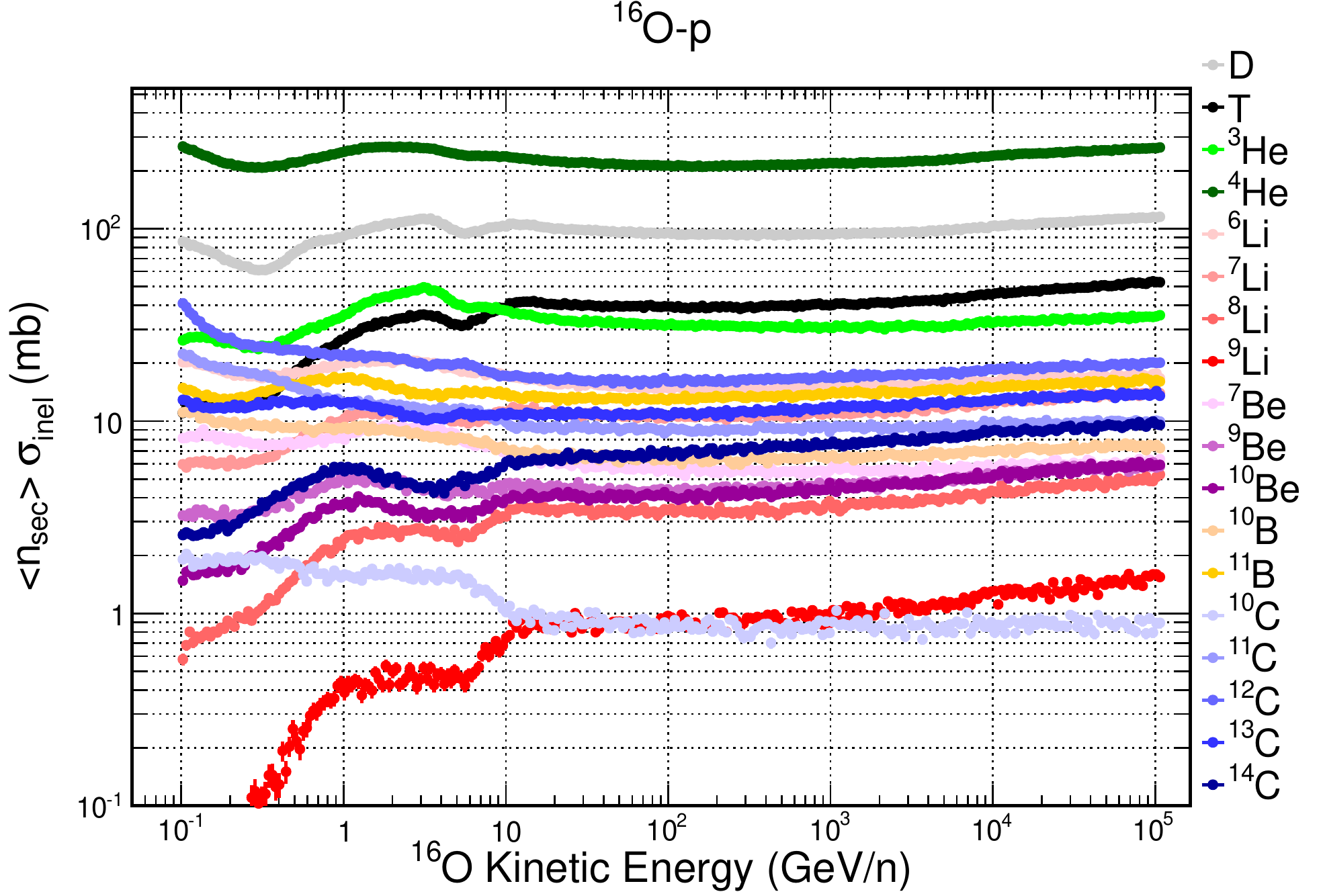} &
\includegraphics[width=0.99\columnwidth,height=0.225\textheight,clip]{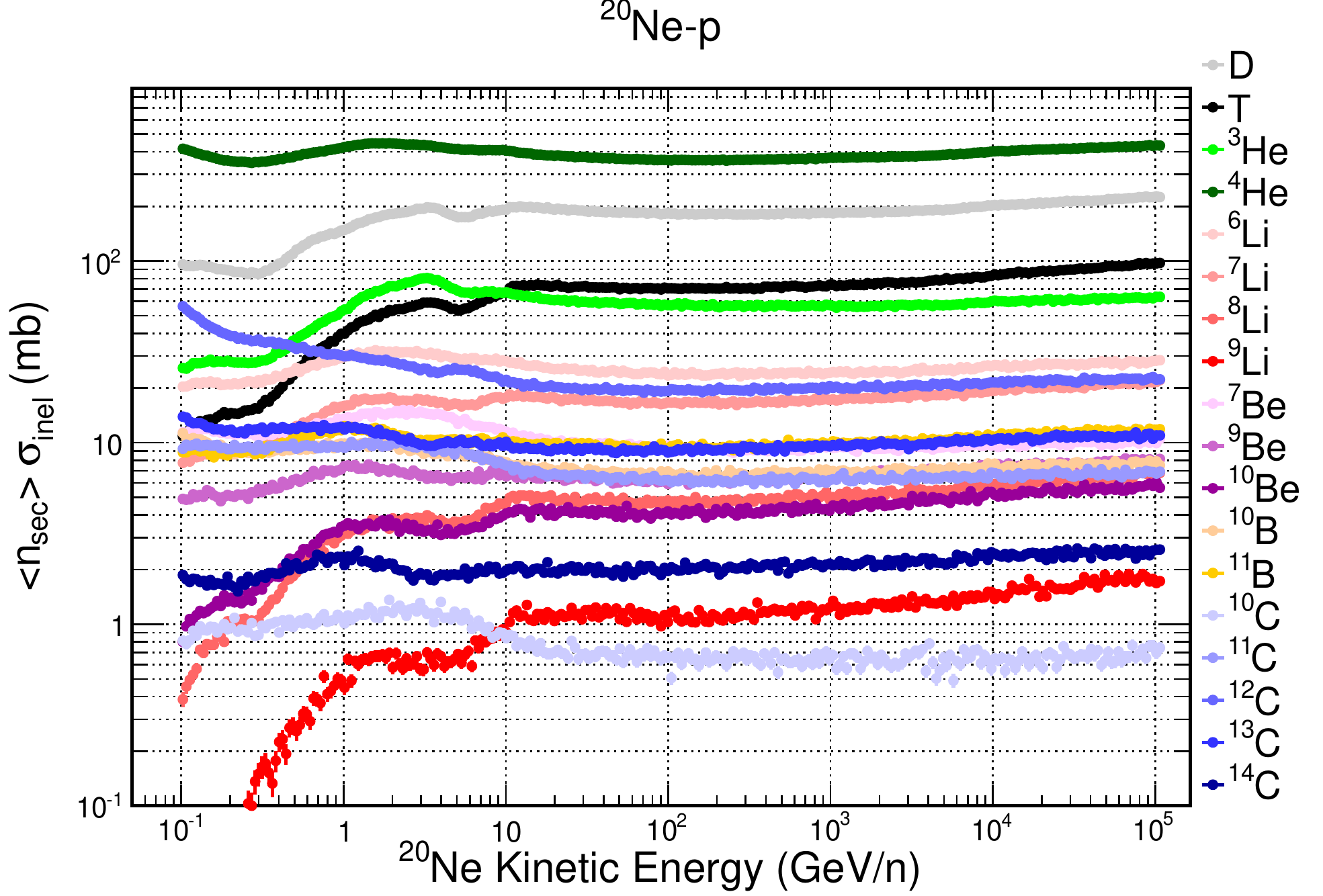} \\
\includegraphics[width=0.99\columnwidth,height=0.225\textheight,clip]{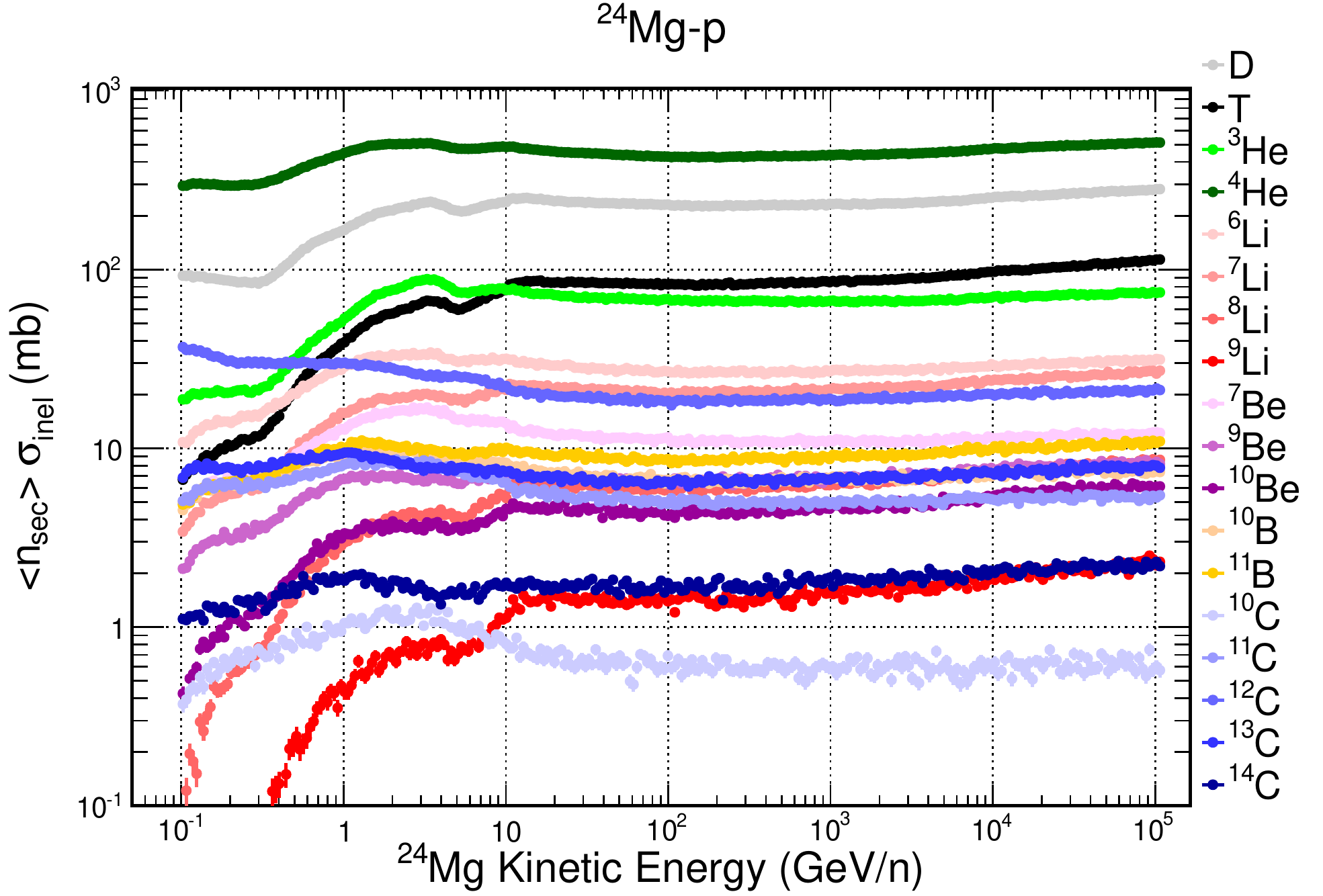} &
\includegraphics[width=0.99\columnwidth,height=0.225\textheight,clip]{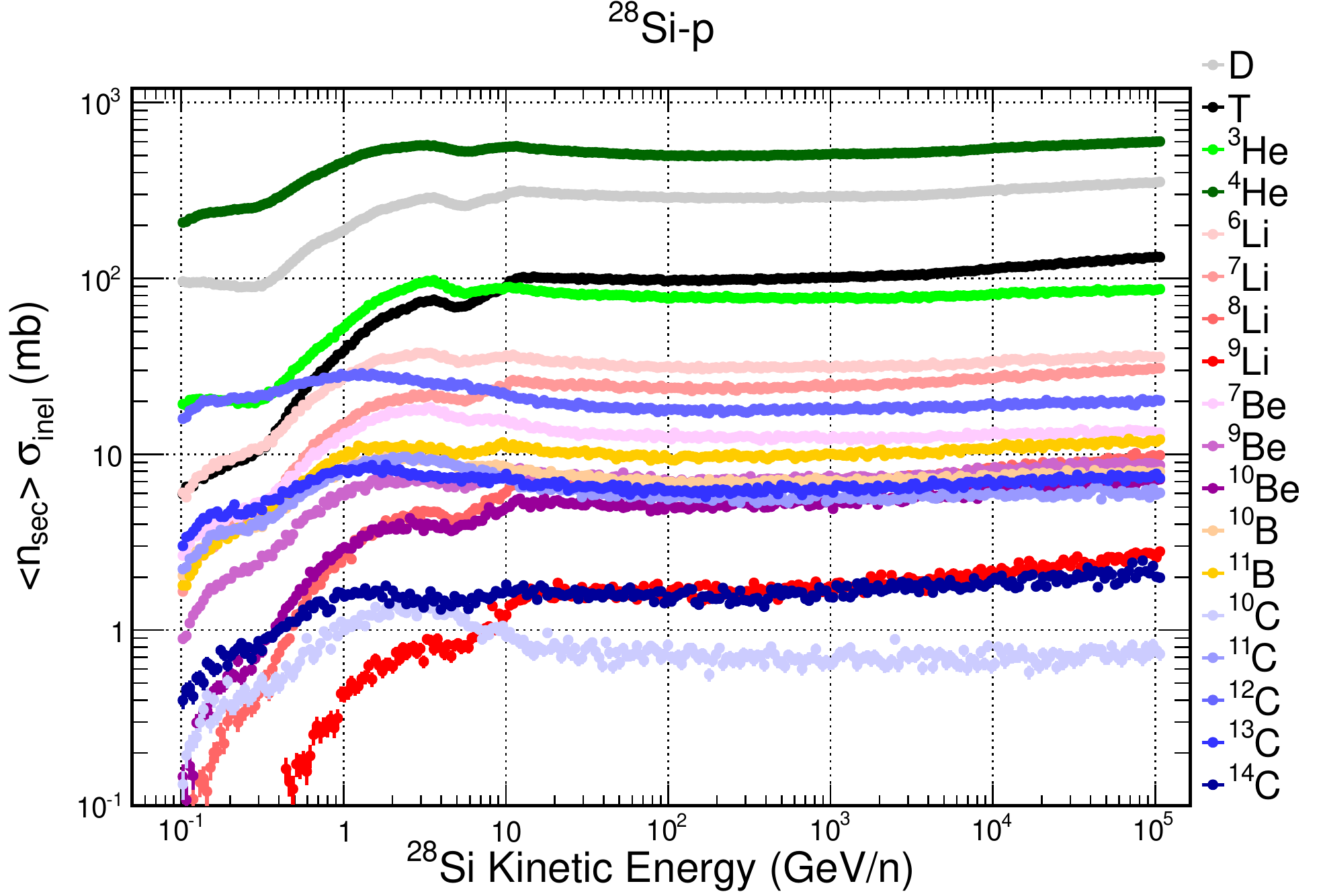} \\
\includegraphics[width=0.99\columnwidth,height=0.225\textheight,clip]{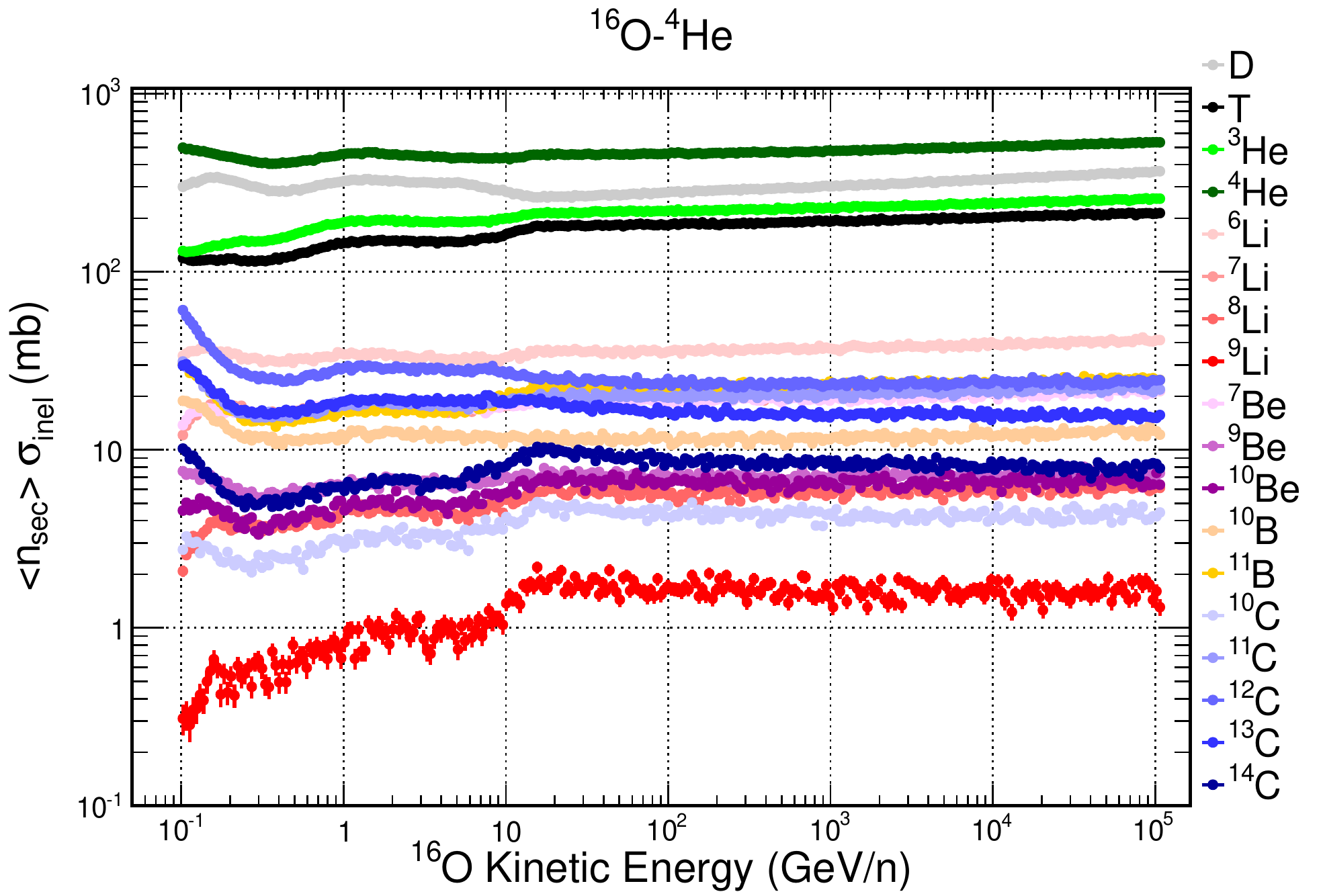} &
\includegraphics[width=0.99\columnwidth,height=0.225\textheight,clip]{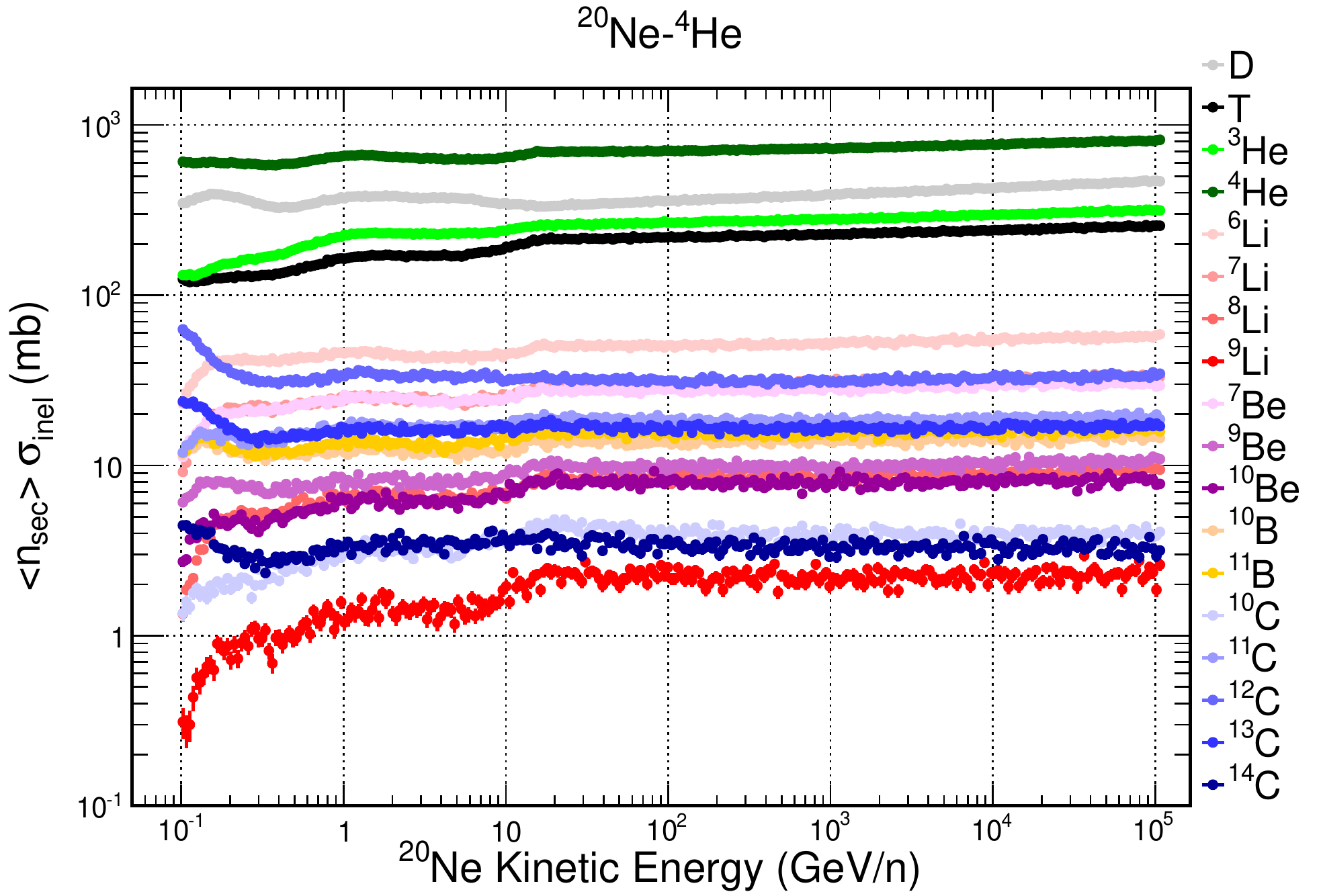} \\
\includegraphics[width=0.99\columnwidth,height=0.225\textheight,clip]{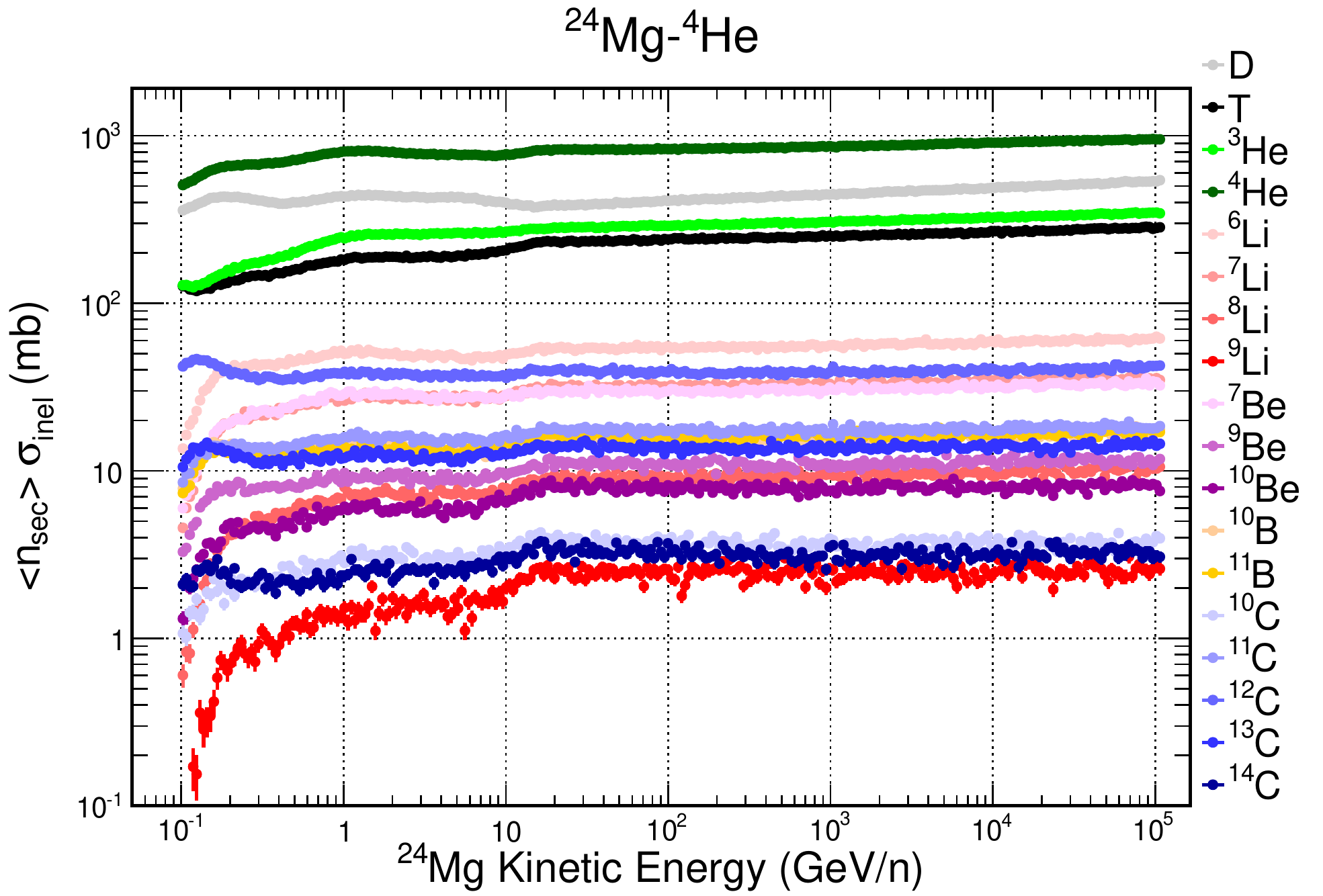} &
\includegraphics[width=0.99\columnwidth,height=0.225\textheight,clip]{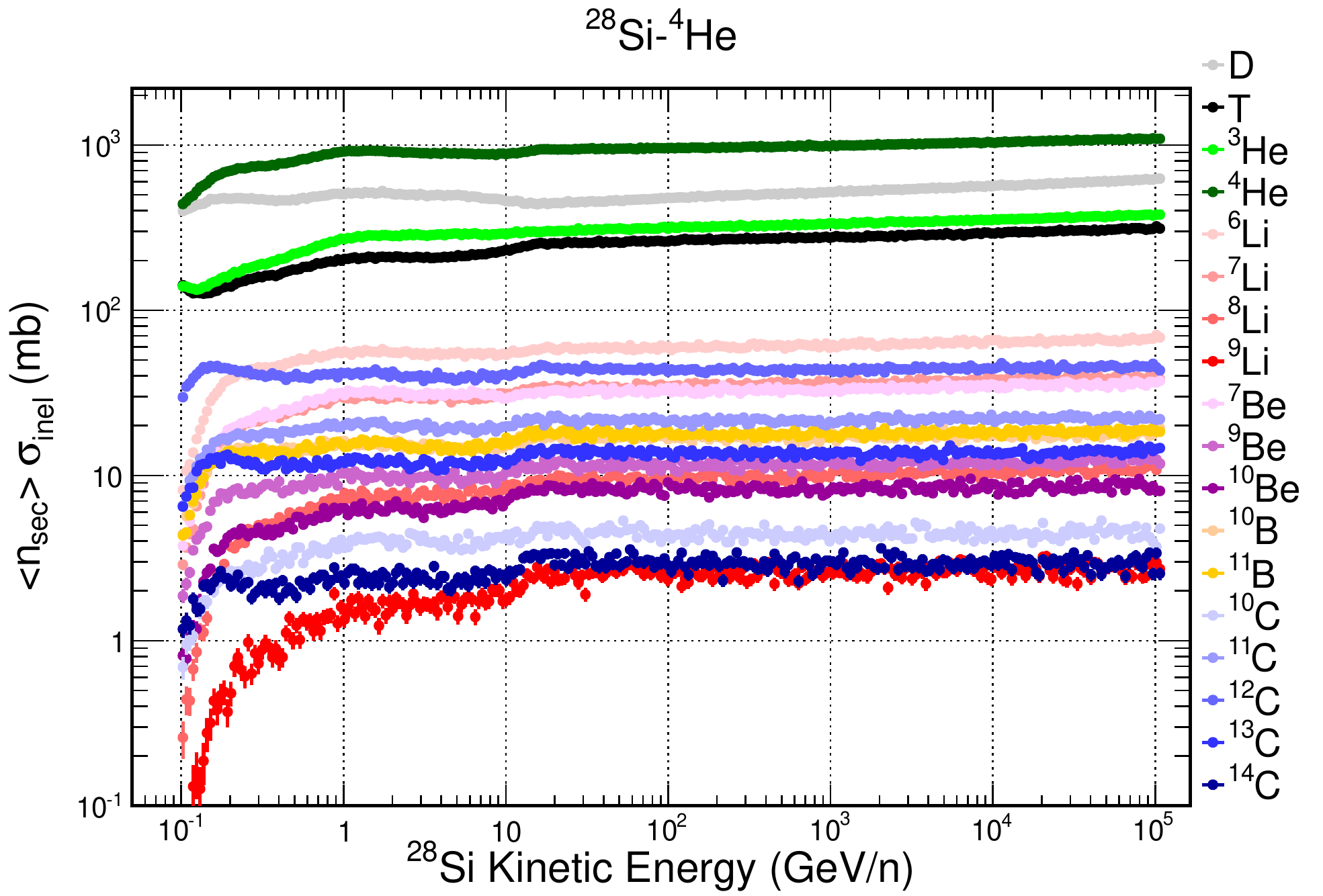} \\
\end{tabular}
\caption{Inclusive cross sections for the production of spallation nuclei in collisions of
$^{16}O$, $^{20}Ne$, $^{24}Mg$, and $^{28}Si$ with p and $^{4}He$ nuclei.
The plots show the cross sections for the production of Deuteron (gray markers), Triton (black markers) 
and for the isotopes of $He$ ($^{3}He$ and $^{4}He$, green markers), 
$Li$ ($^{6}Li$, $^{7}Li$, $^{8}Li$ and $^{9}Li$, red markers), $Be$ ($^{7}Be$ and $^{9}Be$, magenta markers), 
$B$ ($^{10}B$ and $^{11}B$, orange markers) and 
$C$ ($^{10}C$, $^{11}C$, $^{12}C$, $^{13}C$ and $^{14}C$, blue markers). 
Lighter (darker) color shades correspond to lighter (heavier) isotopes.}
\label{FigSiHeXsecDPMApx}
\end{figure*}


\begin{thebibliography}{99}
\bibitem{Putze:2010zn}
  A.~Putze, L.~Derome and D.~Maurin,
  Astron.\ Astrophys.\  {\bf 516} (2010) A66
  [arXiv:1001.0551 [astro-ph.HE]].

\bibitem{DiBernardo:2009ku} 
  G.~Di Bernardo, C.~Evoli, D.~Gaggero, D.~Grasso and L.~Maccione,
  Astropart.\ Phys.\  {\bf 34}, 274 (2010)
  [arXiv:0909.4548 [astro-ph.HE]].

 
\bibitem{Trotta:2010mx} 
  R.~Trotta, G.~Johannesson, I.~V.~Moskalenko, T.~A.~Porter, R.~R.~de Austri and A.~W.~Strong,
  Astrophys.\ J.\  {\bf 729}, 106 (2011)
  [arXiv:1011.0037 [astro-ph.HE]].

 
\bibitem{Atwood2009} W.B.~Atwood at al. (The Fermi LAT Collaboration), Astrophysical Journal 697 (2009) 1071


\bibitem{hess} F. Aharonian et al. (The H.E.S.S. Collaboration), Astroparticle Physics 22 (2004) 109
\bibitem{magic} J. Aleksi\'{c} et al. (The MAGIC Collaboration), Astroparticle Physics 35 (2012) 435
\bibitem{veritas} V.A.~Acciari et al. (VERITAS Collaboration), Astrophysical Journal 679 (2008) 1427
\bibitem{cta} M.~Actis et al. (The CTA Consortium), Exp. Astron. 32 (2011) 193

\bibitem{FermiLAT:2012aa}
  M.~Ackermann {\it et al.} [Fermi-LAT Collaboration],
  Astrophys.\ J.\  {\bf 750} (2012) 3

\bibitem{gradient} C.Evoli et al., Physical Review Letters 108, 21 (2012) 211102
\bibitem{variabledelta} D.Gaggero et al., submitted to Physical Review Letters [arXiv:1411.7623 [astro-ph]]

\bibitem{Ackermann:2014ula} 
  M.~Ackermann {\it et al.}  [Fermi-LAT Collaboration],
  Phys.\ Rev.\ Lett.\  {\bf 112}, 151103 (2014)
  [arXiv:1403.5372 [astro-ph.HE]].

\bibitem{Adriani:2011cu}
  O.~Adriani {\it et al.}  [PAMELA Collaboration],
  Science {\bf 332} (2011) 69
  [arXiv:1103.4055 [astro-ph.HE]].

\bibitem{Aguilar:2015ooa} 
  M.~Aguilar [AMS Collaboration],
  Phys.\ Rev.\ Lett.\  {\bf 114}, no. 17, 171103 (2015).
  
\bibitem{ams02he}  M.~Aguilar [AMS Collaboration], Phys.\ Rev.\ Lett.\ {\bf 115}, 211101 (2015).

\bibitem{Adriani:2008zr}
  O.~Adriani {\it et al.}  [PAMELA Collaboration],
  Nature {\bf 458} (2009) 607
  [arXiv:0810.4995 [astro-ph]].

\bibitem{Adriani:2013uda}
  O.~Adriani {\it et al.}  [PAMELA Collaboration],
  Phys.\ Rev.\ Lett.\  {\bf 111} (2013) 8,  081102
  [arXiv:1308.0133 [astro-ph.HE]].

\bibitem{Adriani:2013as} 
  O.~Adriani, G.~C.~Barbarino, G.~A.~Bazilevskaya, R.~Bellotti, M.~Boezio, E.~A.~Bogomolov, M.~Bongi and V.~Bonvicini {\it et al.},
  Astrophys.\ J.\  {\bf 765}, 91 (2013)
  [arXiv:1301.4108 [astro-ph.HE]].

\bibitem{Adriani:2013tif}
  O.~Adriani {\it et al.}  [PAMELA Collaboration],
  Astrophys.\ J.\  {\bf 770} (2013) 2
  [arXiv:1304.5420 [astro-ph.HE]].

\bibitem{Adriani:2011xv}
  O.~Adriani {\it et al.}  [PAMELA Collaboration],
  Phys.\ Rev.\ Lett.\  {\bf 106} (2011) 201101
  [arXiv:1103.2880 [astro-ph.HE]].

\bibitem{Adriani:2010rc}
  O.~Adriani {\it et al.}  [PAMELA Collaboration],
  Phys.\ Rev.\ Lett.\  {\bf 105} (2010) 121101
  [arXiv:1007.0821 [astro-ph.HE]].

\bibitem{Aguilar:2013qda}
  M.~Aguilar {\it et al.}  [AMS Collaboration],
  Phys.\ Rev.\ Lett.\  {\bf 110} (2013) 14,  141102.


\bibitem{Accardo:2014lma} 
  L.~Accardo {\it et al.}  [AMS Collaboration],
  Phys.\ Rev.\ Lett.\  {\bf 113}, 121101 (2014).

\bibitem{Aguilar:2014mma}
  M.~Aguilar {\it et al.}  [AMS Collaboration],
  Phys.\ Rev.\ Lett.\  {\bf 113} (2014) 121102.

\bibitem{ams02_icrc2013} http://www.ams02.org/2013/07/new-results-from-ams-presented-at-icrc-2013/
\bibitem{ams02_cern2015} http://www.ams02.org/2015/04/ams-days-at-cern-and-latest-results-from-the-ams-experiment-on-the-international-space-station/ \\
https://indico.cern.ch/event/381134/other-view?view=standard\#20150415


\bibitem{stecker1970} F.W.~Stecker, Astrophysics and Space Science 6 (1970) 377
\bibitem{dermer1986} C.D.~Dermer, Astrophysical Journal 307 (1986) 47
\bibitem{moska1998} I.V.~Moskalenko and A.W.~Strong, Astrophysical Journal 493 (1998) 694
\bibitem{pdg} K.A. Olive et al. (Particle Data Group), Chin. Phys. C, 38 (2014) 090001
\bibitem{huang2007} C.-Y.~Huang et al, Astroparticle Physics 27 (2007) 429
\bibitem{mori1997} M.~Mori, Astrophysical Journal 225 (1997) 225
\bibitem{kamae2006} T.~Kamae et al., Astrophysical Journal 708 (2006) 647
\bibitem{kelner2006} S.R.~Kelner et al., Phys. Rev. D 74 (2006) 034018
\bibitem{mori2009} M. Mori, Astroparticle Physics 31 (2009) 341
\bibitem{kache2012} M.~Kachelrie\ss{} and S.~Ostapchenko, Phys. Rev. D 84 (2012) 043004
\bibitem{pythia} T. Sjostrand, Computer Physics Commun. 82 (1994) 74

\bibitem{dpmjetfluka} S. Roesler, R. Engel and J. Ranft, \textit{The Monte Carlo event generator DPMJET-III} Proc. Monte Carlo 2000 Conference, Lisbon, October 23-26 2000, A. Kling, F. Bar~ao, M. Nakagawa, L. T'avora, P. Vaz eds., Springer-Verlag Berlin, p. 1033-1038, 2001


\bibitem{sybill} R. S. Fletcher, T.K. Gaisser, P. Lipari, and T. Stanev, Phys. Rev. D 50 (1994) 5710
\bibitem{qgsjet} S.~Ostapchenko, Phys. Rev. D 83 (2011) 014018


\bibitem{fluka2} A. Ferrari, P.R. Sala, A. Fass\`{o}, and J. Ranft, \textit{FLUKA: a multi-particle transport code} CERN-2005-10 (2005), INFN/TC\_05/11, SLAC-R-773

\bibitem{flukaweb} \url{http://www.fluka.org}

\bibitem{bohlen} T. Bohlen, F. Cerutti, M.P.W. Chin, A. Fass\`{o}, A. Ferrari, P.G.
Ortega, A. Mairani, P.R. Sala, G. Smirnov, and V. Vlachoudis, \textit{The FLUKA
Code: Developments and Challenges for High Energy and Medical Applications},
Nuclear Data Sheets 120 (2014) 211-214

\bibitem{ballarini} F. Ballarini et al., \textit{Nuclear models in FLUKA: a review}, Proceedings 10th International Conference on 
Nuclear Reaction Mechanisms, Varenna, Italy, June 9-13, 2003, edited by E. Gadioli, Ric. Scient. ed Educ. Perm. S122, Univ. degli
Studi di Milano (2003) 579-591, \url{www.mi.infn.it/~gadioli/Varenna2006/Proceedings/Ferrari_A.pdf}

\bibitem{batt2007} G. Battistoni et al., \textit{The FLUKA code: Description and benchmarking}, Proceedings of the Hadronic Shower Simulation Workshop 2006, 
Fermilab 6-8 September 2006, M. Albrow, R. Raja eds., AIP Conference Proceeding 896 (2007) 31-49


\bibitem{batt2015} G. Battistoni, T. Boehlen, F. Cerutti, P. Wai Chin, L.
S. Esposito, A. Fass\`{o}, A. Ferrari, A. Lechner, A.
Empl, A. Mairani, A. Mereghetti, P. G. Ortega, J.
Ranft, S. Roesler, P. R. Sala, V. Vlachoudis, and G.
Smirnov, \textit{Overview of the FLUKA code}, Annals of Nuclear Energy 82 (2015)
10-18



\bibitem{peanut1} A. Fass\`{o}, A. Ferrari, J. Ranft, P.R. Sala, \textit{FLUKA: status and prospective for hadronic applications}, in: A. Kling, F. Bar\~{a}o, M. Nakagawa, L. T\'{a}vora, P. Vaz (Eds.) Proceedings of the Monte Carlo 2000 Conference, Lisbon, October 23–26 2000, Springer-Verlag, Berlin, 2001, p. 955–960 
\bibitem{peanut2} G. Battistoni et al. \textit{Recent developments in the FLUKA nuclear reaction models} Proc. 11th Int. Conf. on Nucl. React. Mech. (Varenna, Italy, 12–16 June 2006) 483
\bibitem{ferrari1996} A. Ferrari and P.R. Sala \textit{The Physics of High Energy Reactions},
 Proc. Workshop on Nuclear Reaction Data and Nuclear Reactors Physics, Design and Safety, International Centre for Theoretical Physics, 
Miramare-Trieste, Italy, 15 April-17 May 1996, Ed. A. Gandini and G. Reffo, World Scientific, p. 424 (1998) \url{https://cds.cern.ch/record/682497}

\bibitem{batt2002} G. Battistoni, A. Ferrari, T. Montaruli and P.R. Sala, Astroparticle Physics 17 (2002) 477
\bibitem{batt2003} G. Battistoni, A. Ferrari, T. Montaruli and P.R. Sala, Astroparticle Physics 19 (2003) 269
\bibitem{batt2005} G. Battistoni, A. Ferrari, T. Montaruli and P.R. Sala, Astroparticle Physics 23 (2005) 526

\bibitem{Ferriere:2001rg} 
  K.~M.~Ferriere,
  Rev.\ Mod.\ Phys.\  {\bf 73}, 1031 (2001)
  [astro-ph/0106359].
\bibitem{ramaty1979} R.~Ramaty, B.~Kozlovsky and R.E.~ Lingenfelter, Astrophysical Journal Supplement Series 40 (1979) 487 


\bibitem{ander2004} V. Andersen et al., Advances in Space Research 34 (2004) 1302
\bibitem{Sor89} H.~Sorge, H. Stoecker and W. Greiner, Ann. Phys. 192 (1989) 266
\bibitem{Sor89a} H.~Sorge, H. Stoecker and W. Greiner, Nucl. Phys. A498 (1989) 567
\bibitem{Sor95} H.~Sorge, Phys. Rev. C52 (1995) 3291

\bibitem{sihver} L.~Sihver et al., \textit{Benchmarking of calculated projectile fragmentation cross-sections using the 3-D, 
MC codes PHITS, FLUKA, HETC-HEDS, MCNPX HI, and NUCFRG2 }, Acta Astronautica 63 (2008) 865

\bibitem{braun} H.H.~Braun et al., \textit{Hadronic and electromagnetic fragmentation of ultrarelativistic heavy ions at LHC}, 
Phys. Rev. Special Topics Accelerators and Beams 17, 021006 (2014)

\bibitem{fed2015} A. Fedynitch and R. Engel, \textit{Revision of the high energy hadronic interaction models PHOJET/DPMJETIII}, 
Proceedings of the 14th International Conference on Nuclear Reaction Mechanisms, edited by F. Cerutti, M.
Chadwick, A. Ferrari, T. Kawano and P. Schoofs, CERN-Proceedings-2015-001 (CERN, Geneva, 2015), 291-299, \url{https://cds.cern.ch/record/2114737}

\bibitem{Alt:2006fr}
  C.~Alt {\it et al.} [NA49 Collaboration],
  Eur.\ Phys.\ J.\ C {\bf 49} (2007) 897
  doi:10.1140/epjc/s10052-006-0165-7
  [hep-ex/0606028].

\bibitem{Anticic:2010yg}
  T.~Anticic {\it et al.} [NA49 Collaboration],
  Eur.\ Phys.\ J.\ C {\bf 68} (2010) 1
  doi:10.1140/epjc/s10052-010-1328-0
  [arXiv:1004.1889 [hep-ex]].


\bibitem{na49}  B.~Baatar {\it et al.} [NA49 Collaboration],
  Eur.\ Phys.\ J.\ C {\bf 73} (2013) no.4,  2364
  doi:10.1140/epjc/s10052-013-2364-3
  [arXiv:1207.6520 [hep-ex]].

\bibitem{Ballarini:2004ai}
  F.~Ballarini {\it et al.},
  AIP Conf.\ Proc.\  {\bf 769} (2005) 1197.
  doi:10.1063/1.1945222

\bibitem{ferrari1997} A.~Ferrari and P.R.~Sala \textit{Intermediate and High Energy Models in FLUKA: Improvements, Benchmarks and Applications}, 
Proceedings of the International Conference on Nuclear Data for Science and Technology (NDST-97), International Centre for 
Theoretical Physics, Miramare-Trieste, Italy, May 19-24 1997, G. Reffo, A. Ventura and C. Grandi eds., Italian Physical 
Society  publ., ISBN 88-7794-114-6, Bologna, Part I, (1997) 247-253, available at \url{http://www.fluka.org/content/publications/1997_trieste.pdf}

\bibitem{Dermer2012} C.~D.~Dermer, Phys. Rev. Lett. 109 (2012), 091101

\bibitem{Evoli:2008dv}
  C.~Evoli, D.~Gaggero, D.~Grasso and L.~Maccione,
  JCAP {\bf 0810} (2008) 018
  [arXiv:0807.4730 [astro-ph]].

\bibitem{Gaggero:2013rya}
  D.~Gaggero, L.~Maccione, G.~Di Bernardo, C.~Evoli and D.~Grasso,
  Phys.\ Rev.\ Lett.\  {\bf 111} (2013) 2,  021102
  [arXiv:1304.6718 [astro-ph.HE]].

\bibitem{GA}
L.J. Gleeson and W.I. Axford, ApJ 149, L115 (1967) and 
L.J. Gleeson and W.I. Axford, ApJ 154, 1011 (1968).

\bibitem{helioprop}
L.~Maccione,
  Phys.\ Rev.\ Lett.\  {\bf 110} (2013) 8,  081101
  [arXiv:1211.6905 [astro-ph.HE]].
  
  See also: D.~Gaggero, D.~Grasso, L.~Maccione, G.~Di Bernardo and C.~Evoli,
  Phys.\ Rev.\ D {\bf 89} (2014) 083007
  [arXiv:1311.5575 [astro-ph.HE]].


\bibitem{Kachelriess:2014mga}
  M.~Kachelriess, I.~V.~Moskalenko and S.~S.~Ostapchenko,
  Astrophys.\ J.\  {\bf 789} (2014) 136
  [arXiv:1406.0035 [astro-ph.HE]].
  
\bibitem{jm2015} J.M.~Casandjian, Astrophysical Journal \textbf{806} (2015) 240
  [arXiv:1506.00047 [astro-ph.HE]]

\bibitem{DiBernardo:2012zu}
  G.~Di Bernardo, C.~Evoli, D.~Gaggero, D.~Grasso and L.~Maccione,
  JCAP {\bf 1303} (2013) 036
  [arXiv:1210.4546 [astro-ph.HE]]
  
\bibitem{Orlando:2013ysa}
  E.~Orlando and A.~Strong,
  MNRAS {\bf 436} (2013) 2127
  [arXiv:1309.2947 [astro-ph.GA]]
  

\bibitem{Evoli:2012ha} 
  C.~Evoli, D.~Gaggero, D.~Grasso and L.~Maccione,
  Phys.\ Rev.\ Lett.\  {\bf 108}, 211102 (2012)
  [arXiv:1203.0570 [astro-ph.HE]]
  
\bibitem{ams_preliminary}
  \url{http://www.ams02.org/2015/04/}

\bibitem{Evoli:2015}
  C.~Evoli, D.~Gaggero, and D.~Grasso
  [arXiv:1504.05175 [astro-ph.HE]]

%
\bibitem{Tavakoli:2011wz} 
  M.~Tavakoli, I.~Cholis, C.~Evoli and P.~Ullio, JCAP \textbf{01} (2014) id. 017
  arXiv:1110.5922 [astro-ph.HE]
  
%
\bibitem{Cirelli:2014lwa} 
  M.~Cirelli, D.~Gaggero, G.~Giesen, M.~Taoso and A.~Urbano,
  JCAP \textbf{12} (2014) id. 045  arXiv:1407.2173 [hep-ph] 

\bibitem{healpix} K.M. G\'{o}rski, E. Hivon, A.J. Banday, B.D. Wandelt, F.K. Hansen, M. Reinecke, and M. Bartelmann, ApJ \textbf{622} (2005) 759-771; \url{http://healpix.jpl.nasa.gov}

\bibitem{stone2013} E.C.~Stone et al., Science \textbf{341} (2013) 150

\bibitem{acecris} K.A.~Lave et al., ApJ \textbf{770} (2013) 117

\bibitem{heao3} J.J.~Engelmann et al., A\&A \textbf{233} (1990) 96

\bibitem{ams01} M. Aguilar et al. (AMS Collaboration), Phys. Reports 366, 331 (2002)

\bibitem{crdb} D. Maurin, F. Melot and R. Taillet, A\&A \textbf{569} (2014) A32 [arXiv:1302.5525]

\bibitem{galprop_website} \url{http://galprop.stanford.edu/}

\bibitem{galprop1998} A.W.Strong and I.V. Moskalenko, ApJ, \textbf{509}, Issue 1, pp. 212-228 (1998)

\bibitem{galprop2002} I.V. Moskalenko {\it et al.}, ApJ, \textbf{565}, Issue 1, pp. 280-296 (2002)

\bibitem{galprop2007} A.W. Strong, I.V. Moskalenko, and V.S. Ptuskin., Ann.Rev.Nucl.Part.Sci., \textbf{57}:285–327 (2007)

\bibitem{galprop2012} M. Ackermann et al. (Fermi-LAT collaboration), ApJ \textbf{750} (2012)




\end{thebibliography}
\end{document}